\documentclass[a4paper,fleqn,usenatbib]{mnras}

\usepackage{savesym}
\usepackage{txfonts}
\savesymbol{iint}
\savesymbol{iiint}
\savesymbol{iiiint}
\savesymbol{idotsint}

\usepackage[T1]{fontenc}
\usepackage{ae,aecompl}

\usepackage{graphicx}	% Including figure files
\usepackage{subfig}
\usepackage{amsmath}	% Advanced maths commands
\restoresymbol{AMS}{iint}
\restoresymbol{AMS}{iiint}
\restoresymbol{AMS}{iiiint}
\restoresymbol{AMS}{idotsint}
\usepackage{amssymb}	% Extra maths symbols
\usepackage[shortlabels]{enumitem}
\usepackage{xcolor}

\definecolor{darkgreen}{RGB}{0,128,0}

\title[Tidal disruptions by SMBH binaries]{Gravitational interactions of stars with supermassive black hole binaries. I. Tidal disruption events}

\author[S. Darbha et al.]{
Siva Darbha,$^{1}$\thanks{E-mail: siva.darbha@berkeley.edu}
Eric R. Coughlin,$^{2}$\thanks{Einstein fellow}
Daniel Kasen$^{1,2,3}$
and Eliot Quataert$^{1,2}$
\\
$^{1}$Department of Physics, University of California, Berkeley, Berkeley, CA 94720, USA\\
$^{2}$Department of Astronomy and Theoretical Astrophysics Center, University of California, Berkeley, Berkeley, CA 94720, USA\\
$^{3}$Nuclear Science Division, Lawrence Berkeley National Laboratory, Berkeley, CA 94720, USA
}

\date{Accepted XXX. Received YYY; in original form ZZZ}

\pubyear{2018}

\begin{document}
\label{firstpage}
\pagerange{\pageref{firstpage}--\pageref{lastpage}}
\maketitle

\begin{abstract}

\noindent
Stars approaching supermassive black holes (SMBHs) in the centers of galaxies can be torn apart by strong tidal forces. We study the physics of tidal disruption by a binary SMBH as a function of the binary mass ratio $q = M_2 / M_1$ and separation $a$, exploring a large set of points in the parameter range $q \in [0.01, 1]$ and $a/r_{t1} \in [10, 1000]$. We simulate encounters in which field stars approach the binary from the loss cone on parabolic, low angular momentum orbits. We present the rate of disruption and the orbital properties of the disrupted stars, and examine the fallback dynamics of the post-disruption debris in the ``frozen-in'' approximation. We conclude by calculating the time-dependent disruption rate over the lifetime of the binary. Throughout, we use a primary mass $M_1 = 10^6 M_\odot$ as our central example. We find that the tidal disruption rate is a factor of $\sim 2 - 7$ times larger than the rate for an isolated BH, and is independent of $q$ for $q \gtrsim 0.2$. In the ``frozen-in'' model, disruptions from close, nearly equal mass binaries can produce intense tidal fallbacks: for binaries with $q \gtrsim 0.2$ and $a/r_{t1} \sim 100$, roughly $\sim 18 - 40 \%$ of disruptions will have short rise times ($t_\textrm{rise} \sim 1 - 10$ d) and highly super-Eddington peak return rates ($\dot{M}_\textrm{peak} / \dot{M}_\textrm{Edd} \sim 2 \times 10^2 - 3 \times 10^3$).

\end{abstract}

% Select between one and six entries from the list of approved keywords. Don't make up new ones.

%%%%%%%%%%%%%%%%%%%%%%%%%%%%%%%%%%%%%%%%%%%%%%%%%%

%%%%%%%%%%%%%%%%% BODY OF PAPER %%%%%%%%%%%%%%%%%%

\section{Introduction}

A star that nears a galactic supermassive black hole (SMBH) is stretched into a stream of debris when the tidal gravity of the black hole overwhelms the self-gravity of the star \citep{hills75, kochanek94, coughlin16b}. If the star originates beyond the sphere of influence of the SMBH, which we expect for most stars \citep{lightman77}, then roughly half of the debris remains bound and returns to the black hole \citep{lacy82, rees88, evans89}. The returning debris dissipates its kinetic energy through shocks \citep{rees88}, forms an accretion disk \citep{cannizzo90, ramirez-ruiz09, shiokawa15, bonnerot16, hayasaki16b}, and emits radiation that illuminates the center of the host galaxy for months to years. The sudden brightening from these tidal disruption events (TDEs) occurs independently of the gaseous environment in the galactic center, and can thus reveal otherwise-quiescent SMBHs. Observations have revealed dozens of TDES (for a review, see \citealt{komossa15, auchettl17}; for recent detections, see \citealt{blagorodnova17, hung17}), and upcoming wide-field surveys, such as LSST, will likely detect hundreds to thousands more \citep{strubbe09}.

Tidal disruptions by isolated SMBHs have been studied extensively, both analytically (e.g., \citealt{rees88, lodato09, strubbe09, coughlin14, coughlin16b}) and numerically (e.g., \citealt{bicknell83, evans89, frolov94, guillochon13, coughlin15, tejeda17}). In contrast, disruptions by black holes in binary systems have received less attention, even though binary SMBHs are likely pervasive throughout the Universe as a result of galaxy mergers \citep{bell06}, and frequently interact with their field stars \citep{begelman80}. Indeed, the interaction between the two alters the binary orbital parameters, such as the separation and eccentricity \citep{begelman80, mikkola92, quinlan96, yu02, sesana06, sesana07, chen11, wegg11}, and ejects hypervelocity stars (HVSs) \citep{yu03, sesana06}. Some stars should approach close enough to the two black holes to become tidally disrupted, which can produce observationally-distinct TDEs if the binary orbital period is comparable to the fallback time of the debris \citep{coughlin17}.

The few studies of TDEs from binary SMBHs have outlined some of their important features. \citet{chen08} showed that the TDE rate can fall well below the canonical value of $10^{-4} - 10^{-5}$ yr$^{-1}$ galaxy$^{-1}$ \citep{frank76} if the disrupted stars are on hyperbolic orbits, while \citet{chen09} showed that the rate can be greatly enhanced if the stars are bound to the binary, which \citet{wegg11} and \citet{chen11} confirmed with more detailed simulations. \citet{li15} investigated the role of the eccentric Kozai-Lidov mechanism \citep{kozai62, naoz16} in the orbital evolution of bound stars, and deduced that stars could reach nearly-parabolic trajectories leading to a disruption, expanding on earlier work by \citet{ivanov05}. \citet{liu09} and \citet{ricarte16} used N-body simulations to model the fallback of tidally-disrupted debris onto the black holes, and showed that the accretion rate exhibits periodic dips; a signature of this form may have been detected in a quiescent galaxy \citep{liu14}.

Recently, \citet{coughlin17} performed the first end-to-end study of TDEs from SMBH binaries, from the pre-disruption stellar dynamics to the post-disruption hydrodynamics, for a range of binary mass ratios. Distinct from previous investigations, they focused on binaries that have exhausted their bound stars through stellar scattering, and thus modeled the incident stars on initially unbound, parabolic orbits corresponding to the loss cone's  ``pinhole'' (or ``full loss cone'') regime \citep{frank76,lightman77}, similar to many isolated SMBH systems. They followed the long-lived, chaotic orbits of the stars in the binary potential, and found that 1) the disruption rate increases marginally over the single black hole case, and 2) the disrupted energies and angular momenta differ appreciably from the input values. These differences impact the ensuing accretion event, including the time to peak accretion and the total accreted mass. Hydrodynamic simulations revealed that the behavior of these features is distinct from the gravitational perturbations induced by the secondary on the tidally-disrupted debris.

Since most binary SMBHs are likely the product of galaxy mergers, the galactic properties and merger details will determine the loss cone dynamics \citep{milosavljevic01,yu02,merritt05}. In a dry merger, following the dynamical friction phase, the binary and surrounding stars will settle into a steady-state configuration with a fully populated loss cone if the stellar relaxation time is short compared to the binary coalescence time \citep{merritt05}. The merger process is complicated and still a focus of study, though simulations have shown that merging galaxies with collisionally relaxed nuclei can lead to binaries with full loss cones \citep{gualandris12}.

Though \citet{coughlin17} explored a representative range of binary mass ratios, they restricted their attention to a single binary separation, and thus the impact of the separation on the properties of the disrupted stars was unexplored. However, a binary SMBH in a center of a galaxy contracts over its lifetime by ejecting stars \citep{begelman80} and interacting with tidally-disrupted debris \citep{rafikov13, goicovic16, goicovic16b}, and the change in binary separation can influence the rates and features of TDEs.

In this paper, we extend the work of \citet{coughlin17} and investigate the properties of stars disrupted by a binary SMBH over a range of binary separations and mass ratios. We simulate a large number of encounters between a star and a binary SMBH in the point particle limit and under the assumptions of the circular restricted three body problem (CRTBP), with the stars initialized on parabolic trajectories. We compute the disruption rate, explore the orbital properties of the disrupted stars, use a simple analytic model to follow the debris, and summarize the dependence of these on the binary parameters. These results can motivate the initial conditions for hydrodynamic simulations of gas accretion onto binary SMBHs. We also record the properties of the ejected stars, which are the preponderant end-states of our three-body integrations, and use these to calculate the expected lifetime of the binary. In a companion paper, we plan to analyze the ejected stars themselves to compare them with past investigations and to provide an updated sample of statistics.

In Section \ref{sec:setup}, we describe the setup of our simulations. In Section \ref{sec:tdes}, we present the rate of disruption and the orbital properties of the disrupted stars. We then use the ``frozen-in'' approximation to model the dynamics of the post-disruption debris and calculate the rise time and peak in the fallback rate. We synthesize our results by calculating the lifetime of the binary SMBH and the time-dependent disruption rate. We conclude and summarize our findings in Section \ref{sec:conclusion}.

\section{Simulation Setup}
\label{sec:setup}

A binary SMBH becomes ``hard'' at roughly the separation \citep{quinlan96}
\begin{equation}
a_h = \frac{G M_1 M_2}{4 (M_1 + M_2) \sigma^2}
\end{equation}
where $M_1$ and $M_2$ are the masses of the primary and secondary, and $\sigma$ is the one-dimensional velocity dispersion of the stars in the surrounding galaxy core. As the binary nears this distance, dynamical friction ceases to be efficient and further contraction must occur through another mechanism, notably stellar scattering \citep{begelman80,mikkola92,quinlan96,sesana06,sesana07,kelley17}. The binary will eventually deplete its bound stars through scattering or disruption, and further stellar interactions arise from loss cone scattering.

Stars that can potentially be disrupted must approach the binary through the loss cone, which is the energy - (low) angular momentum phase space region in which stars can be tidally disrupted (for an isolated SMBH, see \citealt{frank76,lightman77,cohn78,magorrian99}; for a binary SMBH, see \citealt{yu02,yu03}). Stars that approach the binary from large distances do so in the ``pinhole'' (or ``full loss cone'') regime, in which the loss cone is narrow and has a full phase space, leading to a uniformly distributed range of pericenter distances. For the binary loss cone to remain full, it must be repopulated on a timescale short compared to the binary orbital period, most likely by two-body relaxation (though other processes may be important in some galaxies, such as secular evolution in a nuclear disk, as examined by \citealt{madigan18}). These ``pinhole'' stars are responsible for the majority of tidal disruptions by central black holes with masses $M_\bullet \sim 10^6 M_\odot$ \citep{stone16}; we focus on this domain in our study.

Stars injected in the pinhole regime are initially on mildly hyperbolic orbits and are ``slow intruders'' \citep{hills89}, since their velocity far from the binary is much less than the binary velocity, $v_\infty \sim \sigma \ll v_\textrm{bin}$ (see below for the physical parameters that support this). As a result, one can treat the stars as effectively incident on parabolic orbits \citep{quinlan96}. In addition, the binary will expel most of the incident stars with enhanced velocities and its binding energy will increase \citep{quinlan96}. Since its binding energy is much larger than the energy imparted to the stars, the binary does not evolve appreciably in a single encounter. In this case, the system can be modeled under the assumptions of the circular restricted three-body problem (CRTBP), in which the primary and secondary have masses $M_1$ and $M_2$ and move on circular orbits about their common center of mass, and the incident stars have masses $M_* \ll M_1, M_2$ and do not affect the binary evolution. There are no closed-form analytic solutions for the general motion of a star in the CRTBP, so one must perform numerical scattering experiments to study the tidal disruption statistics.

Although the binary will eject most stars, it will tidally disrupt a subset of them. A star approaching a massive BH is tidally disrupted when the tidal force from the BH equals the gravitational self-force of the star \citep{hills75}. This occurs at the BH's tidal radius $r_t \simeq R_* \left( M_\bullet / M_* \right)^{1/3}$, where $M_\bullet$ is the mass of the BH, and $M_*$ and $R_*$ are the mass and radius of the star. We define this as the criterion for disruption, though there are additional dependences on stellar structure \citep{guillochon13,mainetti17}.

We use Mathematica to simulate stars incident on a binary SMBH in the framework of the CRTBP, in the point particle limit and using Newtonian gravitational potentials. The binary parameters set the relevant scales in the problem, namely the semimajor axis $a$ and the total mass $M = M_1 + M_2$, which lead to the binary specific energy $\epsilon = GM/a$ and specific angular momentum $\ell = \sqrt{GMa}$, and the time scale $T = \sqrt{a^3/GM}$ (which is a factor of $2\pi$ off from the binary orbital period, $P = 2\pi T$). We simulate our scattering experiments in the units $G = M = a = 1$, normalizing to these scales. The binary is then described solely by two dimensionless quantities: the mass ratio $q = M_2/M_1$ and the primary's tidal radius $r_{t1}/a$. 

Table \ref{tab:scales} presents the various scales that we use to non-dimensionalize our variables in different parts of the paper, unless otherwise noted. In short, although we vary $r_{t1} = \tilde{r}_{t1}/\tilde{a}$ and $q$ in our simulations, in our results we ultimately interpret this as varying $a = \tilde{a}/\tilde{r}_{t1}$ and $M_2$ while holding $\tilde{r}_{t1}$ and $M_1$ fixed. In this paper, if the dimensional character of a variable is not clear from the context, then we write dimensioned variables with tildes on top and dimensionless ones without them.

\begin{table*}
\centering
\begin{tabular}{|c|c|c|c|c|c|}
\hline
Section & Length & Mass & Time & Specific energy ($\epsilon$) & Specific ang. mom. ($\ell$) \\
% & & & & $\epsilon = E / M_*$ & $\ell = L / M_*$ \\
\hline
Simulation (\ref{sec:setup}) & $a$ & $M = M_1 + M_2$ & $\sqrt{a^3 / GM}$ & $GM/a$ & $\sqrt{GMa}$ \\
\hline
Disruption (\ref{subsec:tderate}, \ref{subsec:properties}) & $r_{t1}$ & $M_1$ & $\sqrt{r^3_{t1} / G M_1}$ & $G M_1 / r_{t1}$ & $\sqrt{GM_1 r_{t1}}$ \\
 & $\left( a = \tilde{a} / r_{t1} \right)$ & & $\left( t_d = \tilde{t}_d / 2\pi \sqrt{r^3_{t1} / G M_1} \right)$ & & \\
\hline
Post-disruption (\ref{subsec:postdisruption}) & & $M_*$ & $\tau_0 = 2\pi GM_\bullet / (2\Delta\epsilon)^{3/2}$ & & \\
 & & $\left( m = \tilde{m}/M_* \right)$ & $\left( \tau = t / \tau_0 \right)$ & & \\
 \hline
Binary inspiral (\ref{subsec:inspiral}) & $r_{t1}$ & $M_1$ & $t_0 = r^4_{t1} c^5 / G^3 M^3_1$ & $GM_1 / r_{t1}$ & \\
 & $\left( a = \tilde{a} / r_{t1} \right)$ & & $\left( t = \tilde{t} / t_0 \right)$ & & \\
\hline
\end{tabular}
\caption{The quantities used to non-dimensionalize variables (with the dimensions length, mass, time, stellar specific energy, and stellar specific angular momentum) in different sections of the paper, unless stated otherwise. Throughout the paper, if the dimensional character of a variable is not clear from the context, then we write dimensioned variables with tildes on top and dimensionless ones without them. The parentheses show the definitions of some variables used in each section.}
\label{tab:scales}
\end{table*}

The setup and initial conditions for our scattering experiments are similar to those of \citet{coughlin17}. We set the origin of the coordinate system to the center of mass of the binary. A star's initial conditions are given by its specific energy $\epsilon$, specific angular momentum $\ell$ relative to the origin, and position. The stars begin on parabolic (zero energy) orbits with respect to the binary center of mass, and are distributed isotropically over a sphere of radius $r=50$. 
% The initial positional orbital elements $(\iota, \Omega, \omega)$ \citep{gravity14} are sampled to produce an isotropic distribution.
The (square of the) angular momentum of each star, and thus the pericenter distance, is uniformly sampled from the range $\ell^2 \in [0,4]$, corresponding to the pinhole regime \citep{frank76,lightman77,cohn78,magorrian99}. An integration terminates if the star crosses the tidal radius of one of the BHs, if it escapes to $r=100$, or if the simulation time reaches $t = 10^4$. In a departure from \citet{coughlin17}, we record information about both the disrupted \emph{and} ejected stars. We also explore a larger set of points in the parameter space spanned by $q \in [0.01, 1]$ and $r_{t1}/a \in [0.001, 0.1]$ ($a/r_{t1} \in [10, 1000]$), and simulate $5 \times 10^6$ encounters for each point. For comparison, we simulated a smaller number of encounters for a few points in our parameter space with the N-body code REBOUND using the IAS15 integrator \citep{rein12,rein15}, and found close agreement with our results.

For a primary mass $M_1 = 10^6 M_\odot$ and stars with solar parameters, the primary's tidal radius is $r_{t1} = 2.3$ $\mu$pc and the range of separations is $a \in [0.023, 2.3]$ mpc. At these scales ($M \sim 10^6 M_\odot$, $a \sim 10^{-3}$ pc), the binary velocity is $v_\textrm{bin} \sim \sqrt{GM/a} \sim 2000$ km/s, and assuming an environment like the Galactic Center, the stellar velocities in the bulge are $\sigma \sim 100$ km$\cdot$s$^{-1}$ \citep{gultekin09}. The velocities thus satisfy $\sigma \ll v_\textrm{bin}$, validating the use of parabolic orbits. To verify this approximation, we ran additional simulations with stars on slightly hyperbolic orbits with asymptotic velocities $v_\infty = 100$, $200$ km/s for several representative points in our parameter space; we found identical results to our main setup in all areas.

General relativistic (GR) precession may modify the secular dynamics of stars caught on long-lived orbits \citep{will14}. For stars bound to a primary with a low mass secondary, apsidal precession even counteracts secular evolution from the Kozai-Lidov mechanism and suppresses the disruption rate \citep{chen11}. Though the stars in our simulations are initially unbound and have shorter disruption times, we ran additional simulations using Paczy{\'n}sky-Wiita potentials to check the influence of GR precession \citep{paczynsky80}. We found identical results to our main setup for most of our parameter space; for small $q$ and large $a$, we found minor changes, with an increase of $\lesssim 10 \%$ in the disruption rate and a slightly reduced probability of late disruptions (see Section \ref{subsec:properties}).

Although we simulate our scatterings in dimensionless units, physical considerations constrain the domain of applicability of our results. The black holes must have masses $M_\bullet \sim 10^6 M_\odot$ in order for stars in the pinhole regime to be the dominant source of TDEs \citep{stone16}, as discussed above. The Schwarzschild radius $r_S = 2GM_\bullet / c^2$, the location of the event horizon of a nonrotating BH, must be much less than the tidal radius for disruptions to occur and to avoid the effects of strong gravity near the horizon. For a stellar mass-radius relation of $R_* = R_\odot (M_* / M_\odot)^\alpha$, the black hole masses should then satisfy
\begin{equation}
\frac{M_\bullet}{M_\odot} \ll \left(\frac{c^2 R_\odot}{2GM_\odot}\right)^{3/2} \left(\frac{M_*}{M_\odot}\right)^{(3\alpha-1)/2} \simeq 1.1 \times 10^8 \left(\frac{M_*}{M_\odot}\right)^{(3\alpha-1)/2}
\end{equation}
For main sequence stars with $M_* \gtrsim M_\odot$, the parameter $\alpha \simeq 3/4$ \citep{stellar04}. Finally, the binary separation must be small compared to the gravitational sphere of influence of the black hole system in its galactic environment, and large enough to neglect gravitational radiation (we include this contribution in Section \ref{subsec:inspiral}).

\section{Tidal Disruption Events}
\label{sec:tdes}

\subsection{Tidal Disruption Rate}
\label{subsec:tderate}

\begin{figure*}
\centering
\subfloat[Binary disruption rate]{\includegraphics[scale=0.43]{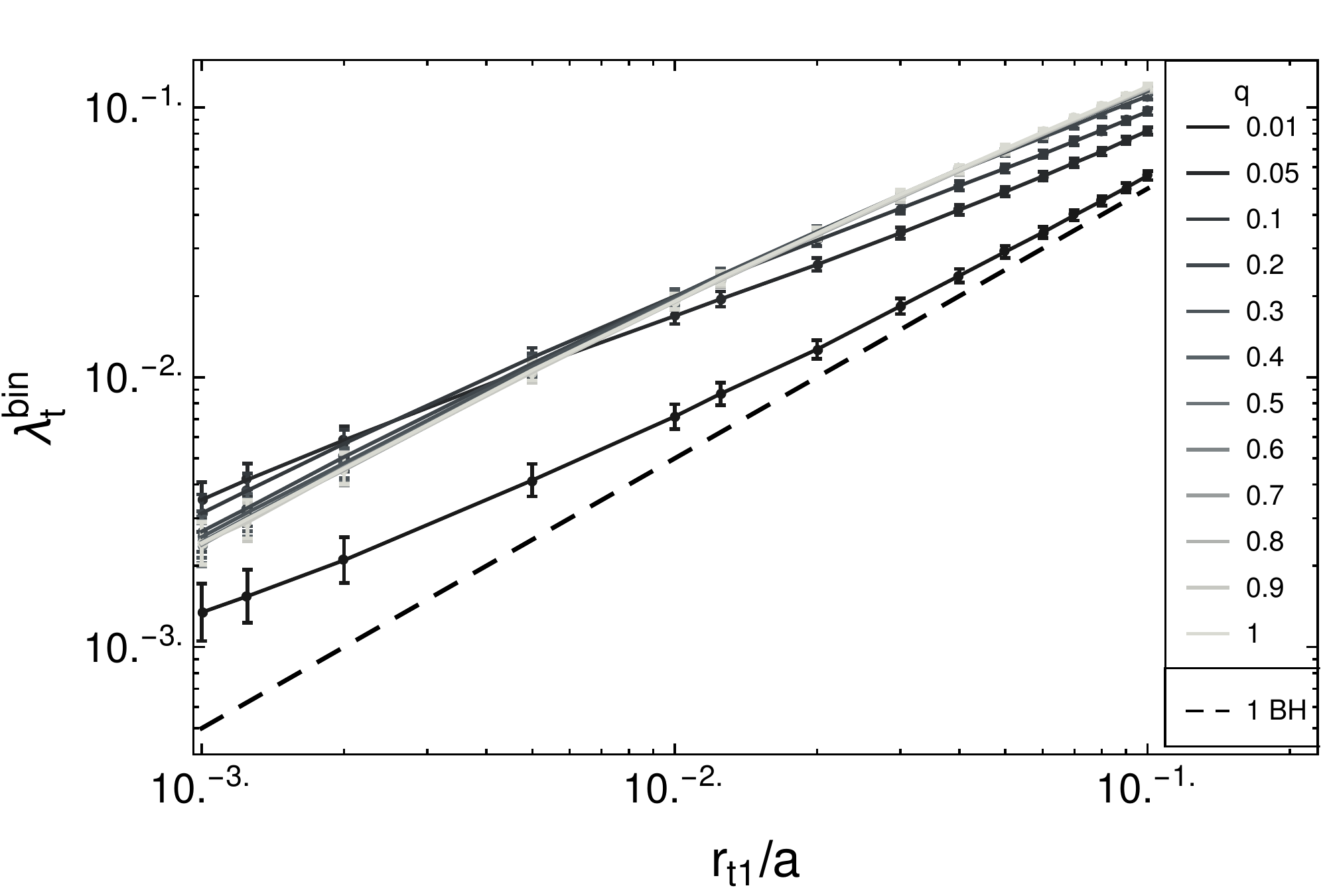}\label{fig:lambdata}}\hfill
\subfloat[Scaled binary disruption rate]{\includegraphics[scale=0.41]{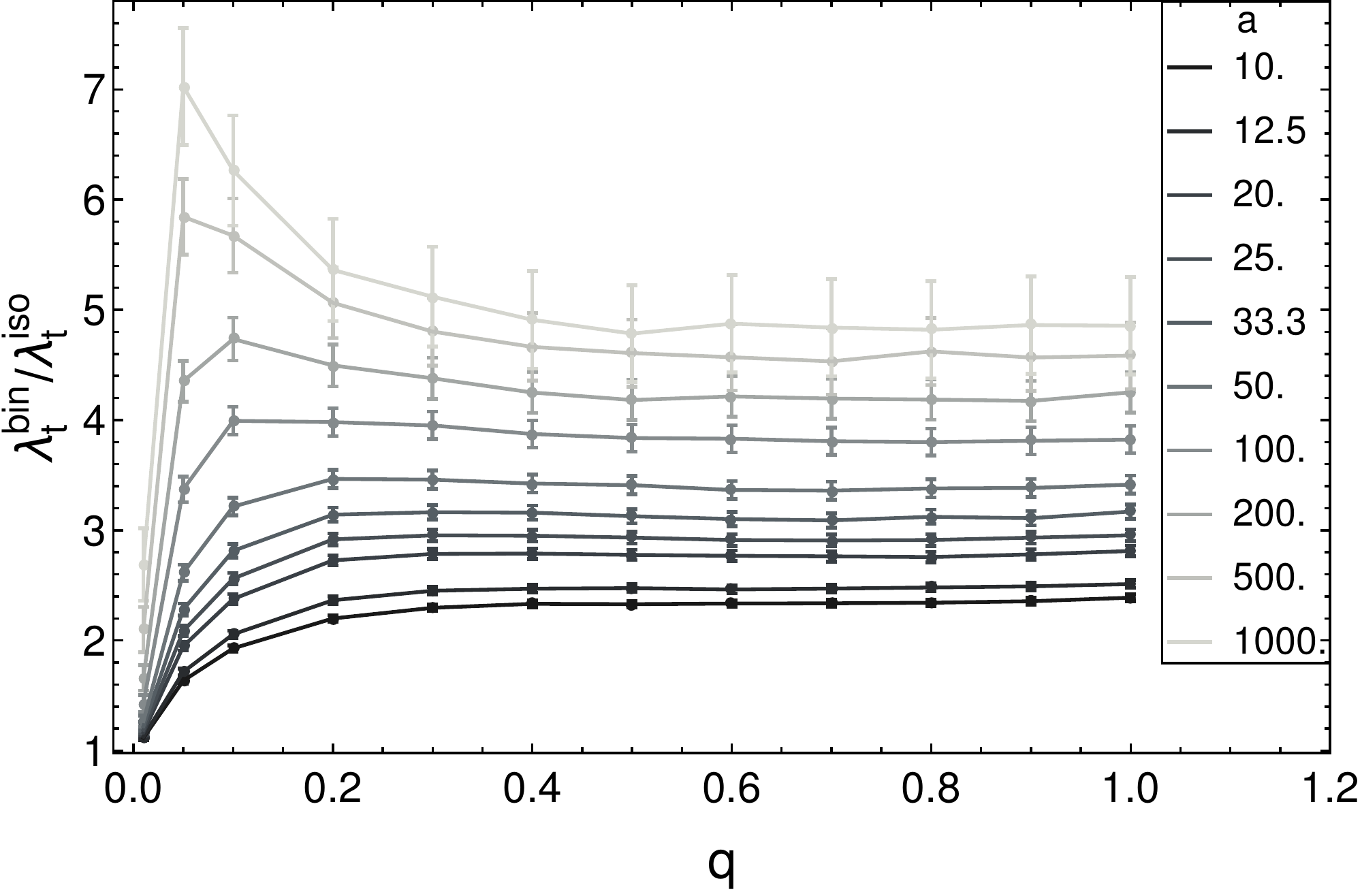}\label{fig:lambdatb}}
\caption{a) The binary BH tidal disruption rate $\lambda^\textrm{bin}_t = N_t / N_e$, where $N_t$ ($N_e = 5 \times 10^6$) is the total number of disruptions (encounters). The horizontal axis gives the primary tidal radius $r_{t1} = \tilde{r}_{t1} / a$ and the legend gives the binary mass ratio $q = M_2 / M_1$. The isolated BH disruption rate (dashed line) is $\lambda^\textrm{iso}_t = r_{t1} / 2a$, and was calculated for a BH with the mass of the primary and stars incident from $r = 50a$ with angular momenta uniformly distributed in $\ell^2 / GM_1 a \in [0, 4]$, where here $a$ is a length scale factor. b) The scaled binary BH disruption rate $\lambda^\textrm{bin}_t/\lambda^\textrm{iso}_t$. The legend gives the separation $a = \tilde{a} / r_{t1}$. In both plots, the error bars have half-width $20\sigma$, where $\sigma = (\lambda^\textrm{bin}_t / N_e)^{1/2}$ are the standard deviations assuming a Poisson distribution.}
\label{fig:lambdat}
\end{figure*}

Figure \ref{fig:lambdata} shows the total tidal disruption rate $\lambda^\textrm{bin}_t= N_t/N_e$ from the binary SMBH, where $N_t$ ($N_e = 5 \times 10^6$) is the total number of disruptions (encounters). Less than 12\% of the encounters end in disruptions over our parameter range, with the system reaching this maximum rate when $q=1$ and $r_{t1}/a = 0.1$. The vast majority of encounters result in stars ejected from the binary, with over 86\% of the simulations ending with this outcome. The end state is inconclusive when stars are placed on weakly bound orbits, neither being disrupted nor expelled by the end time $t = 10^4$; this outcome occurs in less than 2\% of the simulations and thus minimally affects our statistics (see the discussion in \citealt{coughlin17}). Consequently, the escape rate is roughly $\lambda^\textrm{bin}_e \simeq 1 - \lambda^\textrm{bin}_t$.

Figure \ref{fig:lambdatb} shows the scaled tidal disruption rate $\lambda^\textrm{bin}_t/\lambda^\textrm{iso}_t$, which is the rate from the binary black hole normalized to the rate from an isolated black hole with the mass of the primary. The rate exhibits several interesting features in our range of parameters. For low $q$, the rate increases with $q$, and for $q \gtrsim 0.2$, the rate is insensitive to $q$ and is in the range $\sim 2 - 5$. For a given $q$, the rate increases monotonically with $a$, since a larger separation leads to a larger gravitational cross-section for the binary. As $q$ approaches zero, the rate approaches $\lambda^\textrm{bin}_t/\lambda^\textrm{iso}_t = 1$ regardless of the separation, as expected since at $q=0$ the binary system reduces to an isolated BH at the origin with the mass of the primary.

\begin{figure*}
\centering
\subfloat[Primary/secondary disruption rate]{\includegraphics[scale=0.43]{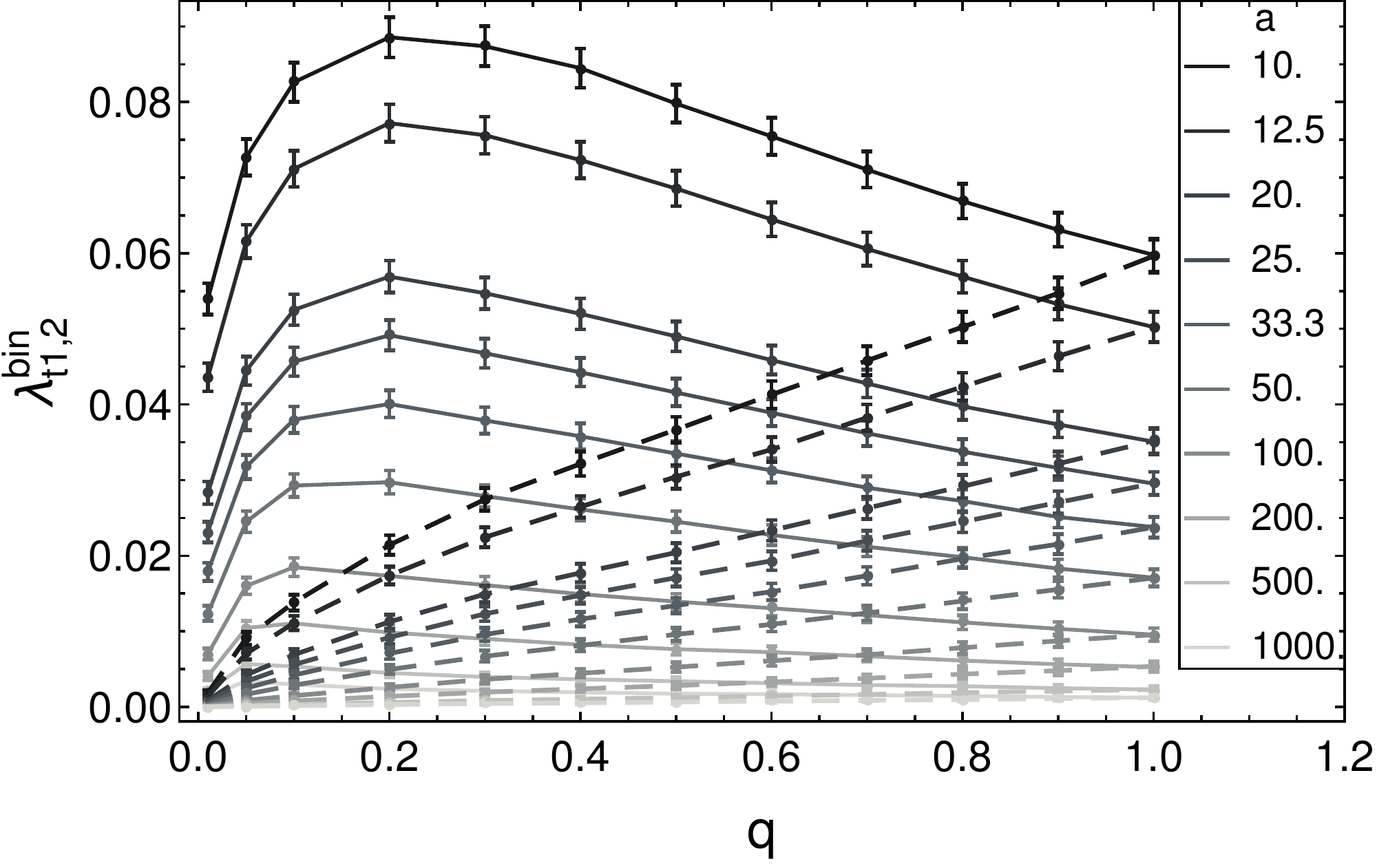}\label{fig:lambdat12a}}\hfill
\subfloat[Normalized primary disruption rate]{\includegraphics[scale=0.41]{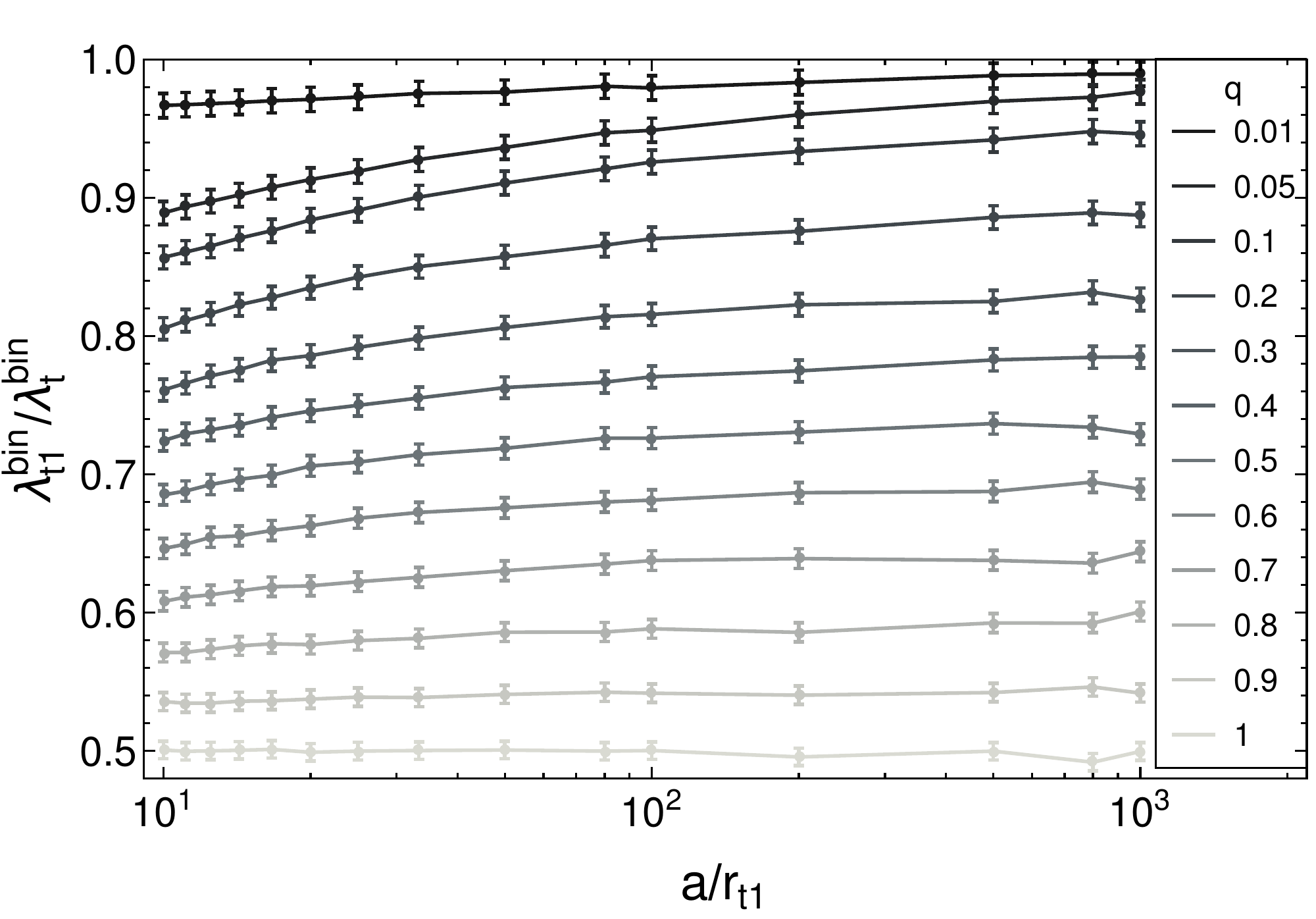}\label{fig:lambdat12b}}
\caption{a) The primary (solid) and secondary (dashed) tidal disruption rates $\lambda^\textrm{bin}_{ti} = N_{ti} / N_e$, where $i = 1 (2)$ refers to the primary (secondary), $N_{ti}$ is the number of disruptions by the corresponding BH, and $N_e = 5 \times 10^6$ is the total number of encounters. The horizontal axis gives the binary mass ratio $q = M_2 / M_1$ and the legend gives the separation $a = \tilde{a} / r_{t1}$. The error bars have half-width $20\sigma_i$, where $\sigma_i = (\lambda^\textrm{bin}_{ti} / N_e)^{1/2}$ are the standard deviations assuming a Poisson distribution. b) The rate of disruption by the primary relative to the total rate of disruption, $\lambda^\textrm{bin}_{t1} / \lambda^\textrm{bin}_t$. The errors bars have half-width $20\sigma$, where $\sigma = (\lambda^\textrm{bin}_{t1} / \lambda^\textrm{bin}_t N_e)^{1/2}$.}
\label{fig:lambdat12}
\end{figure*}

We studied the rates of disruption by the primary and secondary (Figure \ref{fig:lambdat12a}), and interpret these results as follows. When $q$ is small, increasing the mass of the secondary creates larger perturbations on the orbit of the incoming star, which causes the star to remain bound to the binary for longer and increases the likelihood of disruption, mostly for the primary. On the other hand, for $q$ comparable to unity, the secondary mass and tidal radius are more comparable to those of the primary, so the secondary effectively ``steals'' disruptions from the primary with increasing $q$ and the rate remains nearly constant; this agrees with the findings of \citet{coughlin17}, who explored $0.1 \le q \le 1$ for $a/r_{t1} = 100$.

Figure \ref{fig:lambdat12b} shows the primary disruption rate relative to the total binary disruption rate. The primary accounts for nearly all of the disrupted stars for $q = 0.01$ and half of them for $q = 1$, as expected. Between these limits, the rate is well approximated by a linearly decreasing function of $q$ and is largely independent of $a$. Though the relative rate decreases monotonically with increasing $q$, the total primary disruption rate behaves as described above. The relative rate for the secondary is simply $\lambda^\textrm{bin}_{t2}/\lambda^\textrm{bin}_t = 1 - \lambda^\textrm{bin}_{t1}/\lambda^\textrm{bin}_t$.

\subsection{Orbital Properties of Disrupted Stars}
\label{subsec:properties}

In this subsection, we examine the orbital properties of the disrupted stars, including the behavior of the orbits, the disruption timescales, the center of mass energies and eccentricities, and the amount of bound debris. We use this information in the next subsection to study the dynamics of the returning debris, namely how the center of mass energies affect the fallback rate.

The stars in our simulations are initialized on parabolic (zero energy) trajectories about the binary center of mass. A disrupted star typically traces out a chaotic three-body orbit in the binary potential prior to disruption, and can thus acquire a nonzero center of mass energy relative to the disrupting BH. For an isolated SMBH, in contrast, the star's energy is conserved. To quantify this effect, we define a critical energy parameter $\varepsilon_c$ for the (nonrotating) star at the point of disruption:
\begin{equation}
\varepsilon_c = \frac{\epsilon_\textrm{cm}}{\Delta\epsilon}
\label{eq:epsc}
\end{equation}
Here, $\epsilon_\textrm{cm} = v^2 / 2 - GM_\bullet / r_t$ is the star's specific center of mass energy relative to the disrupting BH, where $v$ is its relative velocity and $M_\bullet$ and $r_t$ are the black hole's mass and tidal radius, and
\begin{equation}
\begin{split}
\Delta\epsilon &= \frac{GM_\bullet R_*}{r^2_t} \\
&\simeq 1.9 \times 10^{17} \textrm{erg}\cdot\textrm{g}^{-1} \left(\frac{M_\bullet}{10^6 M_\odot}\right)^{1/3} \left(\frac{M_*}{M_\odot}\right)^{2/3} \left(\frac{R_*}{R_\odot}\right)^{-1}
\end{split}
\label{eq:deltaepsilon}
\end{equation}
is the energy spread across the radius $R_*$ of the star due to the potential of the BH, assuming $R_* \ll r_t$.

If the disrupting black hole is isolated, then $\varepsilon_c$ characterizes the behavior of the resulting debris. In particular, it partitions the state of the debris into three main categories: all of the debris is bound to the BH if $\varepsilon_c < -1$, part of it is bound if $-1 \leq \varepsilon_c < 1$, and all of it is unbound if $\varepsilon_c \geq 1$. In short, a lower value of $\varepsilon_c$ leads to more bound debris. A star approaching an isolated BH on a parabolic orbit would be disrupted with $\varepsilon_c=0$, corresponding to half of the debris remaining bound and half escaping. In Section \ref{subsec:postdisruption}, we discuss in more detail the relation between $\varepsilon_c$ and the behavior of the post-disruption debris.

The value of $\varepsilon_c$ that a star acquires at disruption generally correlates with the type of orbit it follows when approaching the disrupting BH. A star will have $\varepsilon_c \geq 1$ if its velocity is roughly in the opposite direction to that of the BH, leading to a head-on encounter. It will have $\varepsilon_c < -1$ if its velocity is in the same direction as the BH, typically approaching it from behind, leading to a lower relative velocity between the two. This outcome tends to occur when the star becomes bound to the primary on a tight, long-lived, pseudo-circular orbit in the direction of the binary rotation. And finally, it will be disrupted with $-1 \leq \varepsilon_c < 1$ if it approaches in an intermediate way, such as with a combination of a glancing angle or a modest relative velocity.

\begin{figure*}
\centering
\subfloat{\includegraphics[width=0.49\textwidth]{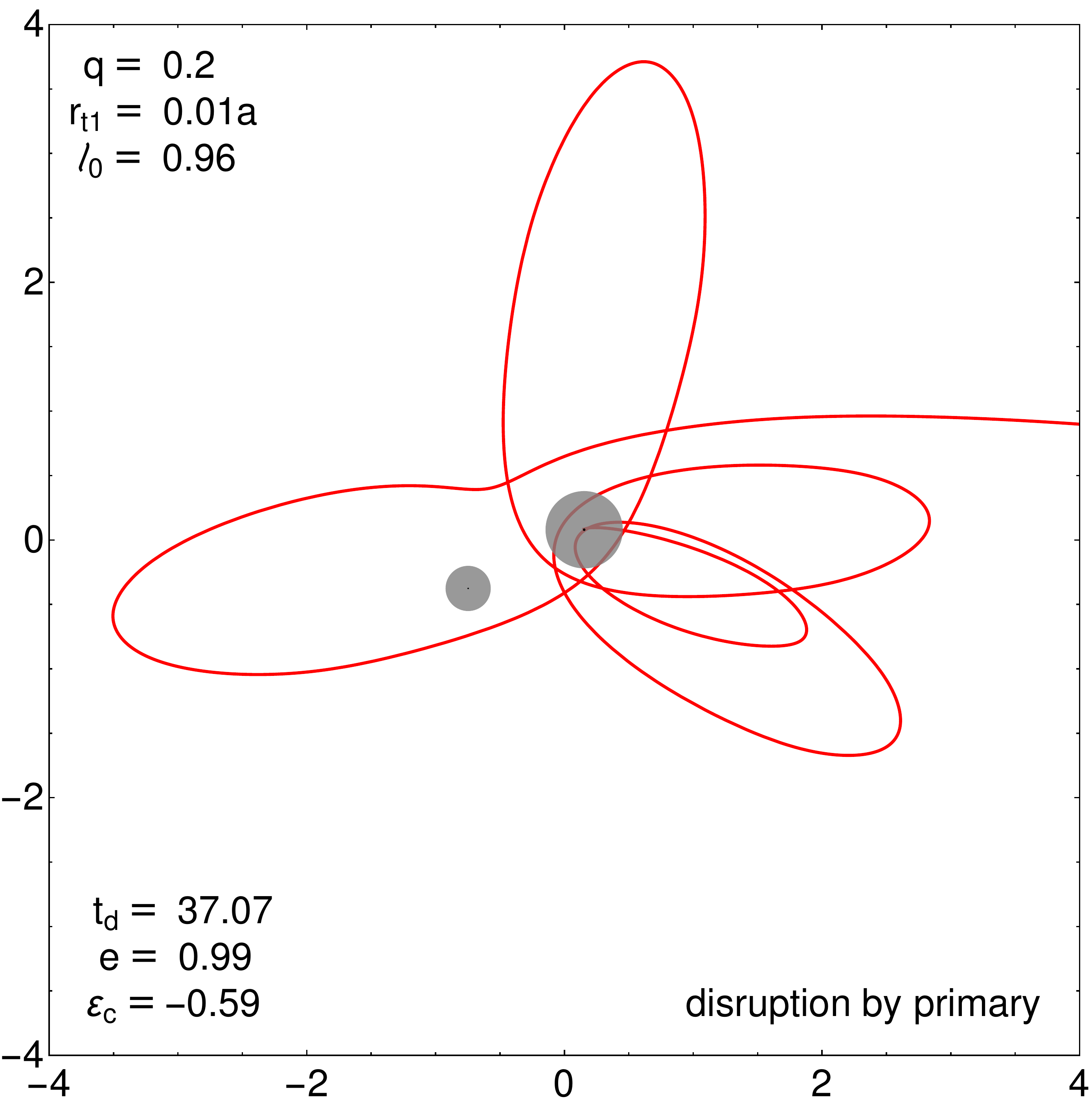}}\hfill
\subfloat{\includegraphics[width=0.49\textwidth]{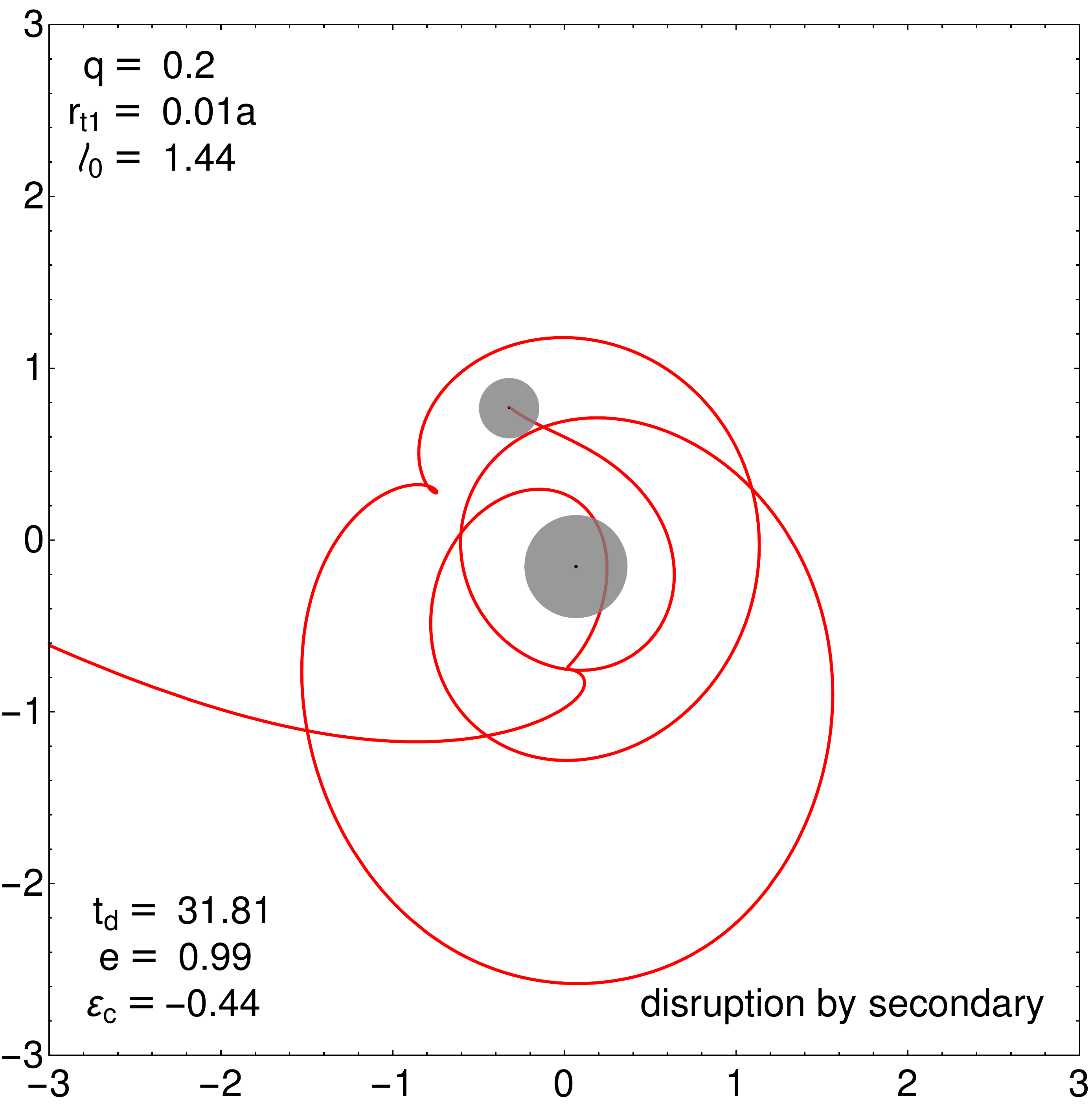}}
\caption{The projection in the $xy$-plane of the orbit for a star disrupted by the primary (left) and secondary (right). The binary has unit separation and parameters $q=0.2$ and $r_{t1}=0.01a$. The binary is rotating counterclockwise and has its center of mass at the origin. The transparent gray circles indicate the locations of the primary (larger) and secondary (smaller), and the black circles at the centers show the tidal radii to scale. The top left corners show the input parameters, including the initial stellar angular momenta $\ell_0 = \tilde{\ell}_0 / GMa$. The bottom left corners show several parameters at the point of disruption, namely the times of disruption $t_d = \tilde{t}_d / 2\pi \sqrt{a^3 / GM}$ (i.e. in binary orbits), and the stars' critical energies $\varepsilon_c$ and eccentricities $e$.}
\label{fig:orbit}
\end{figure*}

Figure \ref{fig:orbit} shows the projection in the $xy$-plane of the orbits of two disrupted stars for $q=0.2$ and $r_{t1}=0.01a$. The orbits give a sense of the motion of a typical star on the path to disruption: the stars are bound to the primary and perturbed by the secondary, both in the far field and in close encounters, until an eventual disruption. The stars shown are disrupted fairly early, fewer than 11 binary orbits after the first encounter (at $\simeq 26.5$ binary orbits, see below). The stars are disrupted with $-1 \leq \varepsilon_c < 1$ since they approach the disrupting BH on an elliptical orbit (primary) or at a glancing angle (secondary) after orbiting the binary several times.

\begin{figure}
\centering
\includegraphics[width=0.47\textwidth]{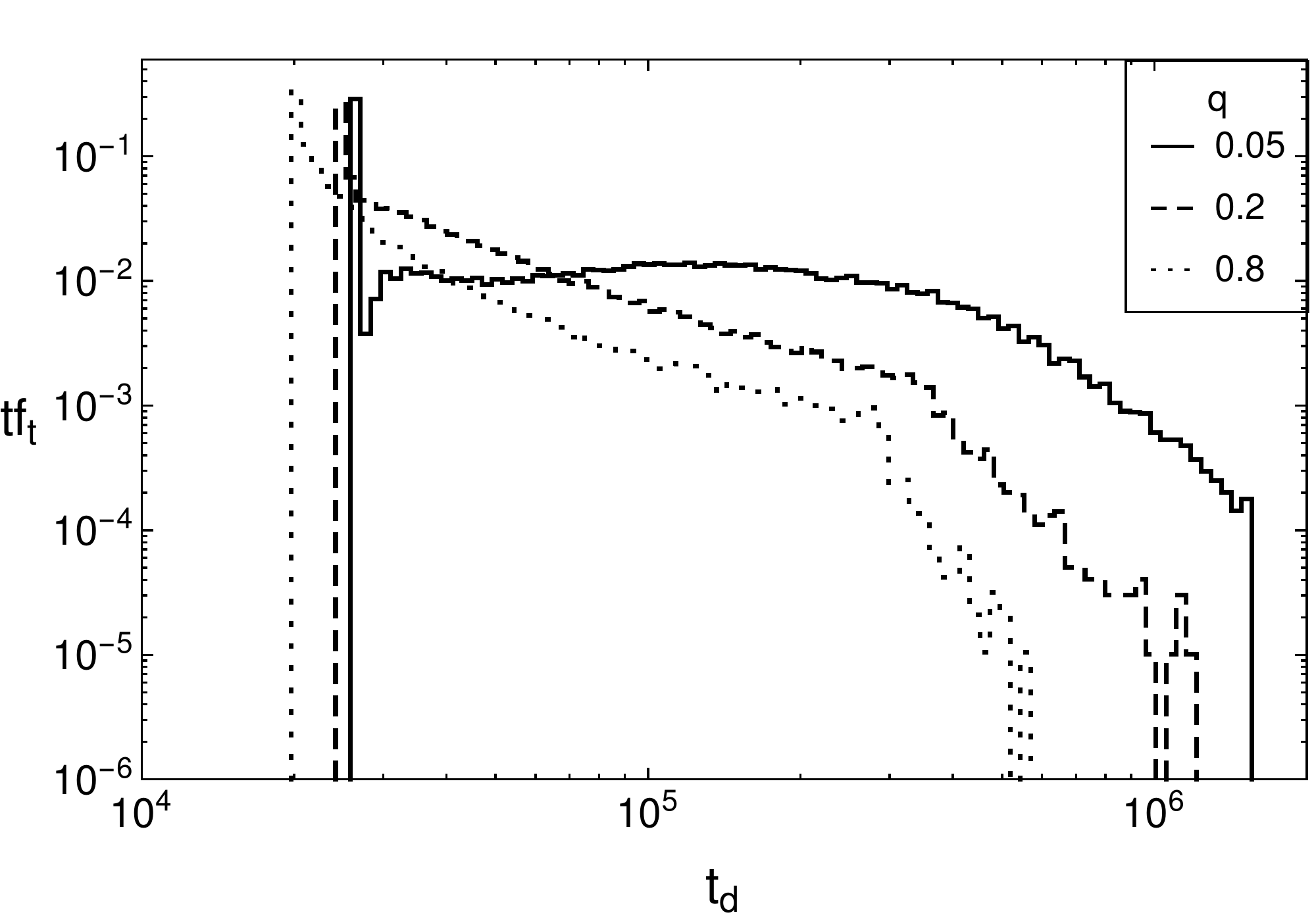}
\caption{The probabilities $tf_t$ for the time of disruption $t_d = \tilde{t}_d / 2\pi \sqrt{r^3_{t1} / GM_1}$, for $a = 100$ and $q=0.05$ (solid), $0.2$ (dashed), and $0.8$ (dotted). The logarithmic bin widths are $\Delta_t = 0.02$ and the heights give the probabilities in each bin. For $M_1 = 10^6 M_\odot$ and Sun-like stars, $t_d = 10^4$ corresponds to roughly $3.2$ years. The leftmost bin in each histogram contains $t_d = 26.5 / \sqrt{(1+q)/a^3}$, which is the time at which the stars first encounter the binary.}
\label{fig:tdis}
\end{figure}

Figure \ref{fig:tdis} shows histograms of the probabilities $tf_t$ for the time of disruption $t_d = \tilde{t}_d / 2\pi \sqrt{r^3_{t1} / GM_1}$ for $a = 100$, where $f_t$ are the probability density functions (PDFs). Over $10\%$ of disruptions occur in a star's first encounter with the binary, which corresponds to the peak at the leftmost point in each of the histograms. For low $q$, the probability drops after the earliest time and then increases to a local maximum at a later time; for $q \gtrsim 0.1$, the disruption probability decreases monotonically with time, most likely approaching a local maximum after the simulation time. The disrupted stars satisfy these trends for all separations that we explored. With apsidal precession included, the probabilities remain unchanged for most of our parameter space; for $q = 0.01$ and $a \gtrsim 100$, the probabilities are slightly lower at late times, leading to an earlier local maximum.

The probability distributions can be understood by considering the typical orbits of disrupted stars. A star takes approximately $50^{3/2} \sqrt{2} / (6\pi) = 26.5$ binary orbits to move from its initial position at $r=50a$ to its first encounter with the binary. The star can be disrupted at this time due to a chance encounter with one of the black holes, more commonly with the primary for lower $q$, so the disruption probability peaks at this earliest time with a value $\gtrsim 10\%$. This time corresponds to $t_d = 26.5 / \sqrt{(1+q)/a^3}$, which is the location of the leftmost bin in each of the histograms. If the star is not disrupted at this earliest moment, it remains bound and follows a chaotic orbit until it is eventually disrupted after many encounters. For low $q$, the star is bound explicitly to the primary and its orbit is gradually perturbed by the smaller secondary, preferentially being disrupted after a number of binary orbits, most often by the primary. As a result, the probability drops after the time of first encounter and then gradually increases to a local maximum. For $q \gtrsim 0.1$, both the primary and secondary have sizable disruption cross sections, so the probability decreases monotonically. Binaries with different separations exhibit similar behavior.

The impact parameter of a disrupted star is given by $\beta = r_t / r_p$ \citep{carter82}, where $r_t$ is the tidal radius of the BH and $r_p$ is the star's pericenter distance relative to the disrupting BH if it were to continue on its trajectory as a point particle. \citet{coughlin17} showed that the PDF of $\beta$ for stars disrupted by a binary BH with $q = 0.2$ and $r_{t1} = 0.01 a$ is well described by $f_\beta = 1 / \beta^2$, and this scaling arises when the pericenters of disrupted stars are uniformly distributed. We find that the PDFs are accurately described by this scaling for most of our parameter space, though they exhibit slightly steeper power laws for close separations and low mass ratios. The black hole will undoubtedly swallow the star if the pericenter is within the black hole's Schwarzschild radius, or equivalently when the impact parameter is
\begin{equation}
\beta \geq \beta_s = \frac{R_* c^2}{2G(M^2_\bullet M_*)^{1/3}} \simeq 23.6 \left(\frac{M_\bullet}{10^6 M_\odot}\right)^{-2/3} \left(\frac{\langle \rho_* \rangle}{\langle \rho_\odot \rangle}\right)^{-1/3}
\end{equation}
where $M_\bullet$ is the BH mass, $M_*$ and $R_*$ are the star's mass and radius, and $\langle \rho_* \rangle = 3M_* / 4\pi R^3_*$ is the star's mean density. This outcome occurs with probability $p \simeq 1/\beta_s$.

\begin{figure*}
\centering
\subfloat{\includegraphics[width=0.49\textwidth]{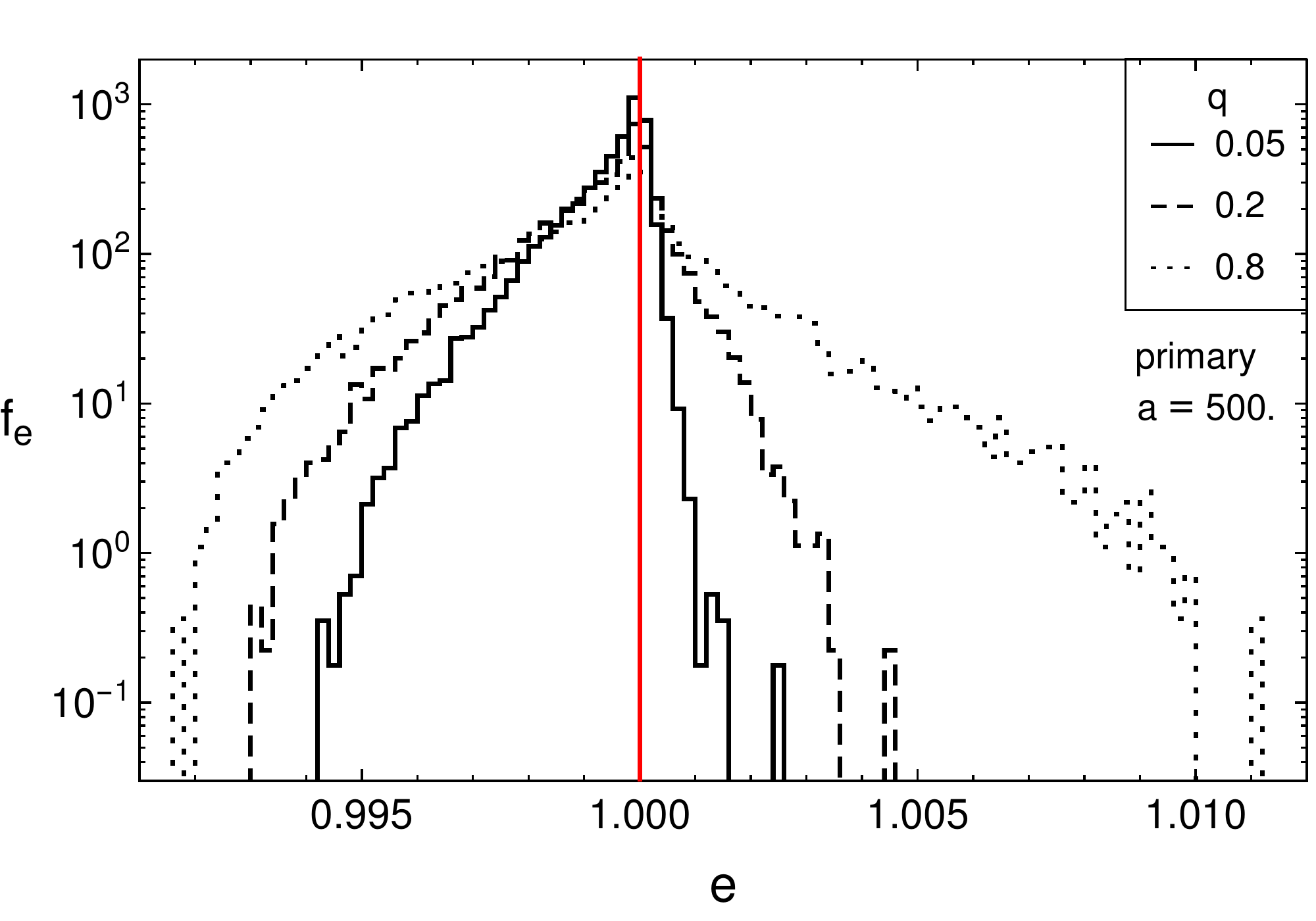}}\hfill
\subfloat{\includegraphics[width=0.49\textwidth]{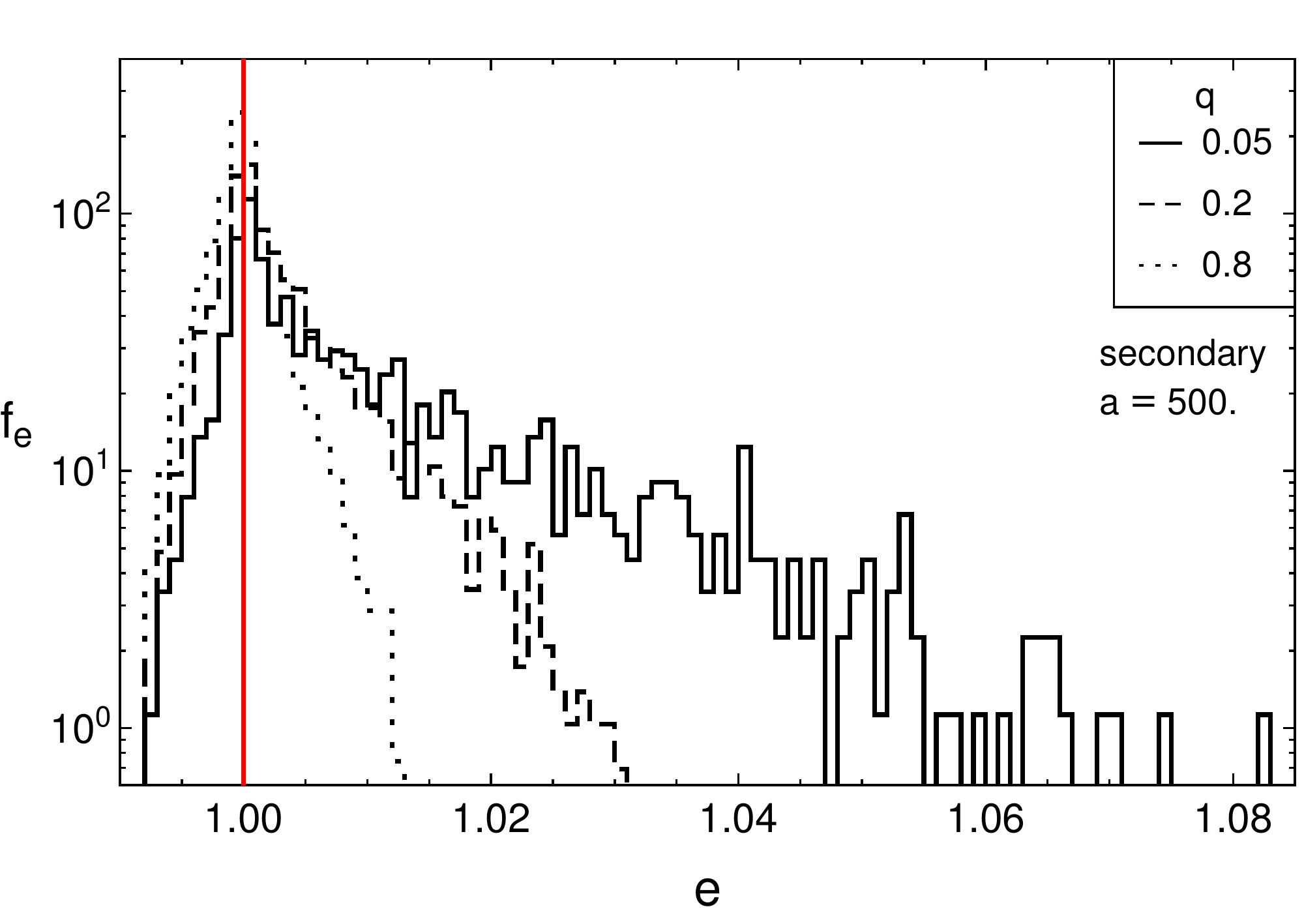}}
\caption{The PDFs for the eccentricity of disrupted stars relative to the disrupting BH calculated at the tidal radius, for $a = 500$ and $q=0.05$ (solid), $0.2$ (dashed), and $0.8$ (dotted).
The left panel shows disruptions by the primary and the right one shows those by the secondary. The bin widths are $2 \times 10^{-4}$ (primary) and $10^{-3}$ (secondary). The relative eccentricity is given in Eq. \ref{eq:relecc}. A star on a parabolic orbit disrupted by an isolated BH has an eccentricity $e=1$, as marked by the red line. For binaries with smaller separations, the PDFs become wider but have similar shapes and retain theirs peaks at $e=1$ (Figure \ref{fig:histeccappendix} in Appendix \ref{sec:appendixa}).}
\label{fig:histecc}
\end{figure*}

Figure \ref{fig:histecc} shows the PDFs for the relative eccentricity of disrupted stars for $a = 500$, given by
\begin{equation}
e = \sqrt{\frac{2 \ell^2_\textrm{cm} \epsilon_\textrm{cm}}{(GM_\bullet)^2} + 1}
\label{eq:relecc}
\end{equation}
where $\epsilon_\textrm{cm}$ and $\ell_\textrm{cm} = \lvert \vec{r}_\textrm{cm} \times \vec{v} \rvert$ are the (dimensioned) specific energy and angular momentum of star's center of mass relative to the BH at its tidal radius, and $M_\bullet$ is the mass of the BH. The distributions are peaked at $e=1$ (a parabolic orbit), which was the initial eccentricity relative to the binary, and are more narrowly centered around this value for the primary ($0.99 \lesssim e \lesssim 1.01$) than the secondary ($0.99 \lesssim e \lesssim 1.08$). As $q$ approaches one, the PDFs for the primary and secondary become similar and symmetric about $e=1$, as expected; as $q$ decreases, the PDF for the primary becomes narrower and favors bound elliptical orbits ($e<1$), and that for the secondary becomes broader and favors unbound hyperbolic orbits ($e>1$). The distributions become broader for smaller separations while preserving their general shape (Figure \ref{fig:histeccappendix}).

\begin{figure*}
\centering
\subfloat{\includegraphics[width=0.49\textwidth]{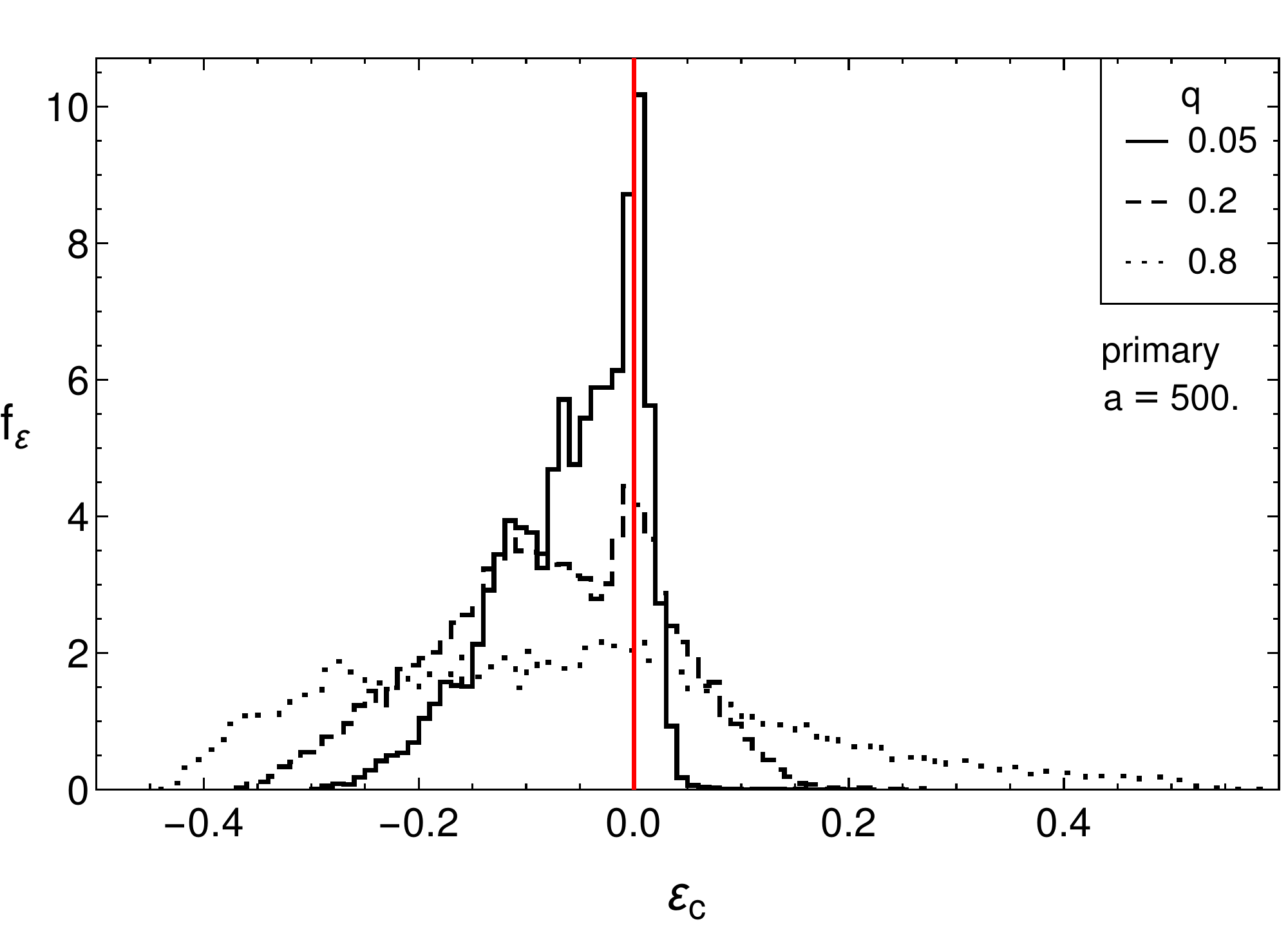}}\hfill
\subfloat{\includegraphics[width=0.49\textwidth]{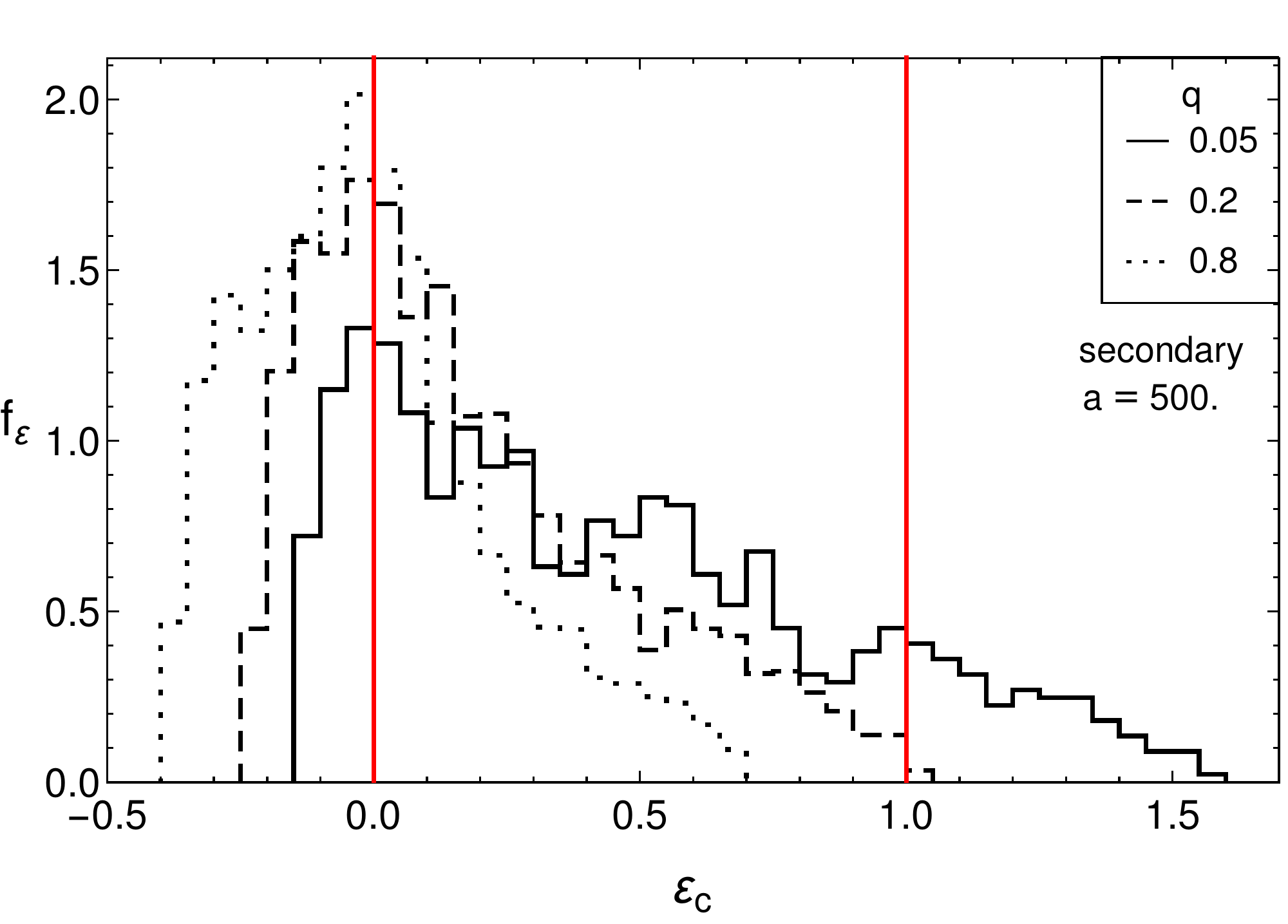}}
\caption{The PDFs for the critical energy $\varepsilon_c$ (Eq. \ref{eq:epsc}) of disrupted stars relative to the disrupting BH calculated at the tidal radius, for $a=500$ and $q=0.05$ (solid), $0.2$ (dashed), and $0.8$ (dotted). The left panels show disruptions by the primary and the right ones show those by the secondary. The bin widths are $\Delta_\varepsilon = 0.01$ (primary) and $0.05$ (secondary). The red lines mark $\varepsilon_c = -1, 0, +1$, if they are in a histogram's domain. The stellar debris is bound to a different extent in the regions the lines delimit: if $\varepsilon_c < -1$, then all of the debris is bound to the BH; if $-1 \leq \varepsilon_c < 1$, then part of it is bound; and if $\varepsilon_c \geq 1$, then all of it is unbound. If $\varepsilon_c=0$, half of the debris remains bound and half escapes, which occurs when a star on a parabolic orbit is disrupted by an isolated BH. For binaries with smaller separations, the PDFs become wider while retaining their peaks at $\varepsilon_c = 0$ (a factor of 10 decrease from $a=500$ to $50$ leads to a factor of 10 increase in the widths); the histograms for the primary become more concentrated at $\varepsilon_c = 0$, and those for the secondary roughly preserve their shape (Figure \ref{fig:histepscappendix} in Appendix \ref{sec:appendixa}).}
\label{fig:histepsc}
\end{figure*}

Figure \ref{fig:histepsc} shows the PDFs for the critical energy parameter $\varepsilon_c$ (Eq. \ref{eq:epsc}) of disrupted stars relative to the disrupting BH for $a = 500$. The PDFs have their global maxima near $\varepsilon_c = 0$. For $q<1$, the primary is spread narrowly about this maximum ($-0.5 \lesssim \varepsilon_c \lesssim 0.6$) and preferentially produces mostly bound debris ($\varepsilon_c < 0$), and the secondary is broader ($-0.5 \lesssim \varepsilon_c \lesssim 1.6$) and produces mostly unbound debris ($\varepsilon_c \geq 0$); the two become more similar and symmetric as $q$ approaches $1$, as required. For $q=0.2$, the PDF for the primary exhibits a double-peaked structure, with one maximum at $\varepsilon_c = 0$ and one slightly lower. The PDFs show similar trends for tighter binaries (Figure \ref{fig:histepscappendix}). The distributions become wider by a factor of 10 as $a$ decreases by a factor of 10 (from $a = 500$ to $50$); the histograms for the primary become more concentrated at $\varepsilon_c = 0$, and those for the secondary roughly preserve their shape.

The general trends in the PDFs for $e$ and $\varepsilon_c$ can be explained by considering the black hole masses and velocities. For low $q$, the primary is much more massive than the secondary, so stars follow bound orbits about the primary that are gradually perturbed by the secondary until a disruption. If a star is disrupted by the primary, which is the more probable outcome, it will thus likely be on a bound orbit ($e < 1$). In contrast, a disruption by the secondary will likely occur due to a chance encounter, and the star will typically be on an unbound orbit relative to the secondary ($e > 1$). As $q$ approaches one, the PDFs for $e$ and $\varepsilon_c$ become more symmetric about their peak values of zero and one, respectively, as they must be equal for $q=1$. The width of the distributions can generally be understood by considering the velocities of the binary constituents, which are $v_1 = q (1+q)^{-1} \sqrt{GM/a}$ for the primary and $v_2 = (1+q)^{-1} \sqrt{GM/a}$ for the secondary. Both velocities scale as $v \sim a^{-1/2}$, so a tighter binary leads to higher black hole velocities and thus a wider spread in the PDFs for $e$ and $\varepsilon_c$. A larger $q$, though, leads to a higher $v_1$ and a lower $v_2$, which causes the PDFs for $v_1$ to widen and the ones for $v_2$ to narrow.

\begin{figure*}
\centering
\subfloat{\includegraphics[width=0.49\textwidth]{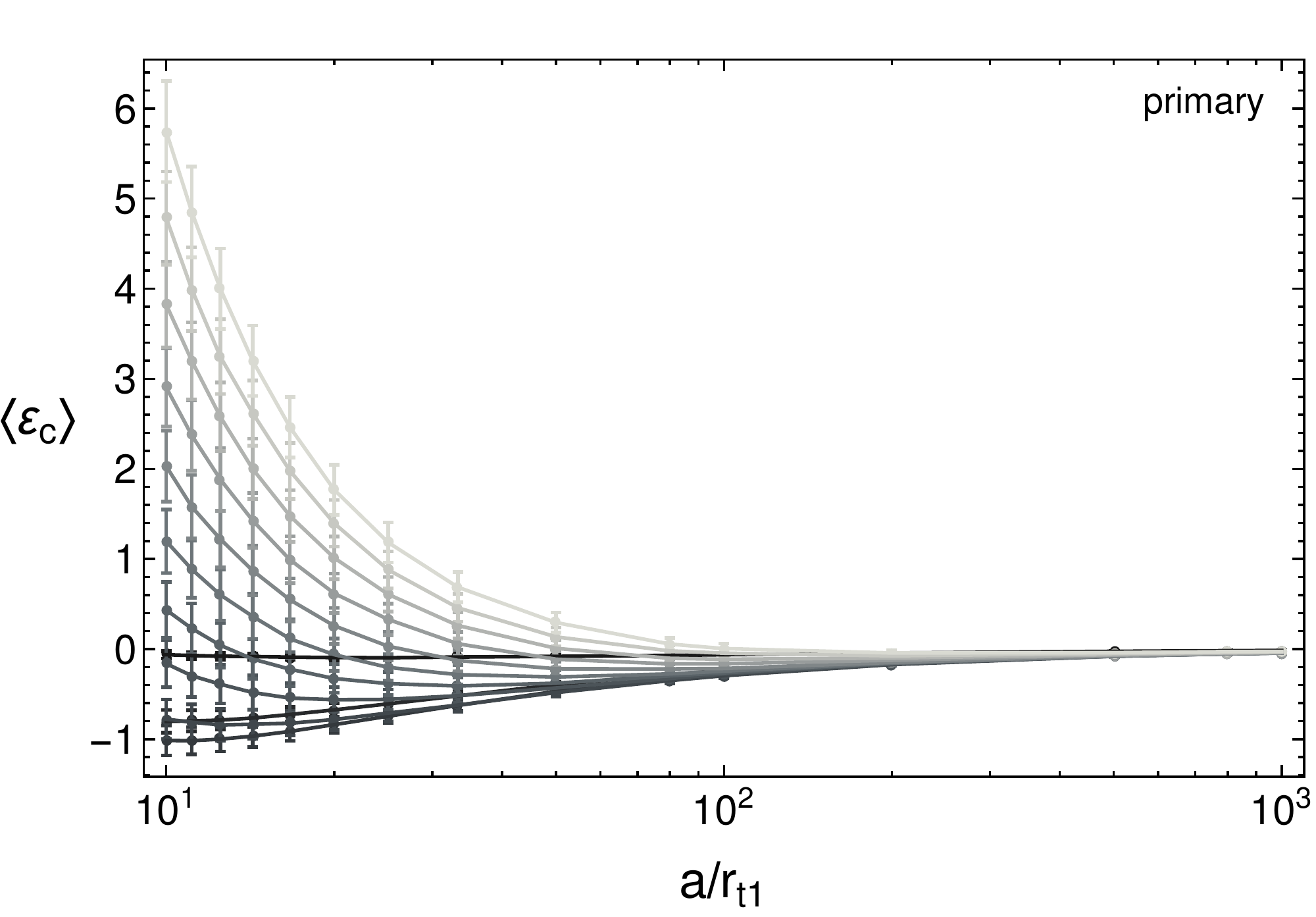}}\hfill
\subfloat{\includegraphics[width=0.49\textwidth]{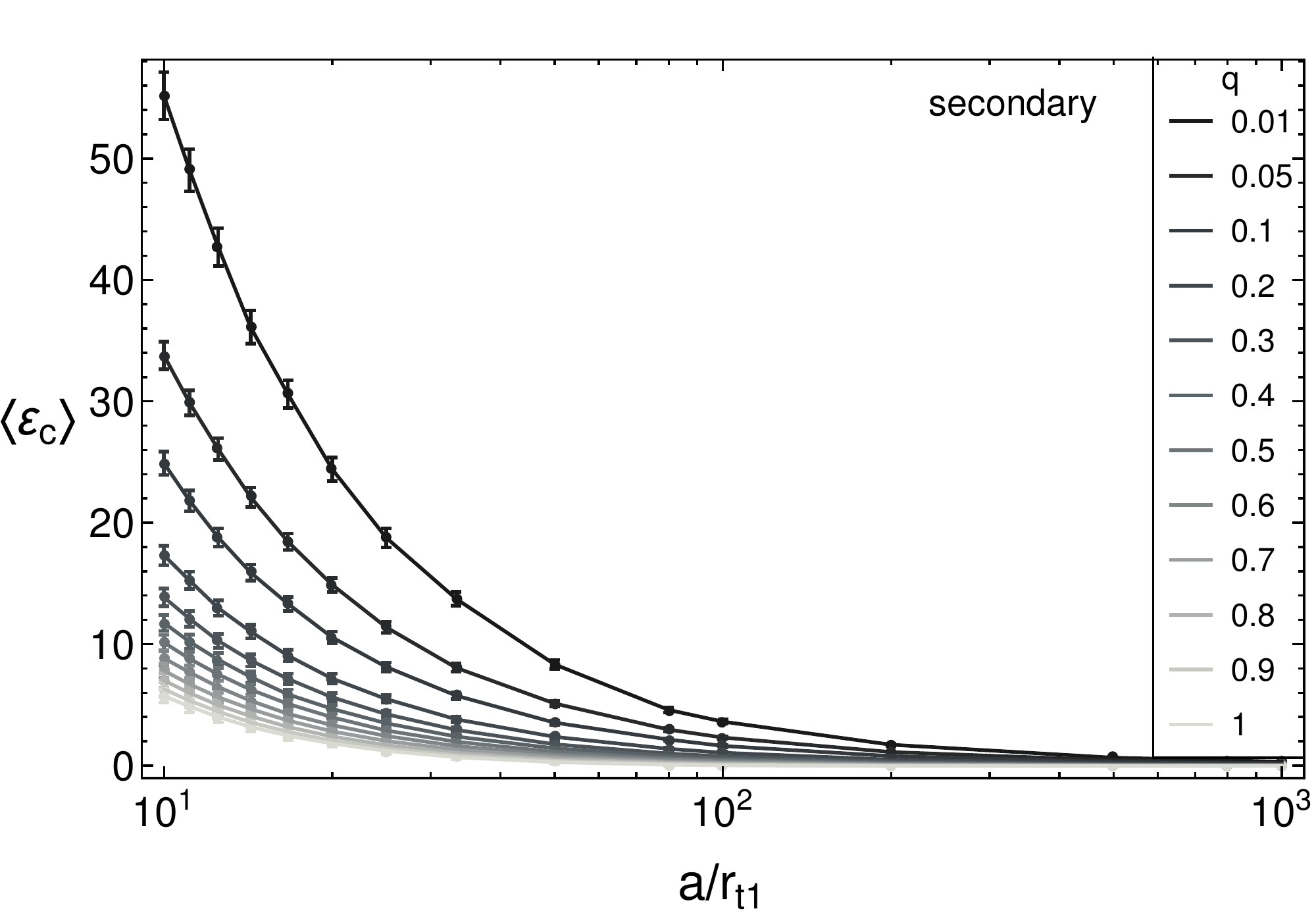}}
\caption{The average critical energy $\langle \varepsilon_c \rangle$ of disrupted stars relative to the disrupting BH calculated at the tidal radius, for stars disrupted by the primary (left) and secondary (right). The value of $\varepsilon_c$ in an individual disruption event determines the general behavior of the debris: if $\varepsilon_c < -1$, then all of the debris is bound to the BH; if $-1 \leq \varepsilon_c < 1$, then part of it is bound; and if $\varepsilon_c \geq 1$, then all of it is unbound. If $\varepsilon_c=0$, half of the debris remains bound and half escapes, which occurs when a star on a parabolic orbit is disrupted by an isolated BH.}
\label{fig:epsc}
\end{figure*}

Figure \ref{fig:epsc} shows the average $\varepsilon_c$ versus $a$ for different $q$. The primary exhibits a complicated behavior. For $q=0.01$, $\langle \varepsilon_c \rangle \simeq 0$ for all separations. For all values of $q$, $\langle \varepsilon_c \rangle$ approaches zero at large separations. For small $a$, it decreases with $q$ until about $q=0.1$, so more than half of the stellar debris will remain bound to the primary on average in this regime. The mean then increases with $q$, and for $q=1$, it decreases monotonically towards zero with increasing $a$. In between these last two values of $q$, there is a trade-off and $\langle \varepsilon_c \rangle$ does not exhibit monotonic behavior. The secondary behaves more straightforwardly. The mean has a very large value at small $a$ and $q$, so much of the debris will be unbound on average in this regime, and it decreases monotonically with increasing $a$ and $q$.

\begin{figure*}
\centering
\subfloat{\includegraphics[width=0.473\textwidth]{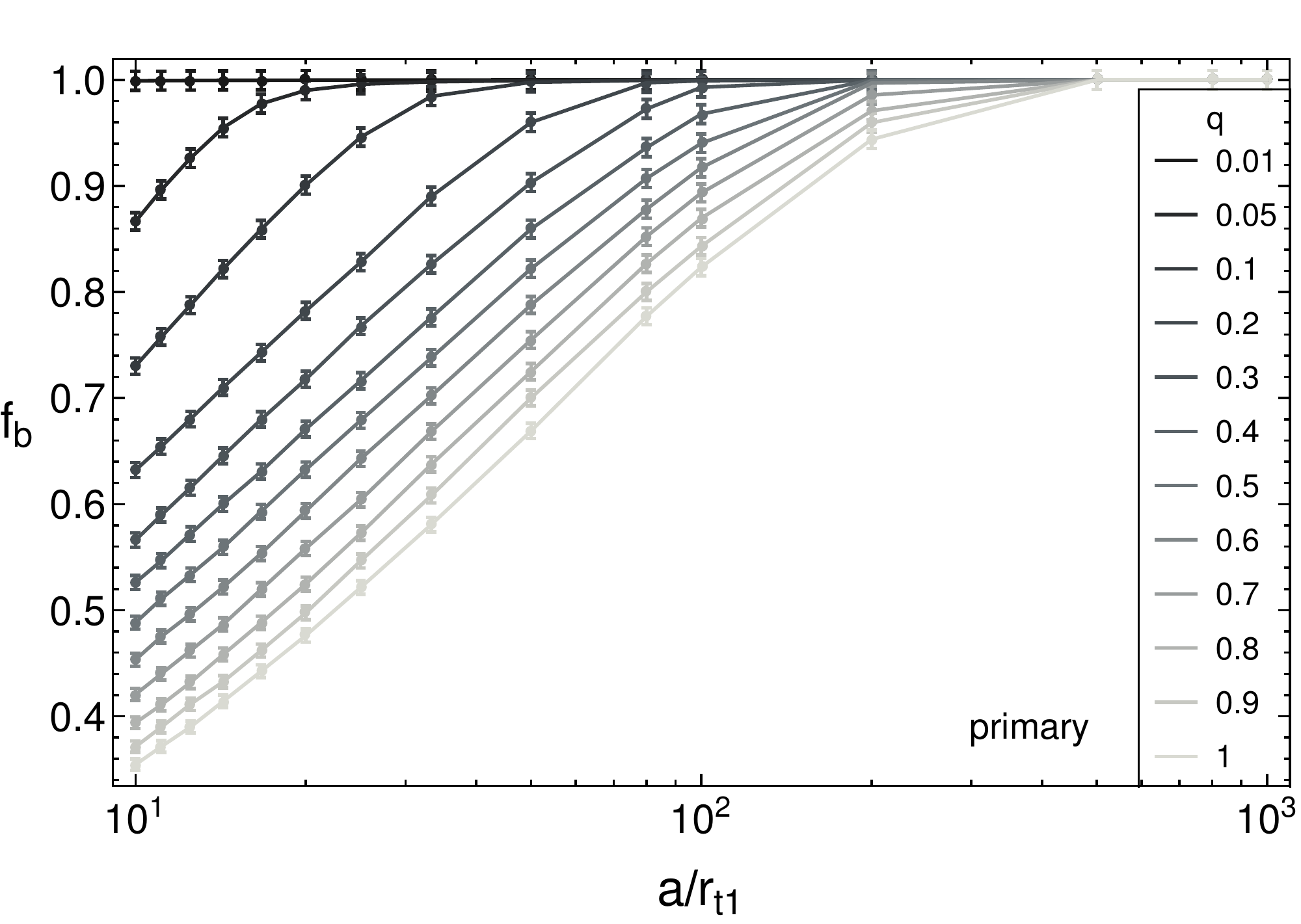}}\hfill
\subfloat{\includegraphics[width=0.5\textwidth]{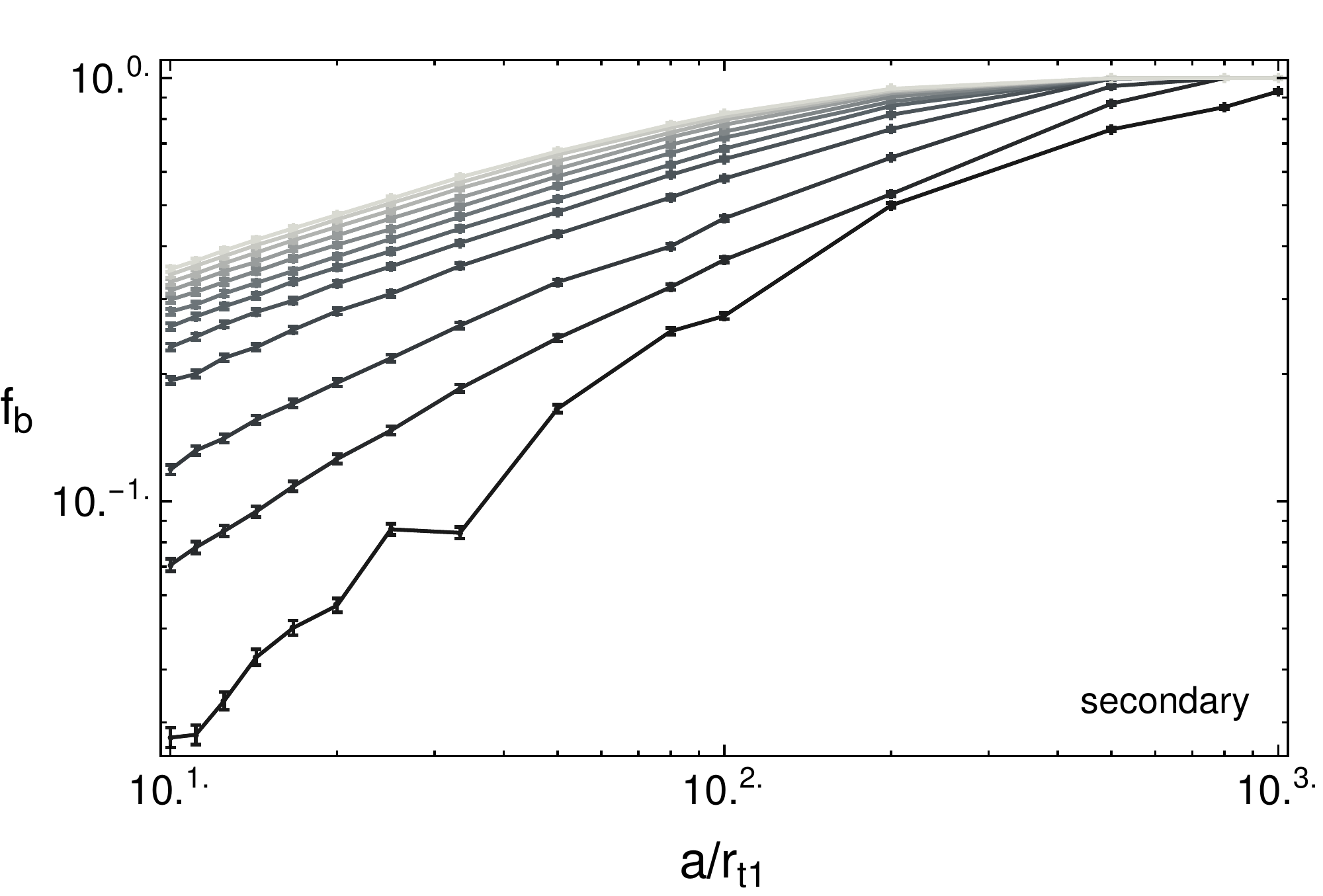}}
\caption{The fraction $f_b$ of stars with at least some debris bound to the disrupting BH ($\varepsilon_c < 1$), for stars disrupted by the primary (left) and secondary (right). The curves for the primary and secondary become equal when $q=1$ (lightest gray), as expected from symmetry.}
\label{fig:fbound}
\end{figure*}

Figure \ref{fig:fbound} shows the fraction of the disrupted stars that have at least some amount of bound debris over our parameter range. For wide separations ($a \gtrsim 200$), most stars disrupted by either the primary ($\gtrsim 95 \%$) or secondary ($\gtrsim 50 \%$) have some bound debris, regardless of the mass ratio. As the binary contracts, the fraction decreases monotonically for both black holes; for the primary this decline is more rapid for larger $q$, and for the secondary it is more rapid for smaller $q$. However, the total TDE rate remains constant for $q \gtrsim 0.2$ (Figure \ref{fig:lambdatb}), so a larger $q$ in this range effectively shifts bound debris from the primary to the secondary. The curves for the primary and secondary become equal when $q=1$, as expected.

\subsection{Return Dynamics of Post-Disruption Debris}
\label{subsec:postdisruption}

The debris from a tidal disruption event can become bound to the binary system. A detailed study of its dynamics would require a simulation in the gravitational potential of the black holes. In lieu of this, we estimate the fallback properties by making the ``frozen-in'' approximation, analytically modeling the debris in the potential of the disrupting BH only. If the debris promptly circularizes into an accretion disk and the radiation diffuses out rapidly, then the accretion rate and lightcurve will track the fallback rate. Analytic studies have argued that the bolometric lightcurve will generally follow the fallback rate \citep{lodato11}, and recent parametric lightcurve fits (coupled with a filter function to estimate the viscous time) have suggested that the time delay may be less than a few days \citep{mockler18}. However, the intermediate steps may be inefficient, and in many cases it may not be possible to directly translate a fallback rate through to a lightcurve \citep{shiokawa15,bonnerot16,hayasaki16b,sadowski16}. We focus on the behavior of the fallback rate in this subsection, noting that its implications depend on the details of the accretion process. We describe our model below, but first give its domain of applicability.

The rise time and peak fallback rate calculated from our model will be accurate when the apocenter of the peak returning debris is within the Roche lobe of the disrupting black hole, $r_\textrm{peak} < r_\textrm{Roche}$. For disruptions by the primary, this condition becomes (in units of $G = M = a = 1$)
\begin{equation}
\frac{\ell^2_\textrm{peak} (1+q)}{1 - [ 2 \epsilon_\textrm{peak} \ell^2_\textrm{peak} (1+q)^2 + 1 ]^{1/2}} < \frac{0.49 q^{-2/3}}{0.6 q^{-2/3} + \ln ( 1 + q^{-1/3} )}
\label{eq:modelcondition}
\end{equation}
where $\epsilon_\textrm{peak}$ and $\ell_\textrm{peak}$ are the peak returning specific energy and angular momentum, and where we used a common expression for the characteristic size of the Roche lobe \citep{eggleton83,accretion02}. The condition for the secondary can be obtained with the replacements $q \rightarrow q^{-1}$ and $r_{t1} \rightarrow r_{t2}$. \citet{coughlin18} additionally examined the timescales over which this single BH assumption remains valid.

\begin{figure*}
\centering
\subfloat{\includegraphics[width=0.49\textwidth]{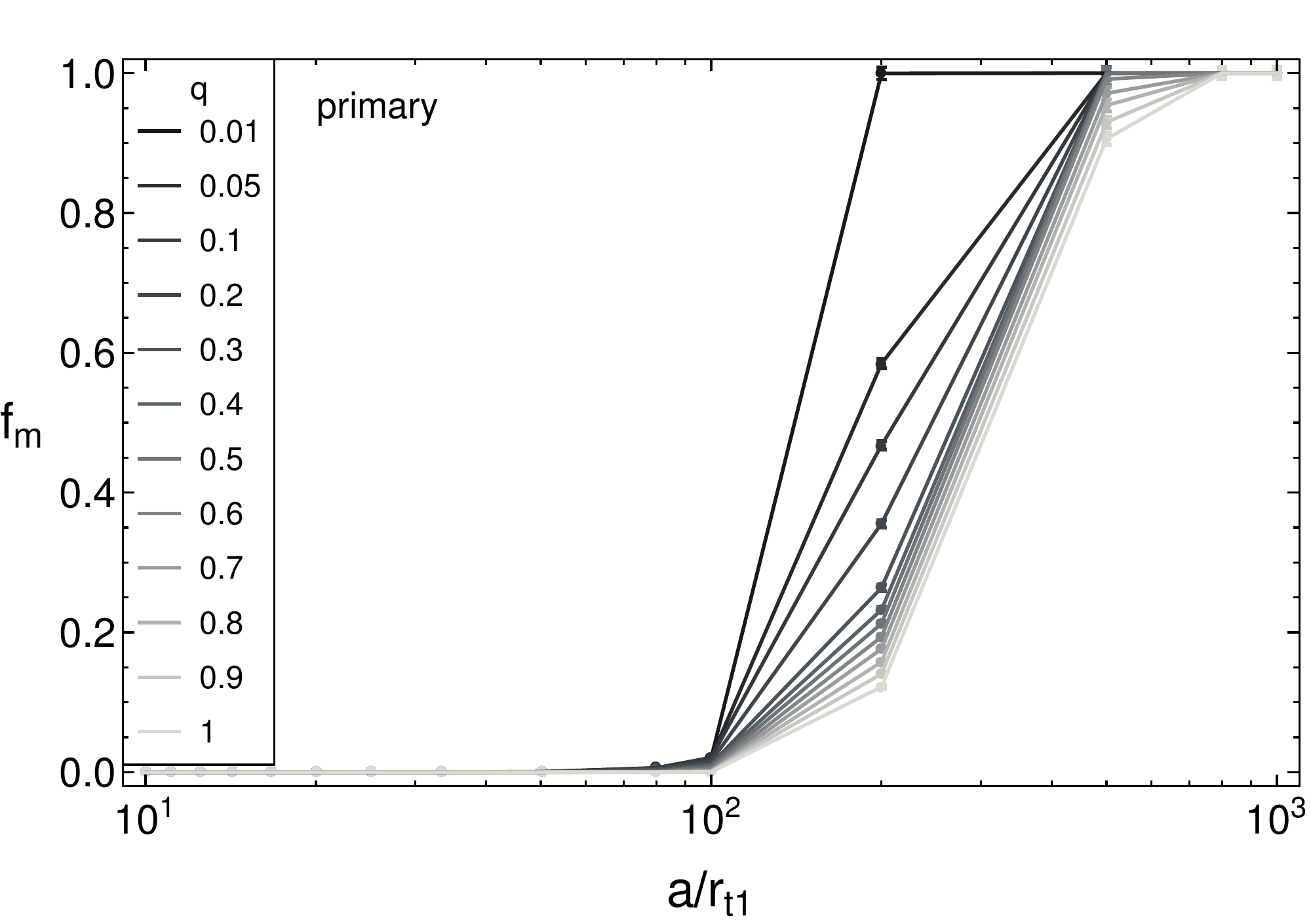}}\hfill
\subfloat{\includegraphics[width=0.49\textwidth]{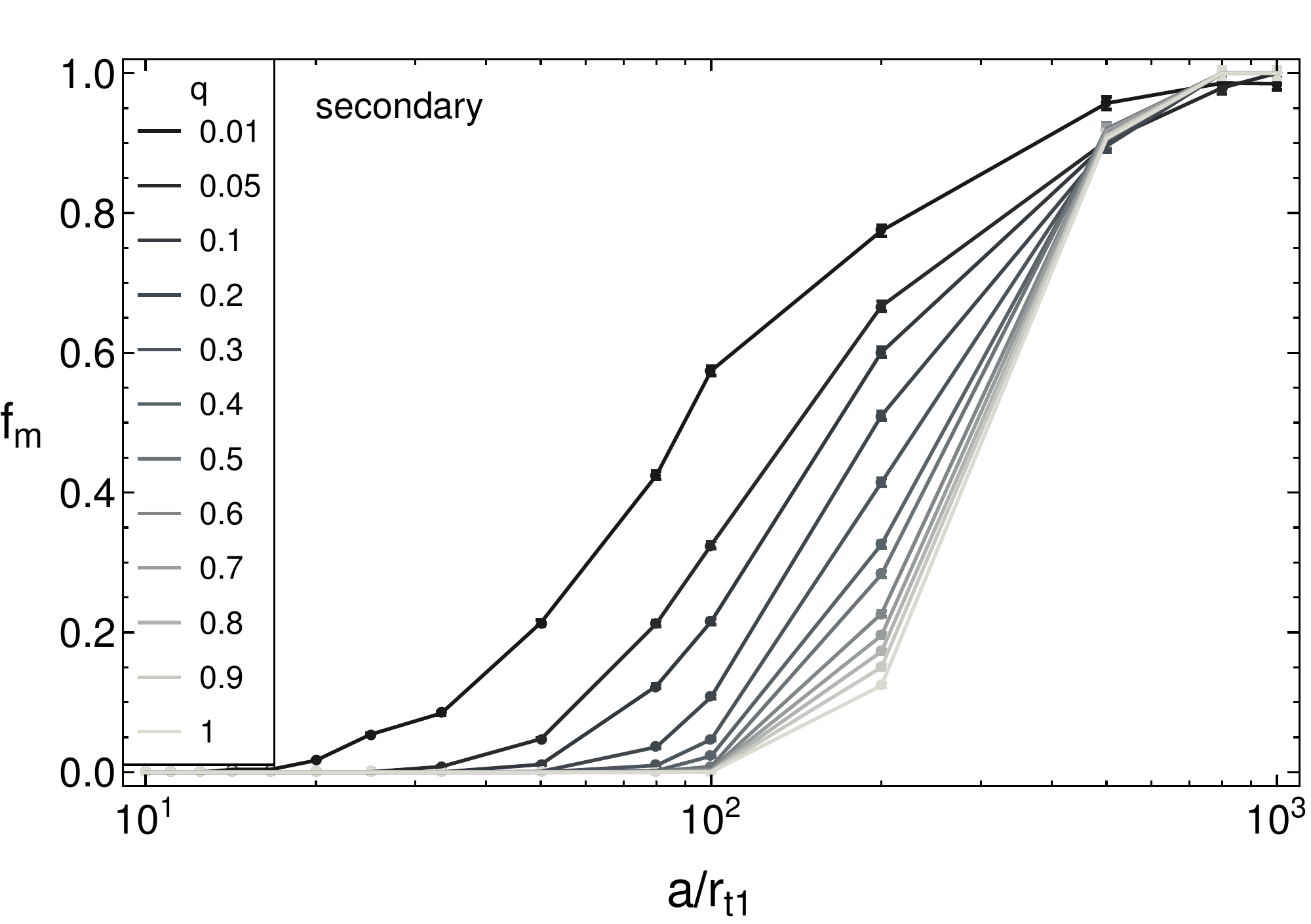}}
\caption{The fraction $f_m$ of bound-debris disruptions that satisfy the criterion for the ``frozen-in'' model (Eq. \ref{eq:modelcondition}), for stars disrupted by the primary (left) and secondary (right). The model is accurate for $500 \lesssim a < 1000$, and represents an average behavior for $a \lesssim 500$ \citep{coughlin17}.}
\label{fig:fmodelvalid}
\end{figure*}

Figure \ref{fig:fmodelvalid} shows the fraction of bound-debris disruptions that satisfy the above constraints. The model is robust for disruptions by wide binaries ($500 \lesssim a < 1000$), and quickly deviates as the binary tightens ($a \lesssim 500$). However, \citet{coughlin17} showed that the single BH model reflects the average accretion behavior observed in hydrodynamic simulations for $a = 100$, and can thus be interpreted in that sense for tight binaries. Importantly, in Section \ref{subsec:inspiral} we show that gravitational radiation dominates the binary inspiral beginning at $a \sim 100$, which suppresses the likelihood of tidal disruptions below this distance. With these qualifications in mind, we present the results from our model over the full range of separations, but focus on the range $100 \lesssim a < 1000$.

The ``frozen-in'' model that we use is similar to that of \citet{lodato09} and \citet{coughlin14}, though with one important difference: a star disrupted by a binary SMBH can have a nonzero center of mass energy (Section \ref{subsec:properties}). To briefly summarize, we use the impulse approximation in which the star moves as a rigid, nonrotating body until it reaches the tidal radius of the disrupting BH, at which point the gravity of the BH overwhelms the self-gravity of the star, and the post-disruption stellar debris elements then travel on independent orbits in the potential of the BH. The energies of the debris elements are thus conserved, and are set solely by their kinetic and potential energy at the tidal radius; the former is the same for all of the elements since the star is not rotating. Since $R_* \ll r_{t1}, r_{t2}$, the star at the tidal radius can be divided into debris slices, each with a constant energy. We assume the stars are Sun-like and have uniform density $\rho_* = 3M_\odot / 4\pi R^3_\odot$. This model allows us to derive simple analytic results for the properties of the debris.

If $\tilde{m}$ is the mass of an individual debris slice and $t$ is the time after disruption, then we can define the dimensionless mass $m = \tilde{m} / M_*$, and time $\tau = t / \tau_0$ using the timescale
\begin{equation}
\tau_0 = \frac{2\pi GM_\bullet}{(2 \Delta\epsilon)^{3/2}} \simeq 41 \textrm{ d } \left(\frac{M_\bullet}{10^6 M_\odot}\right)^{1/2} \left(\frac{M_*}{M_\odot}\right)^{-1} \left(\frac{R_*}{R_\odot}\right)^{3/2}
\label{eq:tauscale}
\end{equation}
where $\Delta\epsilon$ is the energy spread across the star due to the BH (Eq. \ref{eq:deltaepsilon}). The parameter $\tau_0$ contains the properties of the star and black hole involved in the disruption event. For Sun-like stars and $M_\bullet = 10^6 M_\odot$, an interval of $\tau = 1$ corresponds to about 41 days. In these units, the mass fallback rate for the debris is
\begin{equation}
\frac{dm}{d\tau} = \frac{1}{2} \tau^{-5/3} \left( 1 - \varepsilon^2_c - 2\varepsilon_c \tau^{-2/3} - \tau^{-4/3} \right)
\label{eq:dmdtau}
\end{equation}
where the critical energy must be in the range $\varepsilon_c < 1$ for the disruption to produce some bound debris.

For a star disrupted with critical energy $\varepsilon_c$, the most bound debris returns at a time $\tau_\textrm{mb} = (1-\varepsilon_c)^{-3/2}$ and the peak return rate occurs at $\tau_\textrm{peak} = 27 \left[ (4\varepsilon^2_c + 45)^{1/2} - 7 \varepsilon_c \right]^{-3/2}$. We can then define the rise time as
\begin{equation}
\begin{split}
\tau_\textrm{rise} &= \tau_\textrm{peak} - \tau_\textrm{mb} \\
&= 27 \left[ (4\varepsilon^2_c + 45)^{1/2} - 7 \varepsilon_c \right]^{-3/2} - (1-\varepsilon_c)^{-3/2}
\end{split}
\label{eq:taurise}
\end{equation}
For disruptions in which all of the stellar debris is bound ($\varepsilon_c < -1$), the return rate has a finite duration and the least bound debris returns at $\tau_\textrm{lb} = (-1-\varepsilon_c)^{-3/2}$.

\begin{figure*}
\centering
\subfloat[Mass return rate]{\includegraphics[width=0.49\textwidth]{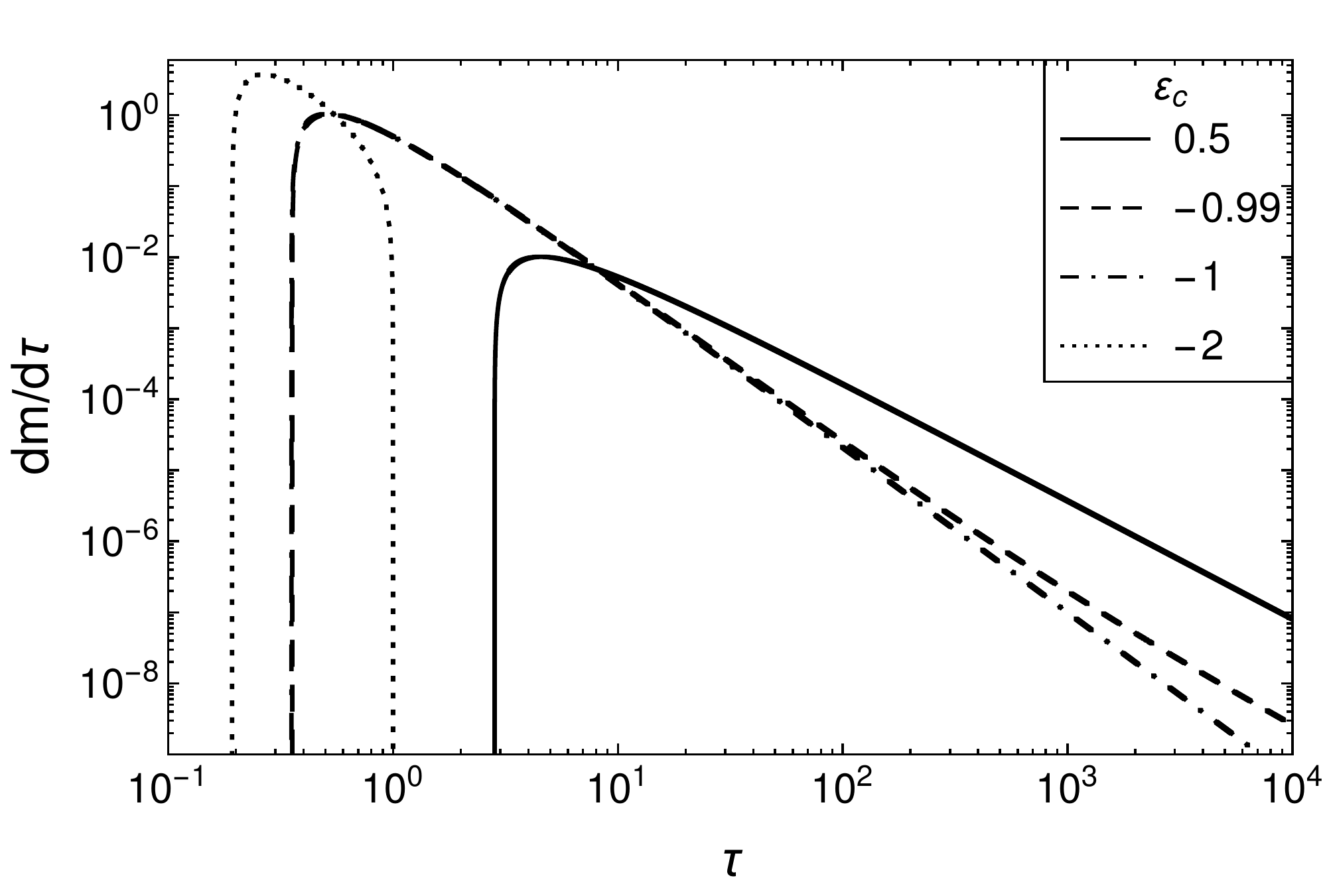}\label{fig:funcsa}}\hfill
\subfloat[Rise time]{\includegraphics[width=0.49\textwidth]{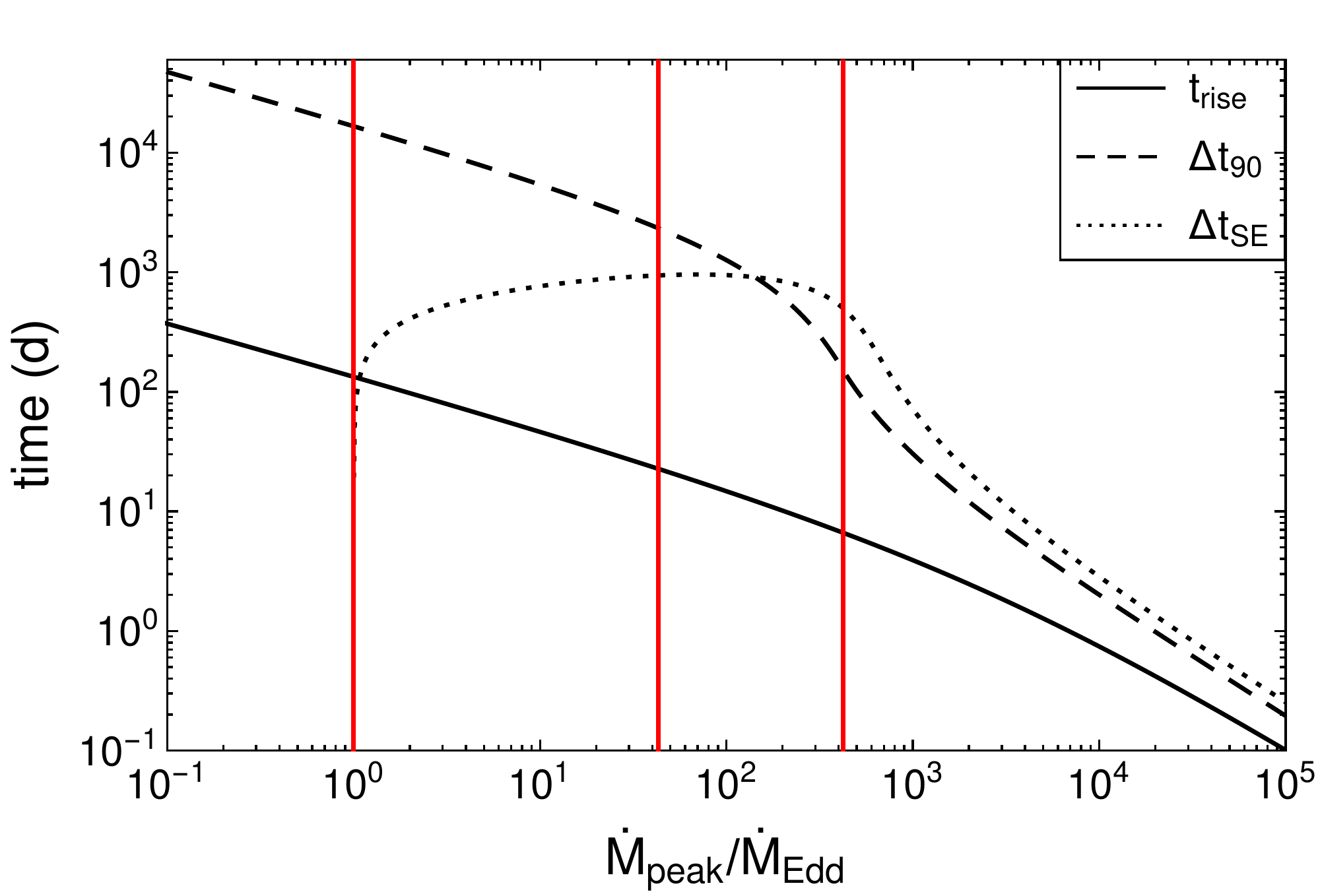}\label{fig:funcsb}}
\caption{a) The mass return rate $\frac{dm}{d\tau}$ (Eq. \ref{eq:dmdtau}; see surrounding text for the definitions of $m$ and $\tau$) for four values of $\varepsilon_c$, corresponding to a partially bound star ($\varepsilon_c = 0.5$), a fully bound star ($\varepsilon_c = -2$), a star exactly at the boundary of the two $(\varepsilon_c = -1)$, and one slightly above it $(\varepsilon_c = -0.99)$. For $M_\bullet = 10^6 M_\odot$ and Sun-like stars, an interval of $\tau = 1$ corresponds to roughly 41 days. b) The rise time $t_\textrm{rise} = \tau_0 \tau_\textrm{rise} = \tau_0 ( \tau_\textrm{peak} - \tau_\textrm{mb} )$, the time $\Delta t_{90} = \tau_0 \Delta \tau_{90} = \tau_0 ( \tau_{90} - \tau_\textrm{mb} )$ for 90\% of the mass to return, and the time $\Delta t_{SE} = \tau_0 \Delta \tau_{SE}$ spent super-Eddington, as a function of the dimensionless peak fallback rate $\dot{M}_\textrm{peak}/\dot{M}_\textrm{Edd}$ (Eq. \ref{eq:mdotpeakovermedd}; see surrounding text for the parameters used), all for $M_\bullet = 10^6 M_\odot$. The red vertical lines correspond to slices in which 1) left: $\dot{M}_\textrm{peak} = \dot{M}_\textrm{Edd}$ ($\varepsilon_c = 0.668$); 2) center: $\varepsilon_c = 0$; and 3) right: $\varepsilon_c = -1$.}
\label{fig:funcs}
\end{figure*}

Figure \ref{fig:funcsa} shows the mass return rate $\frac{dm}{d\tau}$ for four different values of $\varepsilon_c$. The curves all exhibit cutoffs at early times corresponding to the return time of the most bound debris. Fully bound stars ($\varepsilon_c < -1$) also have a finite return time for the least bound debris, shown by the abrupt drop in the curve for $\varepsilon_c = -2$ at $\tau \simeq 1$. Partially bound stars with $-1 < \varepsilon_c < 1$ exhibit the characteristic $\tau^{-5/3}$ scaling at late times derived for disruptions by a single BH (\citealt{rees88}, updated in \citealt{phinney89}). A star at the boundary ($\varepsilon_c = -1$) has $\frac{dm}{d\tau} \sim \tau^{-7/3}$ at late times, slightly steeper than the previous case. The curves for $\varepsilon_c = -0.99$ and $-1$ show these two different return rates, which begin to diverge at roughly $\tau \simeq 20$. The abrupt change arises because we model the stars with uniform density, leading to a density discontinuity at the stellar radius.

The dimensioned peak mass return rate $\dot{M}_\textrm{peak}$ can be expressed in units of the Eddington accretion rate $\dot{M}_\textrm{Edd}$ as
\begin{equation}
\frac{\dot{M}_\textrm{peak}}{\dot{M}_\textrm{Edd}} = \frac{\kappa \varepsilon c}{\sqrt{6\pi^3 G}} \langle \rho_* \rangle^{1/2} \left(\frac{M_\bullet}{M_*}\right)^{-3/2} \left(\frac{dm}{d\tau}\right)_\textrm{peak}
\label{eq:mdotpeakovermedd}
\end{equation}
For Sun-like stars of uniform density, the mean stellar density is $\langle \rho_* \rangle = 3M_\odot / 4\pi R^3_\odot = 1.41 \times 10^3$ kg$\cdot$m$^{-3}$, and the opacity, dominated by ionized hydrogen, is $\kappa = \sigma_T / m_p = 0.04$ m$^2 \cdot$kg$^{-1}$. We use a radiative efficiency of $\varepsilon = 0.1$. The rate $\dot{M}_\textrm{peak} / \dot{M}_\textrm{Edd}$, though dimensionless, depends on the black hole mass and the stellar parameters.

Figure \ref{fig:funcsb} shows the rise time $t_\textrm{rise} = \tau_0 \tau_\textrm{rise}$ (Eq. \ref{eq:taurise}), along with the time $\Delta t_{90} = \tau_0 \Delta \tau_{90} = \tau_0 (\tau_{90} - \tau_\textrm{mb})$ for $90\%$ of the debris to return and the duration $\Delta t_{SE} = \tau_0 \Delta \tau_{SE}$ of super-Eddington fallback, plotted versus $\dot{M}_\textrm{peak} / \dot{M}_\textrm{Edd}$, all for $M_\bullet = 10^6 M_\odot$. The rise time and peak return rate are inversely related: disruptions with lower $\varepsilon_c$ have shorter durations and higher peak return rates. Disruptions with preferentially bound debris ($\varepsilon_c < 0$) have rise times $t_\textrm{rise} \lesssim 23$ d ($\tau_\textrm{rise} \lesssim 0.55$) and super-Eddington fallback rates $\dot{M}_\textrm{peak} / \dot{M}_\textrm{Edd} \gtrsim 43$, where the limits here give the values for the canonical case $\varepsilon_c = 0$. Of particular note, disruptions with fully bound debris ($\varepsilon_c < -1$) exhibit distinguishing behavior, with short rise times $t_\textrm{rise} \lesssim 6.6$ d ($\tau_\textrm{rise} \lesssim 0.16$) and large fallback rates $\dot{M}_\textrm{peak} / \dot{M}_\textrm{Edd} \gtrsim 4.2 \times 10^2$.

In a TDE by an isolated SMBH, the rise time and peak luminosity can probe the black hole mass (through $t_\textrm{rise} \sim M^{1/2}_\bullet$ and $\dot{M}_\textrm{peak} / \dot{M}_\textrm{Edd} \sim M^{-3/2}_\bullet$) if the surrounding stellar population is well constrained, and inform us about the stellar environment (through $t_\textrm{rise} \sim M^{-1}_* R^{3/2}_*$ and $\dot{M}_\textrm{peak} / \dot{M}_\textrm{Edd} \sim \langle \rho_* \rangle^{1/2}$) if the BH mass is known. If the SMBH is in a binary, then for some range of $q$ and $a$ these two indicators may exhibit qualitatively different behavior than in the single BH case, and can then be used to confirm the presence of a binary and characterize its properties. The two depend on the critical energy $\varepsilon_c$ of the disrupted star (Eq. \ref{eq:epsc}) through their dimensionless counterparts $\tau_\textrm{rise}$ and $\left(\frac{dm}{d\tau}\right)_\textrm{peak}$, and $\varepsilon_c$ in turn depends on $q$ and $a$ (Figure \ref{fig:histepsc}, \ref{fig:histepscappendix}).

We investigate the properties of the bound post-disruption debris using a primary mass $M_1 = 10^6 M_\odot$. We examine the distributions in the rise time and peak return rate over our parameter range, and calculate the fraction that are prompt ($t_\textrm{rise} \lesssim 10$ d) and have highly super-Eddington fallback ($\dot{M}_\textrm{peak} / \dot{M}_\textrm{Edd} \gtrsim 2 \times 10^2$) with respect to the disrupting BH, as this behavior is extreme for a single BH. These values correspond to $\varepsilon_c \lesssim -0.6$ for disruptions by the primary ($M_1 = 10^6 M_\odot$), though both will not correspond to a single range for the different secondary masses.

\begin{figure*}
\centering
\subfloat{\includegraphics[width=0.49\textwidth]{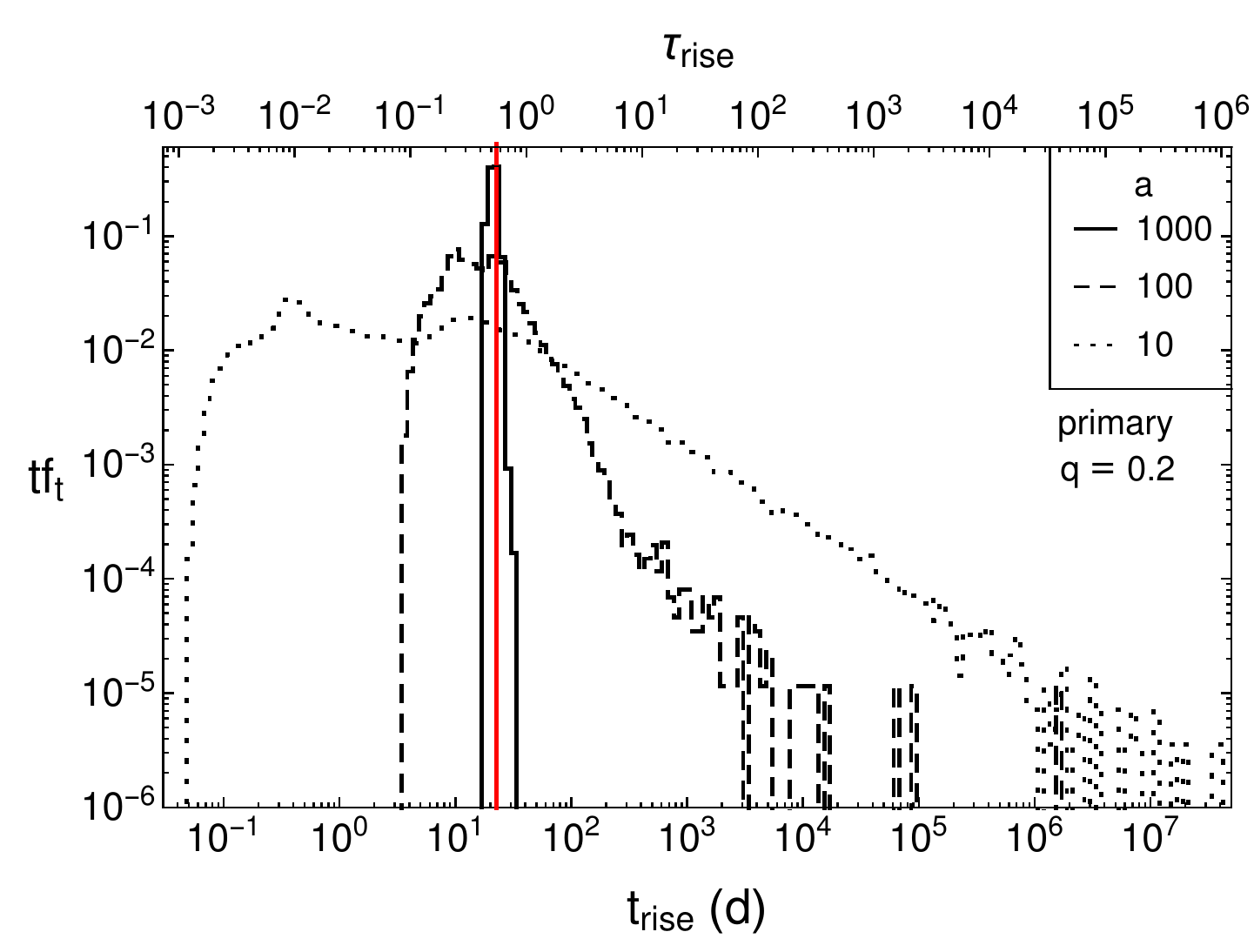}}\hfill
\subfloat{\includegraphics[width=0.49\textwidth]{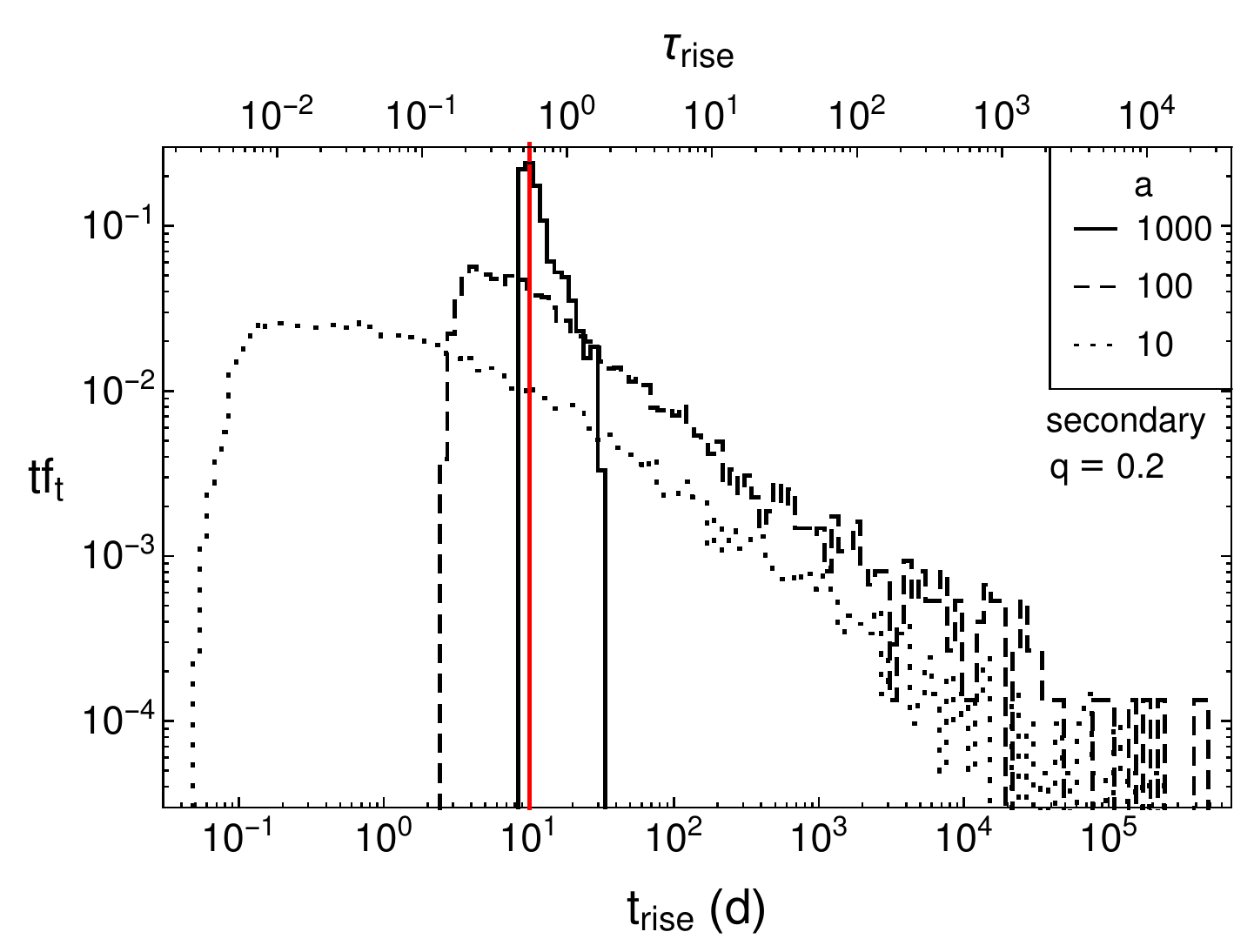}}
\caption{The probability $t f_t$ that the bound debris of the disrupting black hole will produce a rise time $t_\textrm{rise} = \tau_0 \tau_\textrm{rise}$ for a primary mass $M_1 = 10^6 M_\odot$, $q = 0.2$, and $a=1000$ (solid), $100$ (dashed), and $10$ (dotted). The left panel shows disruptions by the primary and the right one shows those by the secondary. The logarithmic bin widths are $\Delta_t = 0.05$ and the heights are the probabilities in each bin. The red line marks the rise time $t_\textrm{rise} = \tau_0 (3\sqrt{3} - 5^{3/4})/5^{3/4}$ for a TDE from a star with $\varepsilon_c = 0$ (parabolic orbit, half of the debris is bound) disrupted by the appropriate black hole. Figure \ref{fig:histtriseappendix} in Appendix \ref{sec:appendixa} presents similar histograms for a range of $q$ in our parameter space.}
\label{fig:histtrise}
\end{figure*}

\begin{figure*}
\centering
\subfloat{\includegraphics[width=0.49\textwidth]{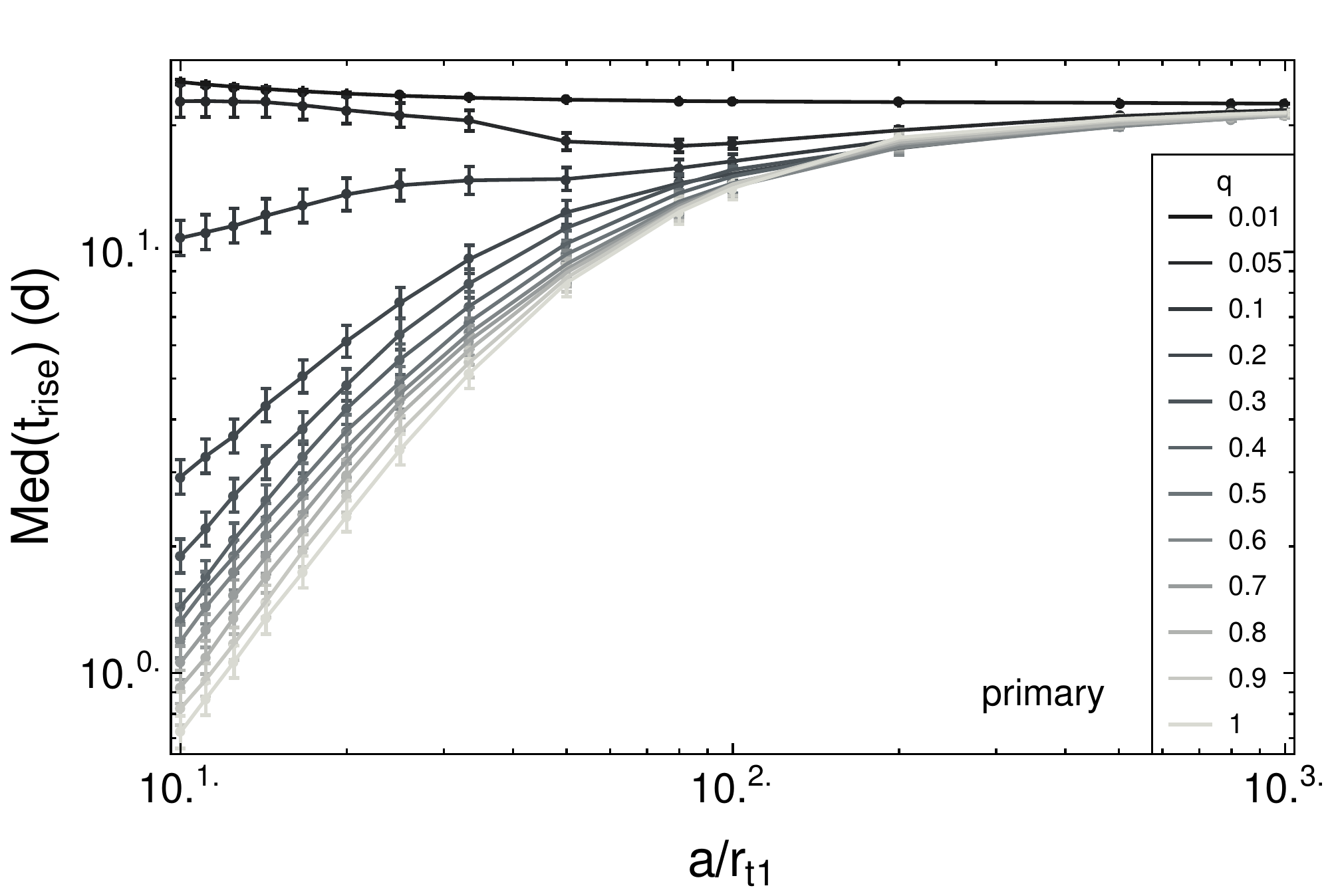}}\hfill
\subfloat{\includegraphics[width=0.49\textwidth]{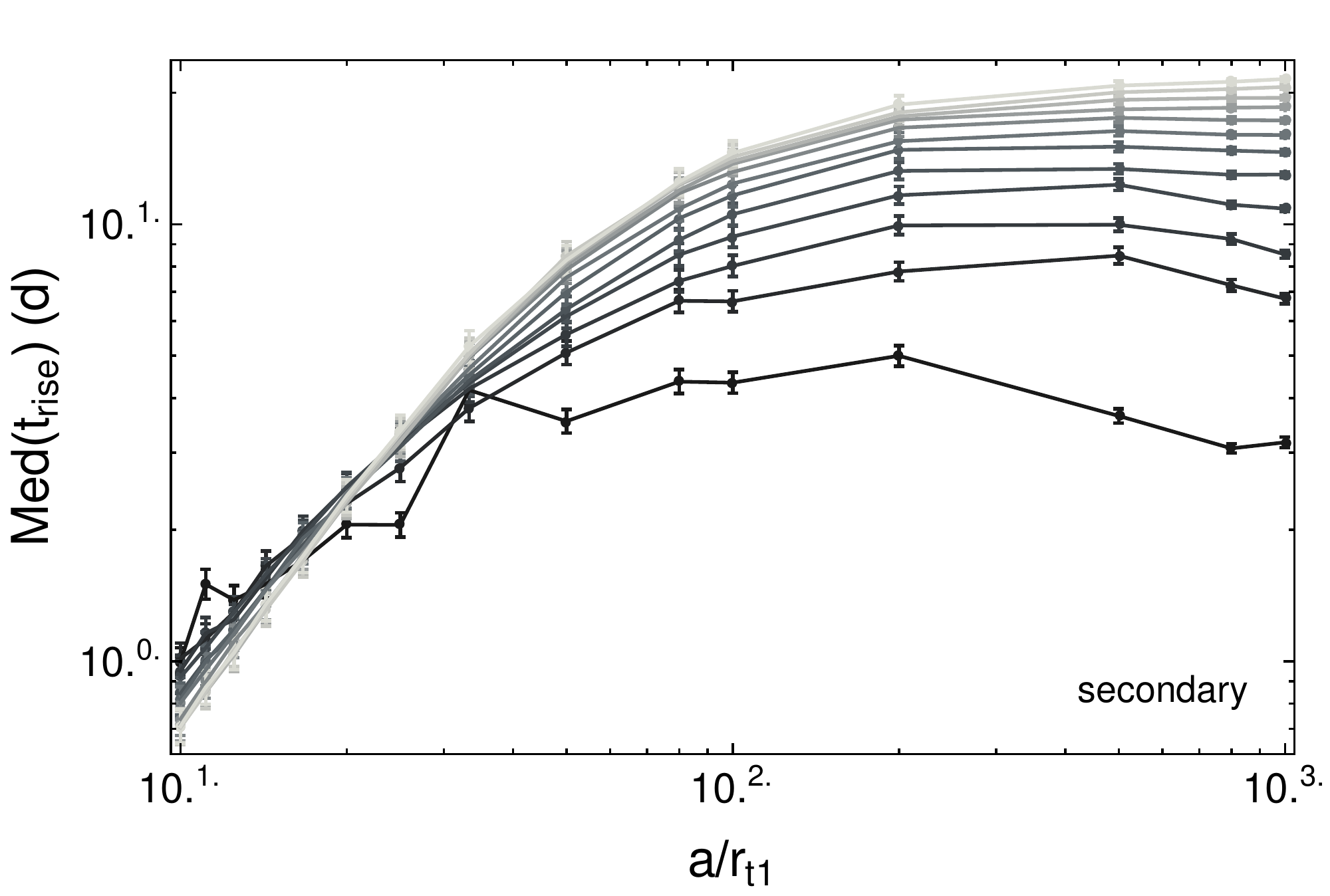}}
\caption{The median rise time $\operatorname{Med}(t_\textrm{rise})$ in the mass return rate for a primary mass $M_1 = 10^6 M_\bullet$. The left panel shows disruptions by the primary and the right panel shows disruptions by the secondary. The error bars have half-width $20\sigma \tau_0$, where $\sigma = \sqrt{f_b / N_e}$ are the standard deviations assuming a Poisson distribution and $N_e = 5 \times 10^6$ is the total number of encounters.
}
\label{fig:taurise}
\end{figure*}

\begin{figure*}
\centering
\subfloat{\includegraphics[width=0.49\textwidth]{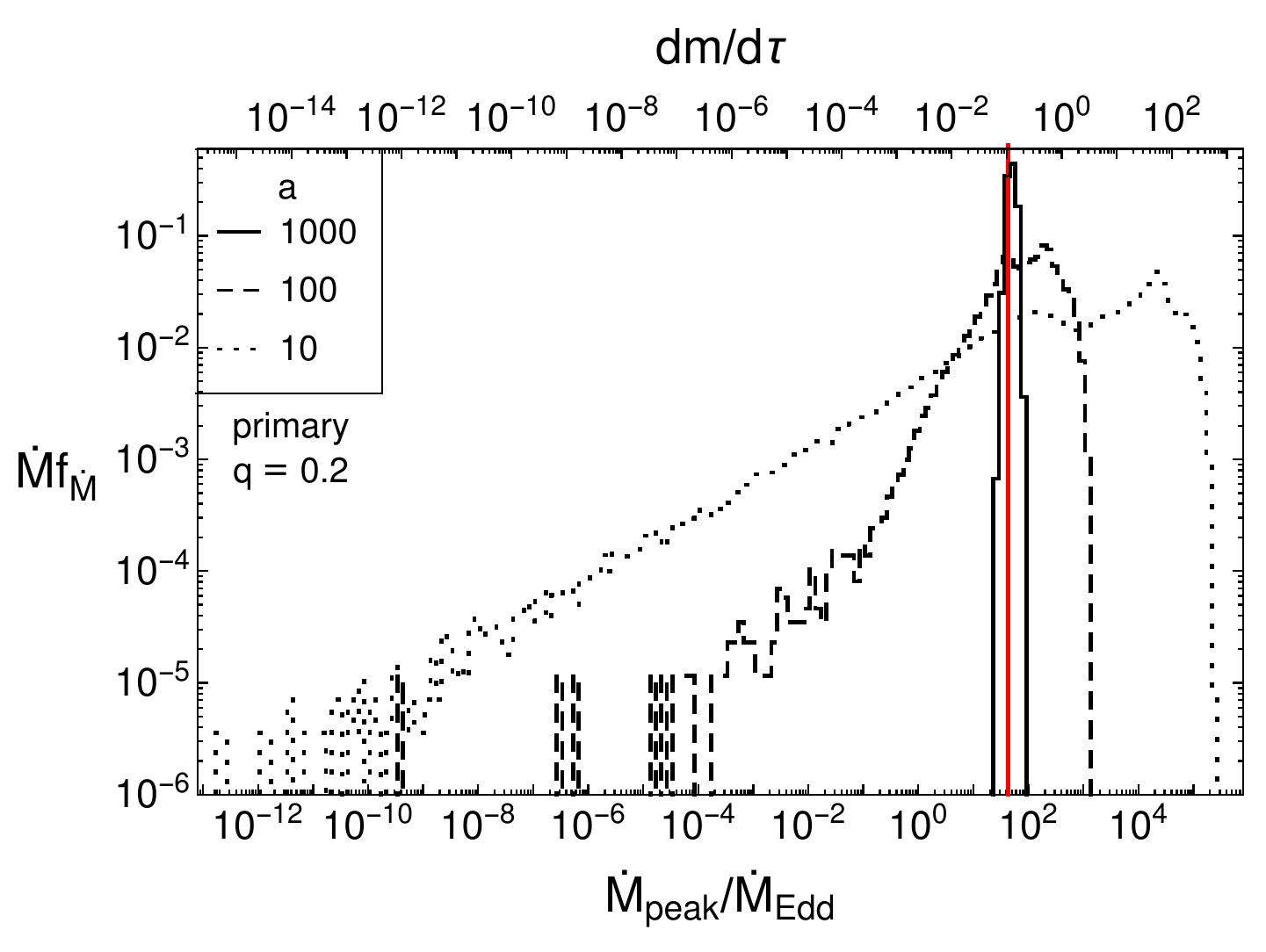}}\hfill
\subfloat{\includegraphics[width=0.49\textwidth]{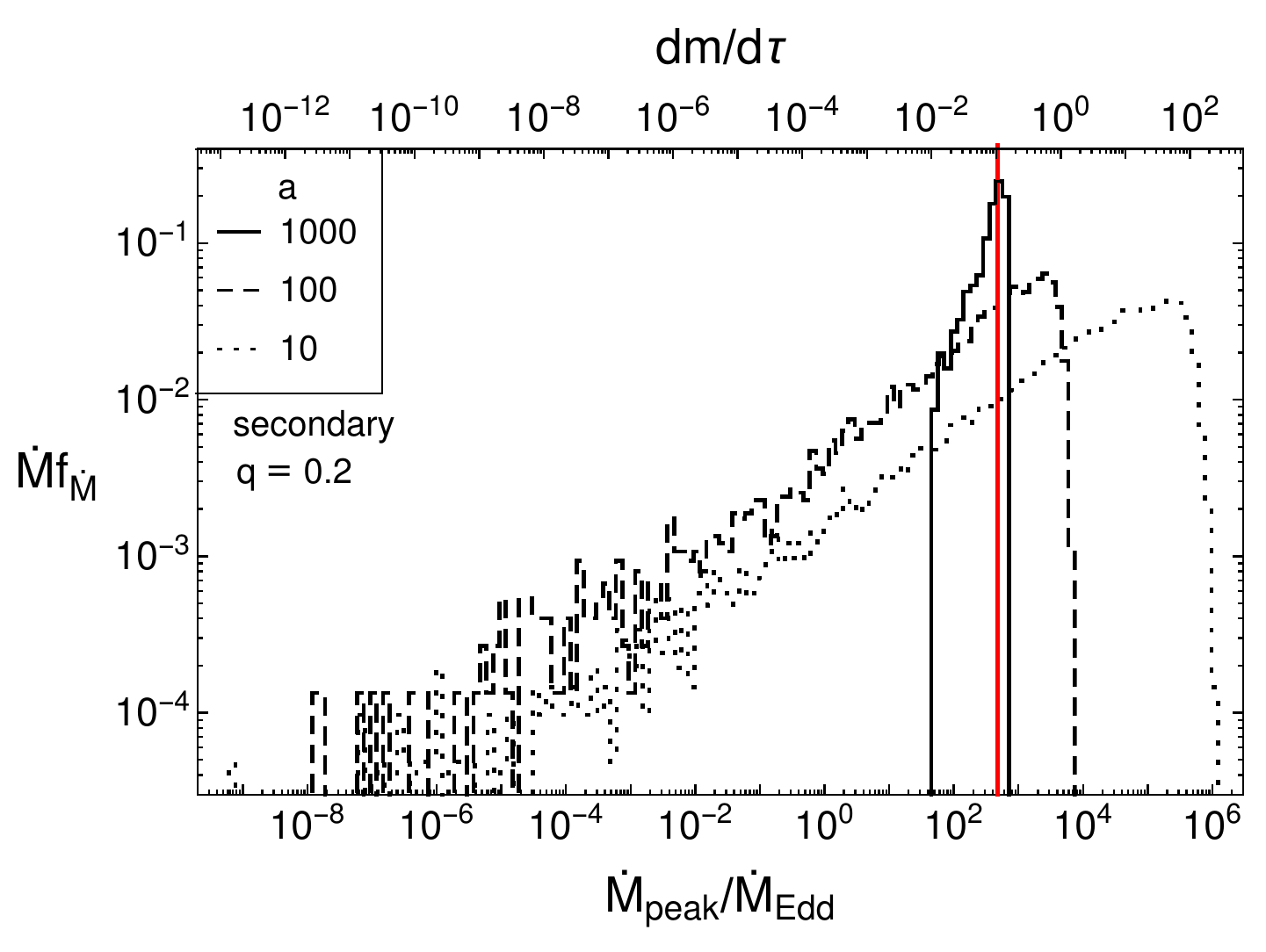}}
\caption{The probability $\dot{M} f_{\dot{M}}$ that the bound debris of the disrupting black hole will produce a peak return rate $\dot{M}_\textrm{peak}/\dot{M}_\textrm{Edd}$ (Eq. \ref{eq:mdotpeakovermedd}; see surrounding text for the parameters) for a primary mass $M_1 = 10^6 M_\odot$, $q = 0.2$, and $a=1000$ (solid), $100$ (dashed), and $10$ (dotted). The left panels show disruptions by the primary and the right ones show those by the secondary. The logarithmic bin widths are $\Delta_{\dot{M}} = 0.1$ and the heights are the probabilities in each bin. The red line shows the peak return rate for a TDE from a star with $\varepsilon_c = 0$ (parabolic orbit, half of the debris is bound) disrupted by the appropriate black hole. Figure \ref{fig:histmdotpeakappendix} in Appendix \ref{sec:appendixa} presents similar histograms for a range of $q$ in our parameter space.}
\label{fig:histmdotpeak}
\end{figure*}

\begin{figure*}
\centering
\subfloat{\includegraphics[width=0.49\textwidth]{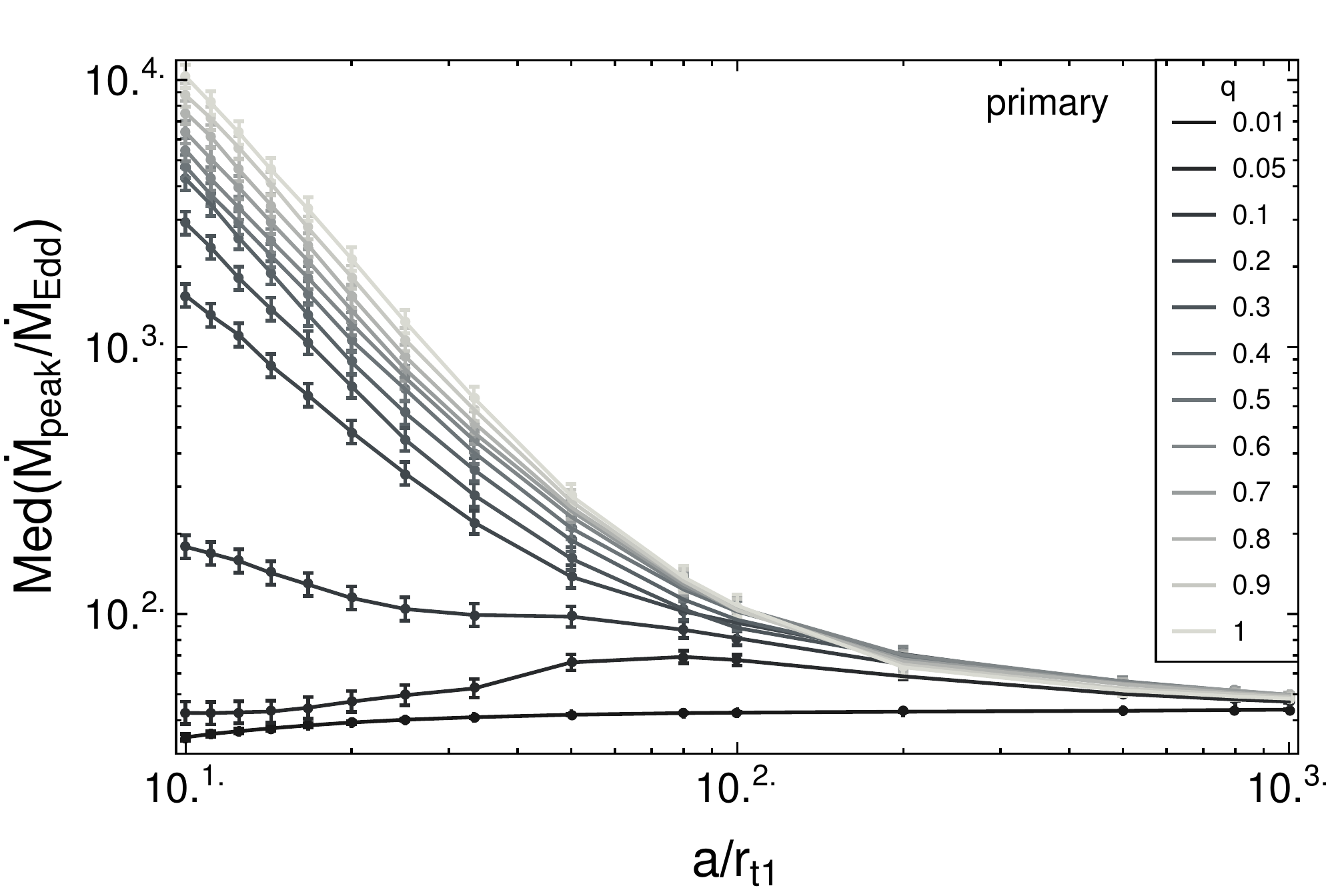}}\hfill
\subfloat{\includegraphics[width=0.49\textwidth]{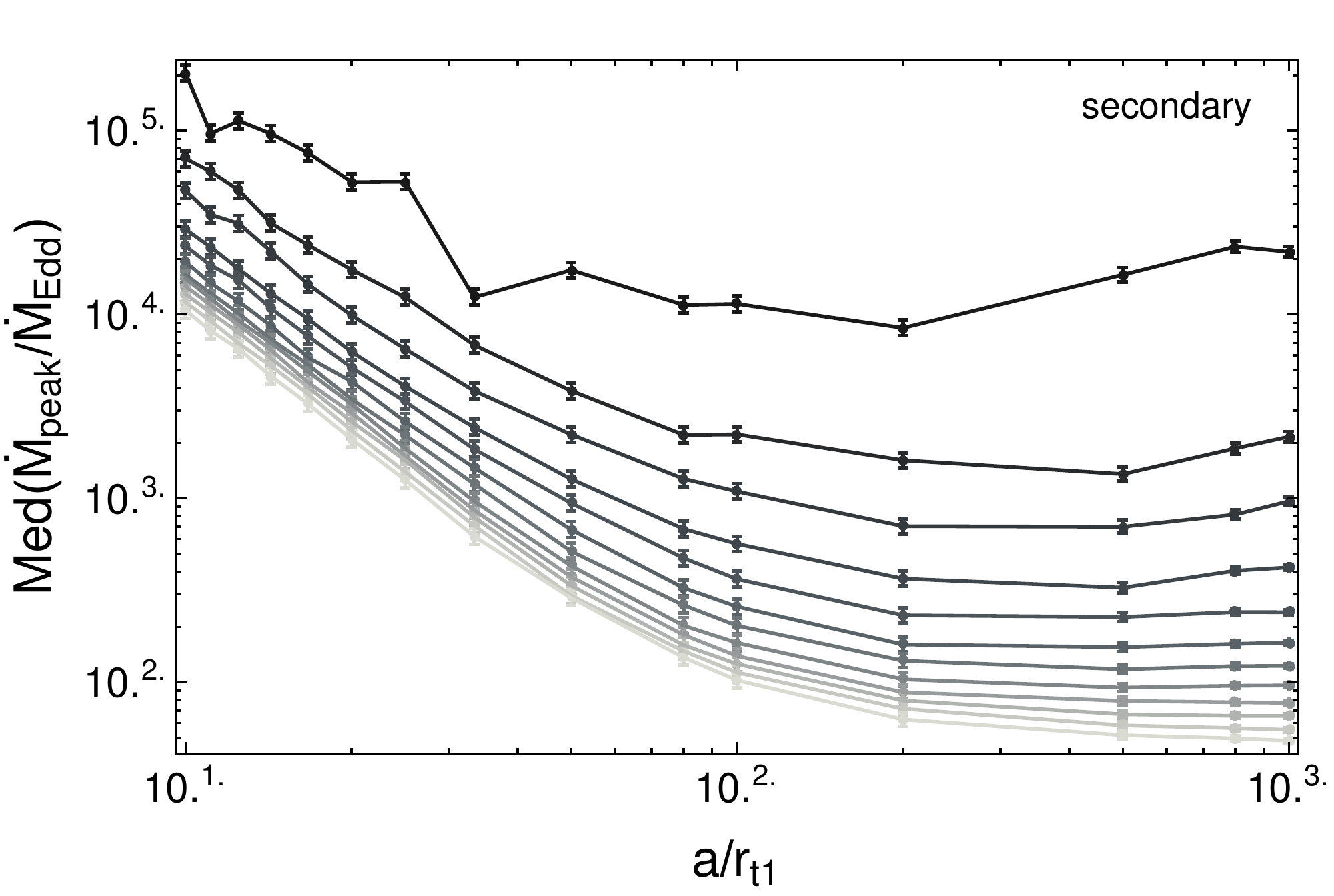}}
\caption{The median peak fallback rate in units of the Eddington rate $\operatorname{Med} (\dot{M}_\textrm{peak}/\dot{M}_\textrm{Edd} )$ (Eq. \ref{eq:mdotpeakovermedd}; see surrounding text for the parameters used) for a primary mass $M_1 = 10^6 M_\odot$. The left panel shows disruptions by the primary and the right panel shows those by the secondary.}
\label{fig:mdotpeak}
\end{figure*}

We can summarize our findings as follows. For tight ($a \lesssim 100$), nearly equal-mass ($q \gtrsim 0.2$) binaries, a sizable portion of the bound-debris disruptions will be short duration, highly super-Eddington TDEs. For tight, unequal-mass binaries, the low mass secondaries also have a large proportion of bound disruptions with this behavior. Wide binaries ($a \gtrsim 500$) nearly always produce bound disruptions with $\varepsilon_c = 0$ for all mass ratios; for unequal masses, this implies that the secondary mostly produces disruptions with short rise times and super-Eddington peak rates. We outline these results in Figures \ref{fig:histtrise} -- \ref{fig:mdotpeak} (and \ref{fig:histtriseappendix} -- \ref{fig:histmdotpeakappendix}, \ref{fig:ptrise1appendix} -- \ref{fig:pmdotpeak1appendix}) and describe them in more detail in the rest of this subsection.

First, though, we note that to connect these results to total disruption rates for a given black hole, one must multiply these probabilities (which assume that a bound disruption has occurred), with the relative probability of disruption (Figure \ref{fig:lambdat12b}) and the probability that a disruption has some bound debris (Figure \ref{fig:fbound}). We present these total probabilities in Figures \ref{fig:histtrisefullprobappendix} -- \ref{fig:histmdotpeakfullprobappendix} and \ref{fig:ptrise1fullprobappendix} -- \ref{fig:pmdotpeak1fullprobappendix}, and save this final analysis for Section \ref{subsec:inspiral} when we study the binary merger and calculate the closest separation at which we expect to observe disruptions.

Figure \ref{fig:histtrise} shows histograms of the probabilities $t f_t$ for the rise time $t_\textrm{rise} = \tau_0 \tau_\textrm{rise}$ (Eq. \ref{eq:taurise}) for $M_1 = 10^6 M_\odot$, $q = 0.2$, and a range of values of $a$. As the binary tightens, the distribution becomes broader and flatter and the peak shifts to lower values. The tight binaries ($a \lesssim 100$) tend to produce prompt TDEs, with $\sim 65 - 25 \%$ ($\sim 85 - 50 \%$) of bound-debris disruptions by the primary (secondary) having $t_\textrm{rise} \sim 4 \times 10^{-2} - 10$ d ($t_\textrm{rise} \sim 4 \times 10^{-2} - 10$ d), where the lower limit depends on $a$. In contrast, the bound-debris disruptions by wide binaries ($a \gtrsim 500$) are nearly always similar to the canonical case from an isolated black hole ($\varepsilon_c = 0$, half of the debris remains bound), namely with $\tau_\textrm{rise} \simeq 0.55$ ($\simeq 23$ d for $M_1 = 10^6 M_\odot$, $\simeq 10$ d for $M_2 = 0.2 M_1$). In our model, the rise time of a disrupted star is determined by the mass of the disrupting black hole (Eq. \ref{eq:tauscale}) and by the value of $\varepsilon_c$ (Eq. \ref{eq:taurise}, Fig. \ref{fig:funcsa}). Therefore, the trends for tight and wide binaries arise due to the $t_\textrm{rise} \sim M^{1/2}_\bullet$ scaling (which locates the peak) and due to the distribution of $\varepsilon_c$ discussed previously (which determines the shape) (Figure \ref{fig:histepsc}).

The rise time distribution varies with the mass ratio (Figure \ref{fig:histtriseappendix}). For tight ($a \lesssim 100$), nearly equal-mass ($q \gtrsim 0.2$) binaries, the primary (secondary) produces $\sim 25 - 40 \%$ ($\sim 55 - 35 \%$) of its bound-debris disruptions with $t_\textrm{rise} \lesssim 10$ d for $a \simeq 100$, and $\sim 65 - 85 \%$ ($\sim 85 - 80 \%$) for $a \simeq 10$. In this range of $q$, the secondary begins ``stealing'' disruptions from the primary and accounts for an increasing fraction of the total disruption rate (Figure \ref{fig:lambdat12a}), though as the binary tightens a decreasing fraction of disruptions by both black holes will have bound debris (Figure \ref{fig:fbound}), leading to a tradeoff.

Tight, unequal mass binaries have a lower fraction of bound-debris TDEs that are prompt on average. For $q \sim 0.05$ over the range $a \lesssim 100$, the primary (secondary) produces $\sim 33 - 15 \%$ ($\sim 85 - 62 \%$) of its bound-debris disruptions with $t_\textrm{rise} \lesssim 10$ d, though the high fraction for the secondary is diminished in the observed rate, since for this value of $q$, the primary produces the bulk of disruptions (Figure \ref{fig:lambdat12}), and most debris from the secondary is unbound (Figure \ref{fig:fbound}). However, if the secondary disrupts a star in a chance encounter, then it has a high probability of producing a TDE with a rise time several orders of magnitude shorter than one by the primary.

Wide binaries ($a \gtrsim 500$) produce bound-debris disruptions peaked and concentrated near $\tau_\textrm{rise} \simeq 0.55$ (corresponding to $\varepsilon_c = 0$) over the entire range of $q$, though for lower $q$ the rise time distributions for the secondary are skewed towards higher values since their distributions of $\varepsilon_c$ are skewed towards more partially bound ($0 < \varepsilon_c < 1$) debris. Nearly all disruptions by both the primary and secondary will have some bound debris for any $q$ (Figure \ref{fig:fbound}). For $q \gtrsim 0.2$, both black holes have a sizable disruption rate (Figure \ref{fig:lambdat12}). For $q \lesssim 0.2$, this value of $\tau_\textrm{rise}$ for the secondary yields $t_\textrm{rise} = \tau_0 \tau_\textrm{rise} \lesssim 10$ d, but its relative rate of disruption is small (Figure \ref{fig:lambdat12}), so again these disruptions are rare, but a chance encounter with the secondary will most likely result in a TDE with a short rise time.

The median rise time $\operatorname{Med} (\tau_\textrm{rise})$ reflects the dependence of the distribution on $q$ and $a$, and is presented in Figure \ref{fig:taurise}. The curves clearly illustrate the trend described in the preceding paragraphs: for nearly equal-mass black holes in a close binary, most bound disruptions will produce TDEs with durations several times shorter than those by isolated black holes. In unequal-mass binaries, most disruptions by the secondary with bound debris can also produce such events for all separations. Wide binaries will nearly always produce disruptions with $\varepsilon_c = 0$; for a low mass ratio, most bound-debris disruptions by the secondary will also have a short rise time.

Figure \ref{fig:histmdotpeak} shows histograms of $\dot{M} f_{\dot{M}}$ for the peak return rate divided by the Eddington rate (Eq. \ref{eq:mdotpeakovermedd}; see text for the parameters used) with respect to the disrupting BH, for $M_1 = 10^6 M_\odot$, $q = 0.2$, and a range of values of $a$. Figure \ref{fig:histmdotpeakappendix} shows these histograms for different values of $q$. The rise time and peak rate are inversely related (Figure \ref{fig:funcsb}), so the distributions exhibit analogous behavior to those for $\tau f_\tau$, but with high peak rates instead of short rise times. For tight binaries ($a \lesssim 100$) with $q \gtrsim 0.2$, the primary (secondary) produces $\sim 25 - 40 \%$ ($\sim 65 - 35 \%$) of its bound-debris disruptions with $\dot{M}_\textrm{peak} / \dot{M}_\textrm{Edd} \gtrsim 2 \times 10^2$ for $a \simeq 100$, and $\sim 65 - 85 \%$ ($\sim 90 - 80 \%$) for $a \simeq 10$. For $q \sim 0.05$ over the range $a \lesssim 100$, the primary (secondary) produces $\sim 33 - 16 \%$ ($\sim 90 - 78 \%$) of its bound-debris disruptions with $\dot{M}_\textrm{peak} / \dot{M}_\textrm{Edd} \gtrsim 2 \times 10^2$. Wide binaries ($a \gtrsim 500$) replicate the canonical, isolated SMBH case with $\dot{M}_\textrm{peak} / \dot{M}_\textrm{Edd} \simeq 43$ for $M_1 = 10^6 M_\odot$ ($ \simeq 480$ for $M_2 = 0.2M_1$). Figure \ref{fig:mdotpeak} shows the median of $\dot{M}_\textrm{peak} / \dot{M}_\textrm{Edd}$ as function of $a$ for different $q$, which collects these results over the parameter space.

\subsection{Binary Inspiral}
\label{subsec:inspiral}
The binary SMBH will eject most of the incoming ``pinhole' stars \citep{quinlan96}. The stars are ``slow intruders'' \citep{hills89}, so they will escape with enhanced energies and the binary will contract. The binary shrinks predominantly by this loss cone scattering until it begins to emit gravitational radiation and ultimately merges \citep{begelman80,mikkola92,quinlan96,sesana06,sesana07}. Since we recorded the state of the ejected stars, we now proceed to calculate the binary separation as a function of time from these two sources, and self-consistently calculate the time-dependent disruption rate. Our simulations contain only low angular momentum stars from the loss cone since they alone can produce TDEs, but those with high angular momentum can still extract energy from the binary; we exclude this latter contribution as it is likely small.

In this subsection, we write the binary separation and stellar specific energies and angular momenta in the units $G = M_1 = r_{t1} = 1$, since these parameters are fixed for changing $q$ and $a$. We define the dimensionless time $t = \tilde{t}/t_0$ using the timescale
\begin{equation}
t_0 = \frac{r^4_{t1} c^5}{G^3 M^3_1} \simeq 7.7 \times 10^{-1} \textrm{ y} \left(\frac{M_1}{10^6 M_\odot}\right)^{-5/3} \left(\frac{M_*}{M_\odot}\right)^{-4/3} \left(\frac{R_*}{R_\odot}\right)^4
\end{equation}
We use a primary mass $M_1 = 10^6 M_\odot$ and stars with solar parameters where necessary.

The total inspiral rate in these variables is
\begin{equation}
\frac{da}{dt} = \left(\frac{da}{dt}\right)_\textrm{ss} + \left(\frac{da}{dt}\right)_\textrm{gw}
\end{equation}
where the subscripts denote contributions from stellar scattering (ss) and the emission of gravitational waves (gw). The contraction rate due to gravitational radiation emitted by two point particles on a circular orbit is \citep{peters64}
\begin{equation}
\left(\frac{da}{dt}\right)_\textrm{gw} = -\frac{64}{5} \frac{q(1+q)}{a^3}
\end{equation}
The inspiral rate due to stellar scattering is
\begin{equation}
\left(\frac{da}{dt}\right)_\textrm{ss} = -\frac{2M_*}{M_1} \phi t_0 \frac{a^2}{q} \langle \Delta \epsilon_* \rangle (q,a) \lambda_\textrm{esc}(q,a)
\end{equation}
where $\langle \Delta \epsilon_* \rangle$ is the average change in the specific energy of the ejected stars and $\lambda_\textrm{esc}$ is the stellar escape rate, both found from our simulations; $M_* / M_1$ is the ratio of the stellar mass to the primary mass; and $\phi$ is the stellar injection rate from the binary loss cone (in units of y$^{-1}$).

For an isolated SMBH of mass $M_\bullet = 10^6 M_\odot$ with a full loss cone that remains populated through two-body relaxation, the rate of removal of stars from the loss cone is roughly $\phi_0 \sim 10^{-4}$ yr$^{-1}$ \citep{magorrian99,wang04,stone16}, which we take to be our fiducial rate. For a binary SMBH, if the loss cone is full and stars are injected with uniformly distributed $\ell^2$ even far from the binary center of mass, then the removal rate can be written as $\phi(q,a) = 2a (1+q) \phi_0$. Given our fiducial rate, this is a factor of $\sim 10$ below that calculated by \citet{yu03}, but consistent with their scaling with $a$ and insensitivity to $q$. The time-dependent disruption rate is then simply $\dot{n}_\textrm{bin} = \phi \lambda^\textrm{bin}_t$, where $\lambda^\textrm{bin}_t$ is the disruption rate as a function of $q$ and $a$ (Figure \ref{fig:lambdata}). We calculate $a(t)$ and $\dot{n}_\textrm{bin}$ as the binary contracts from $a = 1000$ ($2.3$ mpc) to $10$ ($0.023$ mpc).

\begin{figure*}
\centering
\subfloat[Critical separation]{\includegraphics[width=0.49\textwidth]{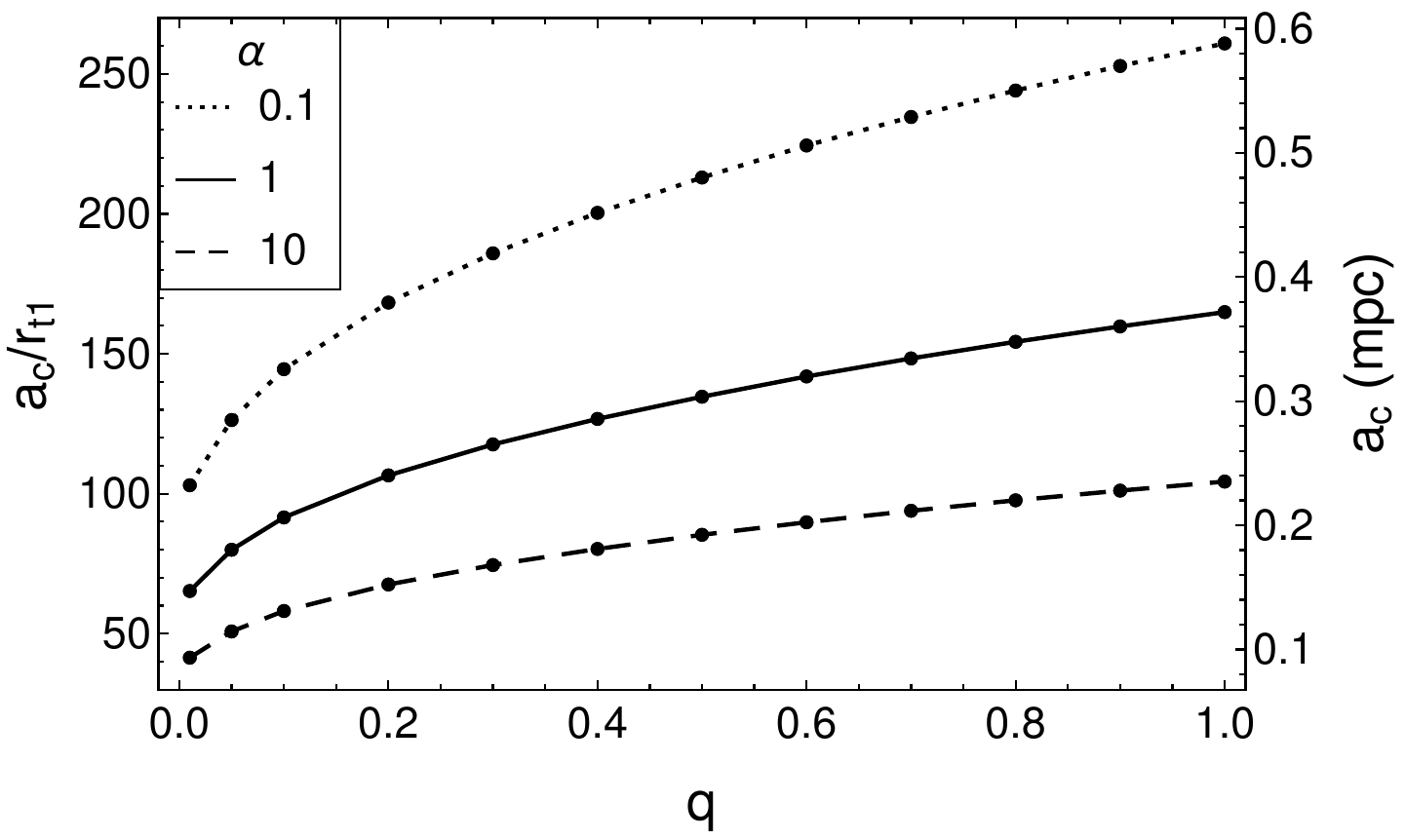}\label{fig:avsta}}\hfill
\subfloat[Inspiral]{\includegraphics[width=0.49\textwidth]{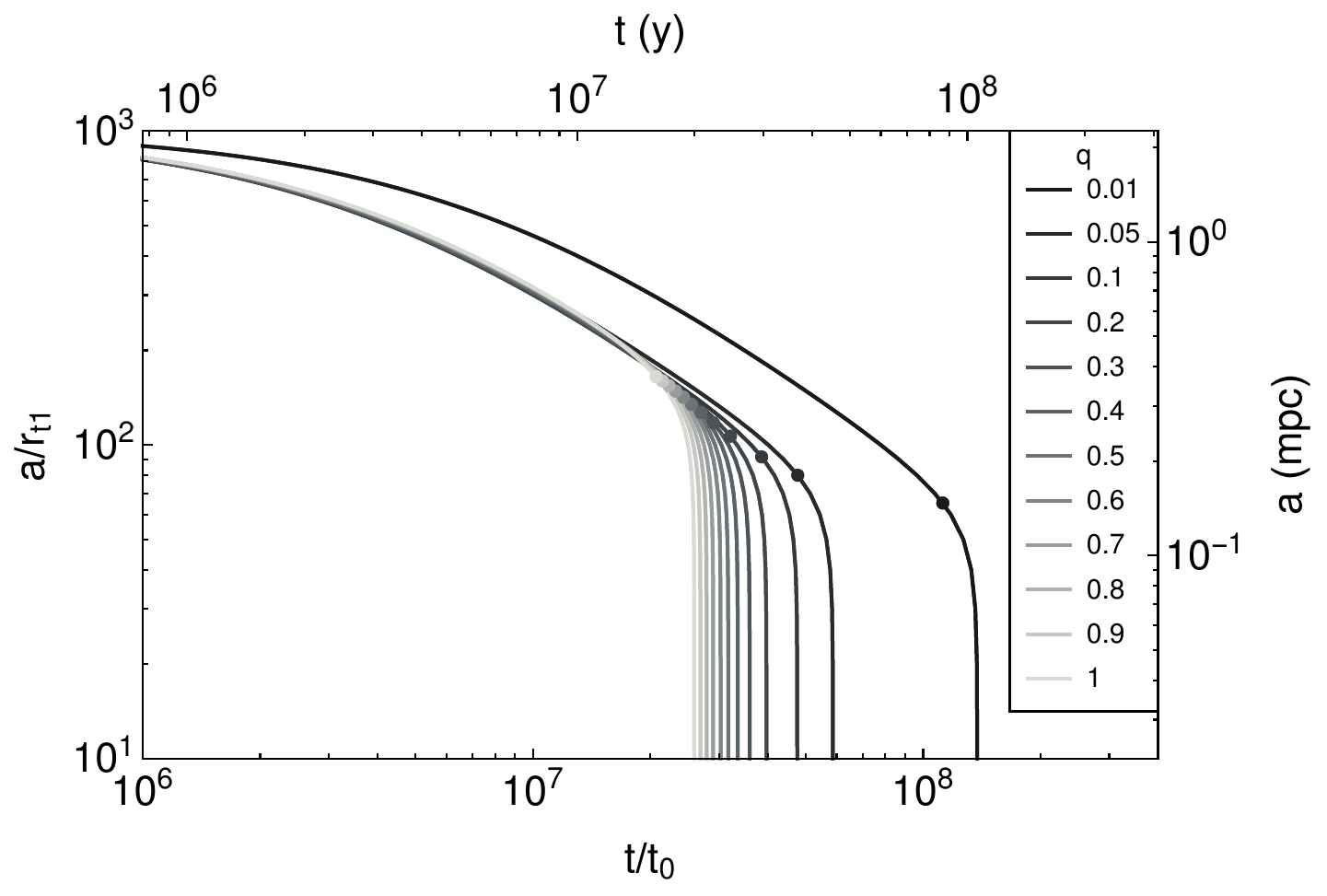}\label{fig:avstb}}
\caption{a) The critical separation $a_c$ at which gravitational radiation overtakes stellar scattering as the dominant inspiral mechanism, for $M_1 = 10^6 M_\odot$ and Sun-like stars. The three curves correspond to stellar injection rates that are $\alpha = 0.1$, $1$, and $10$ times the injection rate $\phi = 2a (1+q) \phi_0$, where $\phi_0 \simeq 10^{-4}$ y$^{-1}$ is the fiducial rate for an isolated BH of mass $M_\bullet = 10^6 M_\odot$. b) The binary separation as a function of time for $M_1 = 10^6 M_\odot$ and Sun-like stars, starting at an initial separation of $a = 1000$ ($2.3$ mpc). The injection parameter is $\alpha = 1$. The points mark the values of $a_c$.}
\label{fig:avst}
\end{figure*}

Figure \ref{fig:avsta} shows the critical separations $a_c$ at which gravitational radiation overtakes stellar scattering as the dominant contraction mechanism for $\alpha = 0.1$, $1$, and $10$ times the injection rate $\phi$. The critical separations are largely insensitive to $q$ and $\alpha$, as a change in these quantities by two orders of magnitude leads to a modest change in $a_c$ of a factor of $\sim 6$. For the fiducial case $\alpha = 1$, they fall in the range $a_c \sim 60 - 170$ ($0.14 - 0.38$ mpc).

Figure \ref{fig:avstb} shows the binary separation as a function of time over our range of $q$ and with an injection parameter $\alpha = 1$. The binaries with lower mass ratios merge more slowly; in particular, a binary with $q = 0.01$ takes roughly five times as long to merge as one with $q=1$. The black holes rapidly merge once the binaries reach the critical separations, as seen in the precipitous drop immediately after these points (marked on the plot).

Since the binary SMBH rapidly merges below $a_c$, we expect to observe TDEs while its separation $a > a_c$. For $\alpha = 1$, the critical separation is roughly $a_c \sim 100$. We use the results from Sections \ref{subsec:tderate} -- \ref{subsec:postdisruption} to estimate the fraction of disruptions with short rise times and high peak fallback rates, as these can act as indicators to distinguish binaries from single black holes. Close binaries ($a \simeq 100$) with nearly equal masses ($q \gtrsim 0.2$) have $\sim 25 - 40 \%$ ($\sim 55 - 35 \%$) of bound-debris disruptions by the primary (secondary) with rise times in the range $t_\textrm{rise} \sim 1 - 10$ d and peak fallback rates with respect to the disrupting BH in the range $\dot{M}_\textrm{peak} / \dot{M}_\textrm{Edd} \sim 2 \times 10^2 - 3 \times 10^3$ ($\sim 2 \times 10^2 - 10^4$), both potentially up to an order of magnitude or two different than for an isolated black hole. For binaries in this parameter range, the primary (secondary) produces $\sim 90 - 50 \%$ ($\sim 10 - 50 \%$) of the disruptions (Figure \ref{fig:lambdat12b}), and $\sim 100 - 80 \%$ ($\sim 55 - 85 \%$) of the disruptions by the primary (secondary) will have some debris bound to the disrupting black hole (Figure \ref{fig:fbound}). All together, $\sim 15 - 25 \%$ ($\sim 3 - 15 \%$) of disruptions for $a \sim 100$ and $q \gtrsim 0.2$ will be short, highly super-Eddington TDEs produced by the primary (secondary), so $\sim 18 - 40 \%$ of all disruptions in this parameter range will exhibit this behavior. Similarly, binaries with $a \simeq 100$ and $q \sim 0.05$ have $\sim 15 \%$ ($\sim 1 \%$) of these disruptions produced by the primary (secondary), or $\sim 16 \%$ of the total number. Wide binaries ($a \simeq 1000$) with nearly equal masses ($q \gtrsim 0.2$) have roughly $\sim 0 \%$ ($\sim 4 - 0 \%$) of the short rise time disruptions and $\sim 0 \%$ ($\sim 9 - 0 \%$) of the high fallback rate disruptions by the primary (secondary). Binaries with $a \simeq 1000$ and $q \sim 0.05$ have $\sim 0 \%$ ($\sim 2 \%$), or $\sim 2 \%$ of the total number of disruptions.

Therefore, near-merger, nearly equal mass binaries (those with $a \sim 100$ and $q \gtrsim 0.2$) have the highest probability of producing distinctive TDEs. As the binary prepares to merge, a disruption from such an event can signal the presence of two black holes. These disruptions could arise from both long-lived disruptions with partially bound debris ($-1 \leq \varepsilon_c < 0$) and finite duration ones with fully bound debris ($\varepsilon_c < -1$). A higher stellar injection rate will lead to a tighter binary before gravitational radiation takes over (Figure \ref{fig:avsta}), and thus more extreme late-time TDEs.

\begin{figure}
\centering
\includegraphics[width=0.48\textwidth]{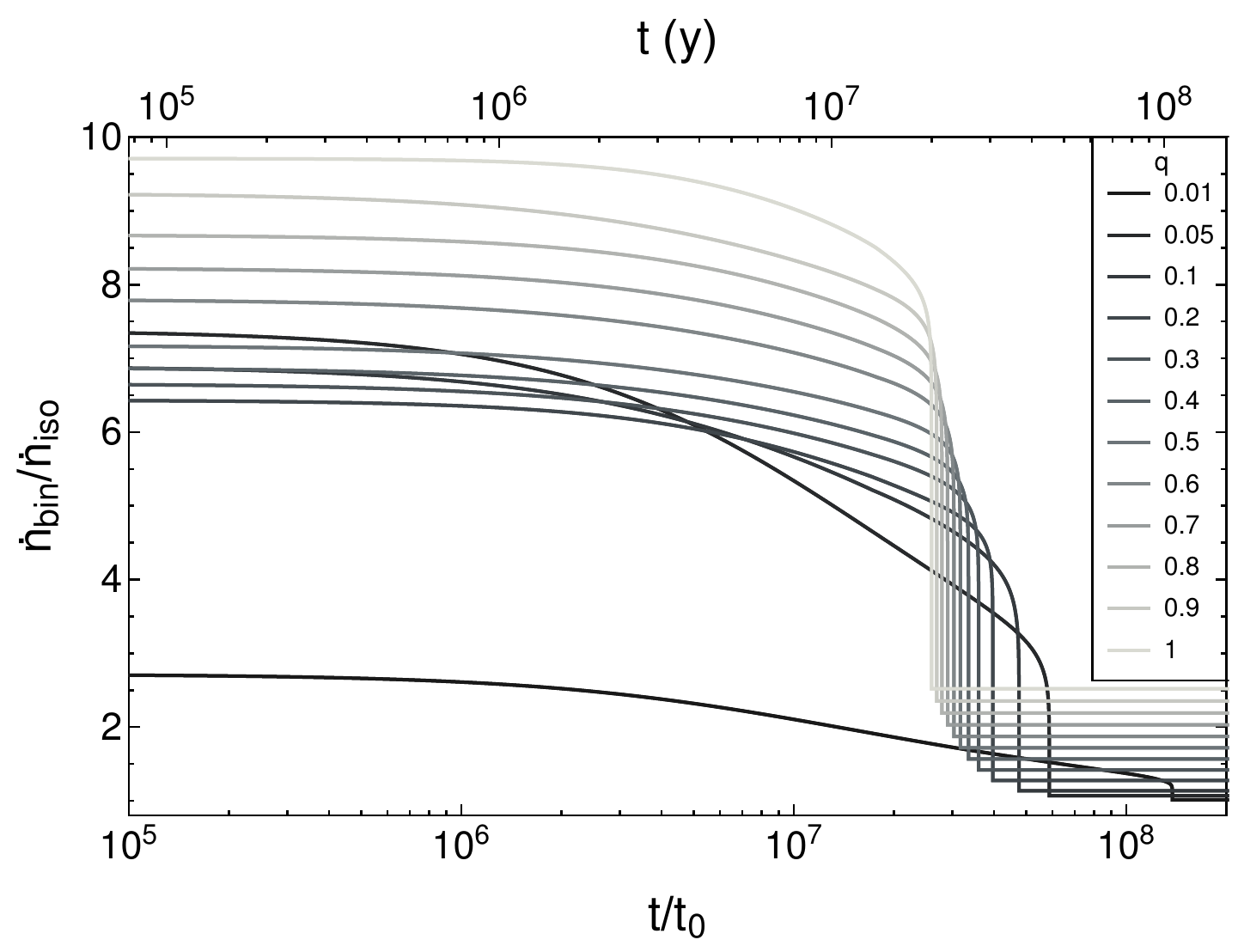}
\caption{The tidal disruption rate for the binary SMBH, for $M_1 = 10^6 M_\odot$ and $M_* = M_\odot$, starting at an initial separation of $a = 1000$ ($2.3$ mpc). The rate is normalized to the disruption rate $\dot{n}_\textrm{iso} = \phi_0 \sim 10^{-4}$ yr$^{-1}$ for an isolated BH of mass $M_\bullet = 10^6 M_\odot$. The injection parameter is $\alpha = 1$. After coalescence, the rate is $\dot{n}_\textrm{iso} = \phi_0 (1+q)^{4/3}$, corresponding to a post-merger black hole of mass $M_\bullet = M_1 (1+q)$, which we include through a piecewise function.}
\label{fig:dndtvst}
\end{figure}

Figure \ref{fig:dndtvst} shows the time-dependent disruption rate for the binary SMBH normalized to that for an isolated BH with the mass of the primary. The latter is simply $\dot{n}_\textrm{iso} = \phi_0$. At early times, stellar scattering is the dominant inspiral mechanism, and the disruption rate can be up to an order of magnitude larger than in the single black hole case; for $q \gtrsim 0.05$, the rate is $\sim 6 - 10$ times larger. At late times, gravitational radiation becomes dominant and the binary rapidly merges, producing an abrupt drop in the rate. In the brief intermediate range where we expect to observe the distinctive TDEs described above, the rate declines to $\sim 5 - 9$ for $q \gtrsim 0.1$. After coalescence, we expect the rate to be that scattered into the tidal radius of the merger remnant of mass $M_\bullet = M_1 (1+q)$, which is $\dot{n}_\textrm{iso} = \phi_0 (1+q)^{4/3}$, so we include this behavior through a piecewise constraint at late times. The time-dependent rate is not quite monotonic in the mass ratio at early times due to the $q$-dependence of $\lambda^\textrm{bin}_t / \lambda^\textrm{iso}_t$ (Figure \ref{fig:lambdatb}). The total lifetimes are all $\sim 30$ Myr, though have a weak monotonic dependence on the mass ratio.

\citet{chen08} also studied tidal disruptions from unbound stars scattering off of a SMBH binary, and found a time-dependent TDE rate that is roughly an order of magnitude smaller than for a single black hole. The discrepancy between our results arises mainly from a different loss cone model. \citet{chen08} focused on nearby galaxies when calculating their disruption rates, and thus modeled their stars in the ``diffusive'' (or ``empty loss cone'') regime, using two-body relaxation as the loss cone refilling process. Under these conditions, the loss cone remains depleted, as stars only intermittently enter the loss cone after many binary orbits and are expelled promptly. In addition, the stars only populate the edge of the loss cone and approach the binary with pericenter distances $r_p \sim a$. The disruption rate is thus strongly suppressed. We note that the discrepancy in our results does not seem to arise from the initial conditions of our scattering experiments.

The binaries for our range of $q$ produce a total of $n^\textrm{bin}_\textrm{tot} \sim (1.5 - 2) \times 10^4$ disruptions over their lifetimes, shrinking from an initial separation of $a = 1000$ ($2.3$ mpc) until the merger in $\sim (2.5 - 15) \times 10^7$ yr. In contrast, \citet{chen11} considered a binary SMBH embedded in a stellar cusp, and found $\sim (2.3 - 6.5) \times 10^4$ disruptions over $10^8$ yr. The TDEs in their study arise from bound stars whose orbits are perturbed until a disruption, thus enhancing the TDE rate, but if the bound stars become depleted, then the loss cone will become the only source of disruptions.
% (Now include a comparison with \citealt{wegg11}; they don't quote this in their paper, you may have to email them to get this).

\section{Summary and Conclusions}
\label{sec:conclusion}

A tidal disruption event occurs when a star is torn apart by the gravitational field of a supermassive black hole. A fraction of the stellar debris becomes bound to the SMBH, and the ensuing accretion provides a means of detecting otherwise-quiescent black holes in galactic nuclei and studying their properties (and, in principle, the properties of the circumnuclear medium; e.g, \citealt{alexander16}). Tidal disruptions by isolated black holes have been studied in detail, but those produced by binary SMBHs -- from which there could arise a number of observationally-distinct phenomena -- have only received attention fairly recently.

In this paper, we studied the physics of tidal disruption from chaotic three-body encounters between stars and a binary SMBH in the pinhole regime \citep{frank76,lightman77,cohn78,magorrian99}, in which stars approach the binary on initially-parabolic, low angular momentum orbits. We found several important features for disruptions by massive black hole binaries in our range of parameters:

\begin{enumerate}[1., leftmargin=*]

\item The disruption rate is a factor of $\sim 2 - 7$ times larger relative to the rate for a single BH, depends only weakly on the semimajor axis $a$, and is almost entirely independent of the mass ratio $q$ for $q \gtrsim 0.2$ (Section \ref{subsec:tderate} and Figure \ref{fig:lambdatb}).

\item Roughly $\sim 10 - 20 \%$ of disruptions occur in a star's first encounter with the binary for all $q$ and $a$; if $q \gtrsim 0.1$, the probability of disruption declines monotonically as a function of time, while systems with $q \lesssim 0.1$ have an increased probability for disruption at later times before falling off rapidly (see Figure \ref{fig:tdis}).

\item For small $q$, disruptions by the primary (secondary) produce preferentially bound (unbound) debris streams, with these preferences becoming less pronounced as $q$ approaches one (Figures \ref{fig:histecc} -- \ref{fig:epsc}).

\end{enumerate}

\noindent
For a primary mass $M_1 = 10^6 M_\odot$:

\begin{enumerate}[1., leftmargin=*, start=4]

\item In our ``frozen-in'' model, close, nearly equal mass binaries ($a/r_{t1} \sim 100$, $q \gtrsim 0.2$) can produce prompt, intense tidal disruptions, with rise times $t_\textrm{rise} \sim 1 - 10$ d and peak fallback rates $\dot{M}_\textrm{peak} / \dot{M}_\textrm{Edd} \sim 2 \times 10^2 - 3 \times 10^3$ with respect to the disrupting BH; in this parameter space, these account for $\sim 18 - 40 \%$ of the total number of disruptions (Figures \ref{fig:histtrise} -- \ref{fig:mdotpeak}, \ref{fig:histtriseappendix} -- \ref{fig:pmdotpeak1fullprobappendix}, \ref{fig:lambdat12}, and \ref{fig:fbound}).

\item Over the lifetime of the binary, the mean time-dependent disruption rate for $q \gtrsim 0.05$ can be a factor of $\sim 6 - 10$ larger than that for a single black hole (Figure \ref{fig:dndtvst}).

\end{enumerate}

In addition to recording disruptions, our simulations also recorded the ejection of stars from the binary; over $86 \%$ of the interactions actually terminate with this outcome. We assessed the inspiral timescales of SMBH binaries using this result along with the average energy of the ejected stars as a function of $q$ and $a$ (Section \ref{subsec:inspiral}). For typical values of the primary mass, the injection rate of stars, and an initial separation of $\simeq 2.3$ mpc, we found that the time taken for gravitational waves to dominate the inspiral is roughly $\sim 30$ Myr; after this point, the binary very rapidly merges (see Figure \ref{fig:avst}). We anticipate performing a more thorough investigation of the properties of the ejected stars in a companion paper.

We compared the properties of our binary-induced TDEs to those from an isolated black hole. In doing so, we assumed the same initial conditions, namely that the stars approach the black hole(s) on parabolic orbits from isotropically-distributed starting points in the pinhole regime (with uniformly-distributed pericenters relative to the origin). However, different astrophysical conditions can alter the dynamics of the incoming stars and modify the disruption properties. The central BH can be embedded in a stellar cusp and stars can approach it on bound orbits, which can greatly enhance the disruption rate \citep{ivanov05,chen11,wegg11}. The loss cone removal rate can be enhanced if it is refilled by mechanisms other than two-body scattering, such as resonant relaxation \citep{madigan11}, the tidal separation of binary stars \citep{amaroseoane12}, and BH recoil following a binary coalescence \citep{stone11}, which can all enhance the TDE rate. Stars can also orbit a central BH in an eccentric nuclear disk, and large amplitude eccentricity oscillations in the inner part of the disk can redirect a subset of stars towards the BH, increasing the TDE rate \citep{madigan18}. %Following a disruption event, the returning debris may not all accrete onto the BH, and upwards of $\sim 75\%$ can eventually become unbound \citep{ayal00}

Recent optical surveys have found that TDEs appear to occur preferentially in post-merger (E+A) galaxies \citep{arcavi14, french16}.
% \cite{french17} examined optical/UV TDEs from six post-starburst galaxies and explored the possibility of a binary BH origin. They concluded that the binary mass ratios are likely greater than $q = 1/12$, largely excluding one model of binary-induced TDEs \citep{chen11} and favoring other, dynamical origins for the rate enhancement (e.g., \citealt{stone17}).
While the influence of a binary SMBH does not likely explain the rate enhancement at the E+A stage, it seems probable that a binary is present in the center of many post-merger galaxies (that is, if each merging galaxy had a central SMBH and their coalescence was not driven by some mechanism, such as a third black hole, as explored by \citealt{silsbee17}). Consequently, the effects described in this paper could also affect some TDEs occurring in E+A galaxies.

Our results concerning the fraction of bound material, the peak fallback rate, and the time to peak fallback (Figures \ref{fig:fbound} -- \ref{fig:mdotpeak}) assumed that the motion of the stellar debris was ``frozen-in'' by the tidal field of the disrupting SMBH. The tidal radius is typically small compared to the binary separation, so this approximation is valid when the apocenter of the peak debris is small compared to the Roche lobe radius of the disrupting BH, leading to an early evolution in which the debris is influenced only by the gravitational field of the disrupting black hole (Section \ref{subsec:postdisruption}). This condition holds rigorously when the black holes are widely separated (Figure \ref{fig:fmodelvalid}) and in an average sense at smaller separations (verified by the simulations in \citealt{coughlin17}). However, for ``general'' disruptions where some portion of the debris stream may become unbound, eventually the material will recede beyond the Roche lobe of the disrupting SMBH. At that point, the total fraction of bound material can be altered by the gravitational potential of the binary and the accretion rate can exhibit different behavior \citep{coughlin18}. We leave an analysis of these effects to a future investigation.

\section*{Acknowledgements}

This research used resources of the National Energy Research Scientific Computing Center, a DOE Office of Science User Facility supported by the Office of Science of the U.S. Department of Energy under Contract No. DE-AC02-05CH11231. ERC was supported by NASA through the Einstein Fellowship Program, Grant PF6-170150. This work was supported by the National Science Foundation under Grant No. 1616754. This work was supported in part by a Simons Investigator award from the Simons Foundation (EQ) and the Gordon and Betty Moore Foundation through Grant GBMF5076.

%%%%%%%%%%%%%%%%%%%%%%%%%%%%%%%%%%%%%%%%%%%%%%%%%%



\bibliographystyle{mnras}
\bibliography{references} % if your bibtex file is called references.bib

%% Alternatively you could enter them by hand, like this:
%% This method is tedious and prone to error if you have lots of references
%\begin{thebibliography}{99}
%\end{thebibliography}

%%%%%%%%%%%%%%%%%%%%%%%%%%%%%%%%%%%%%%%%%%%%%%%%%%

%%%%%%%%%%%%%%%%% APPENDICES %%%%%%%%%%%%%%%%%%%%%

\appendix

\section{Additional Figures}
\label{sec:appendixa}

%If you want to present additional material which would interrupt the flow of the main paper, it can be placed in an Appendix which appears after the list of references.

\begin{figure*}
\centering
\subfloat{\includegraphics[width=0.49\textwidth]{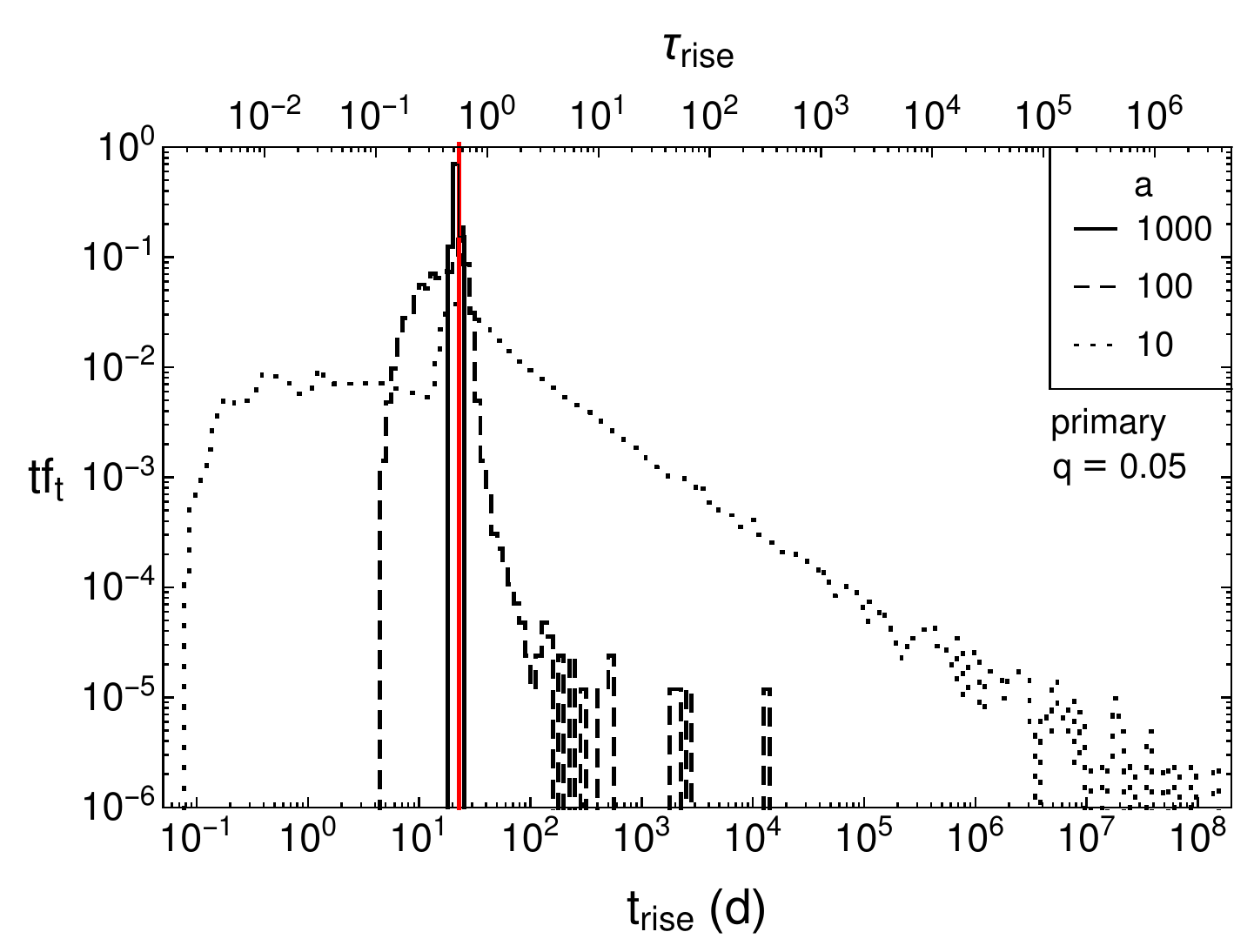}}\hfill
\subfloat{\includegraphics[width=0.49\textwidth]{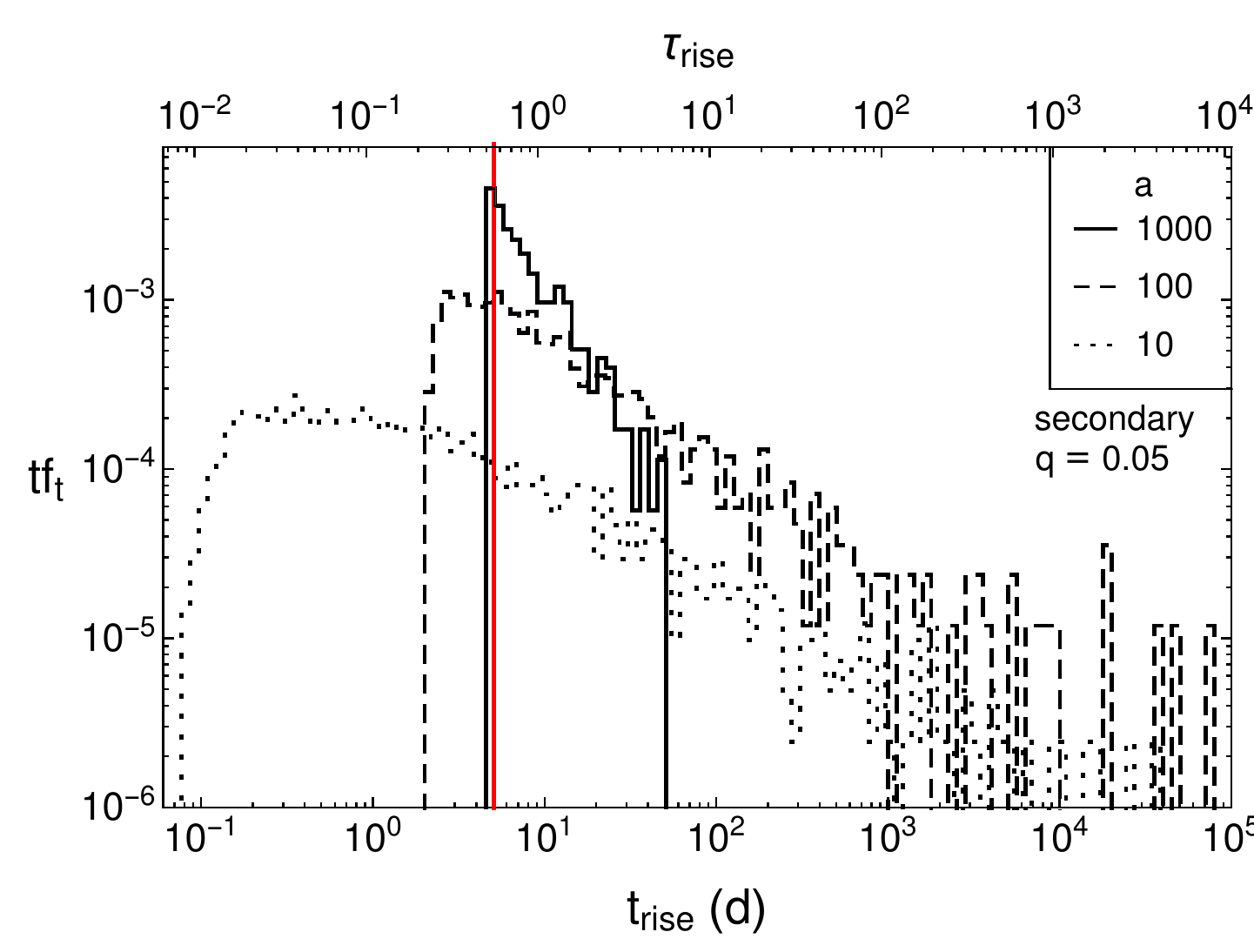}}\\
\subfloat{\includegraphics[width=0.49\textwidth]{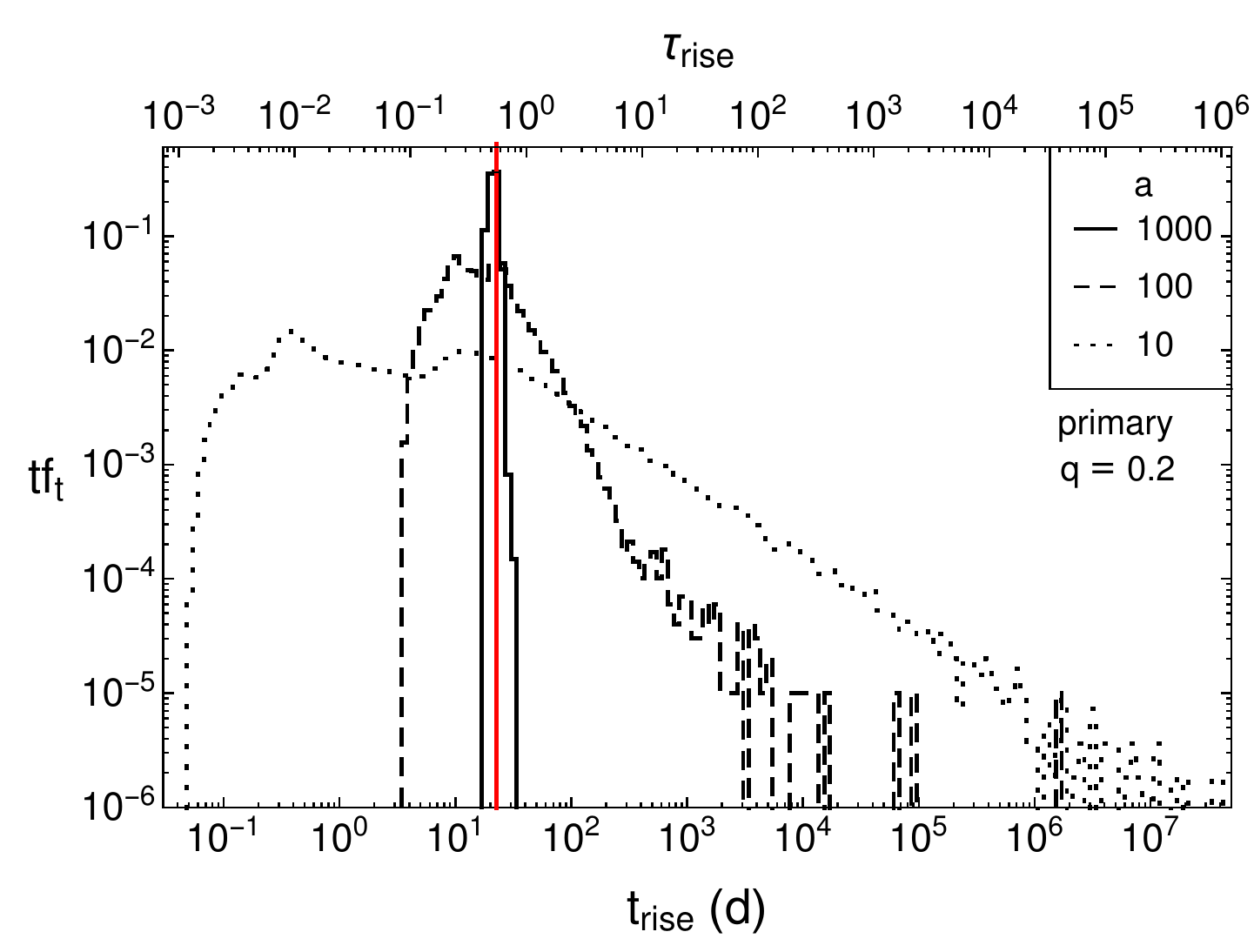}}\hfill
\subfloat{\includegraphics[width=0.49\textwidth]{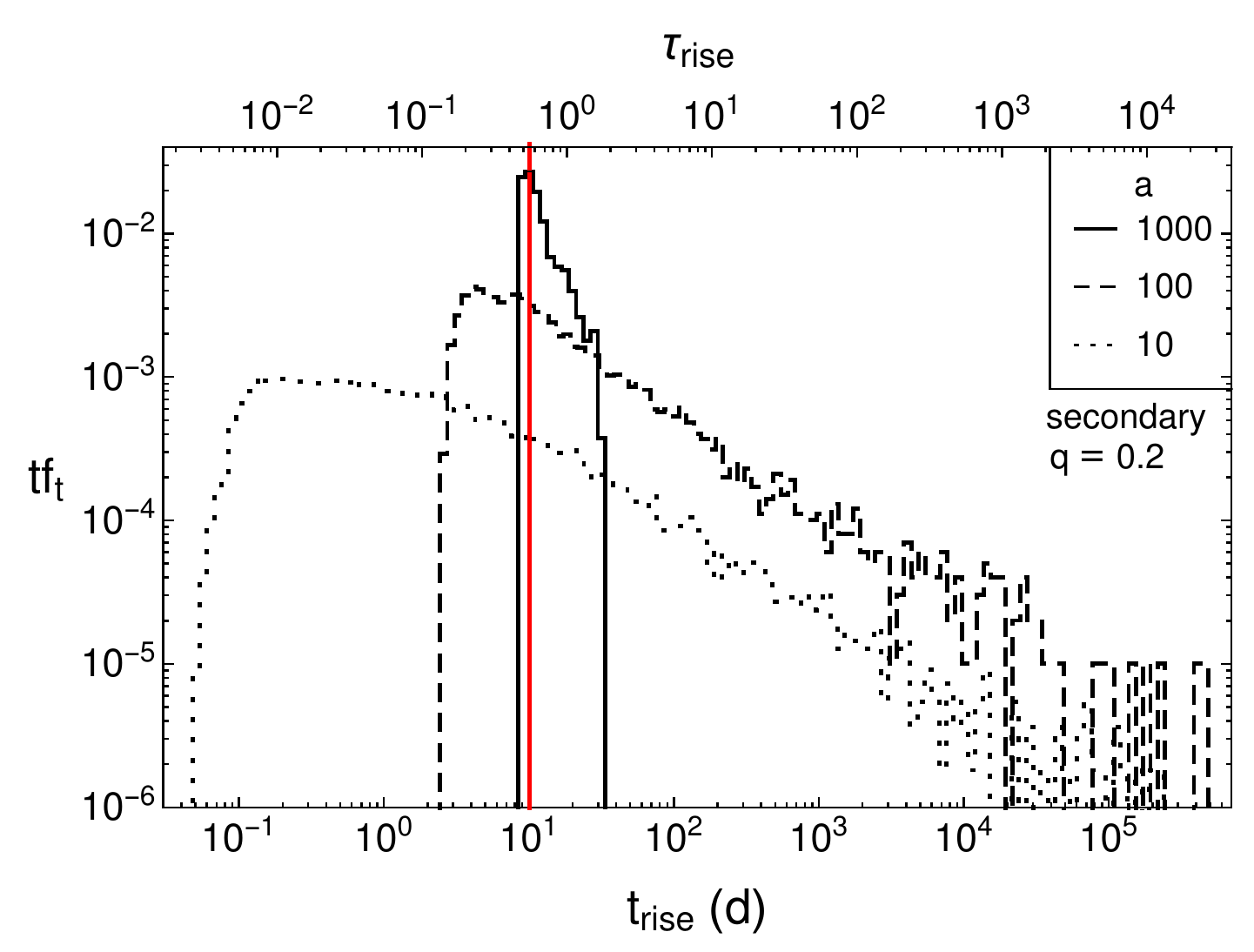}}\\
\subfloat{\includegraphics[width=0.49\textwidth]{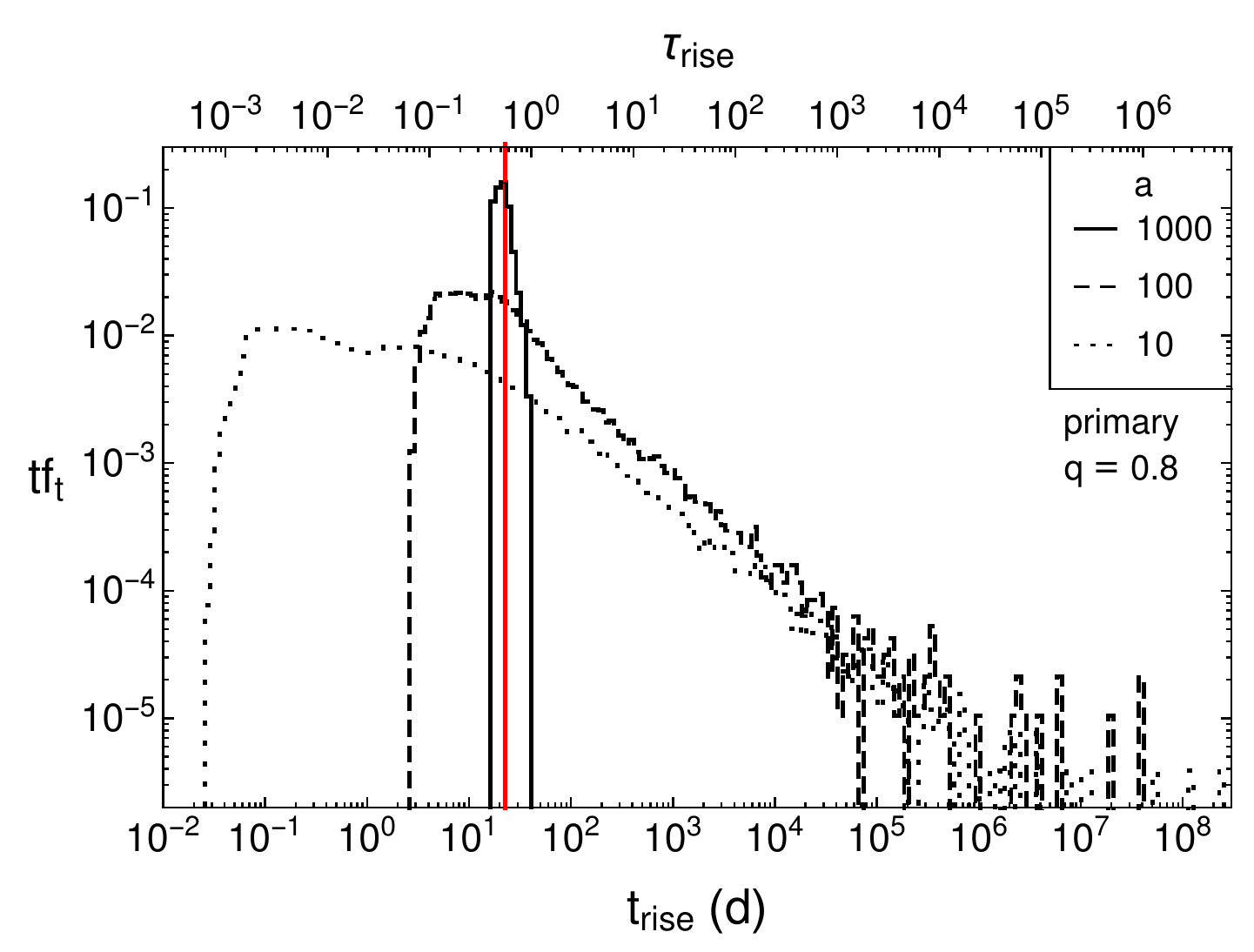}}\hfill
\subfloat{\includegraphics[width=0.49\textwidth]{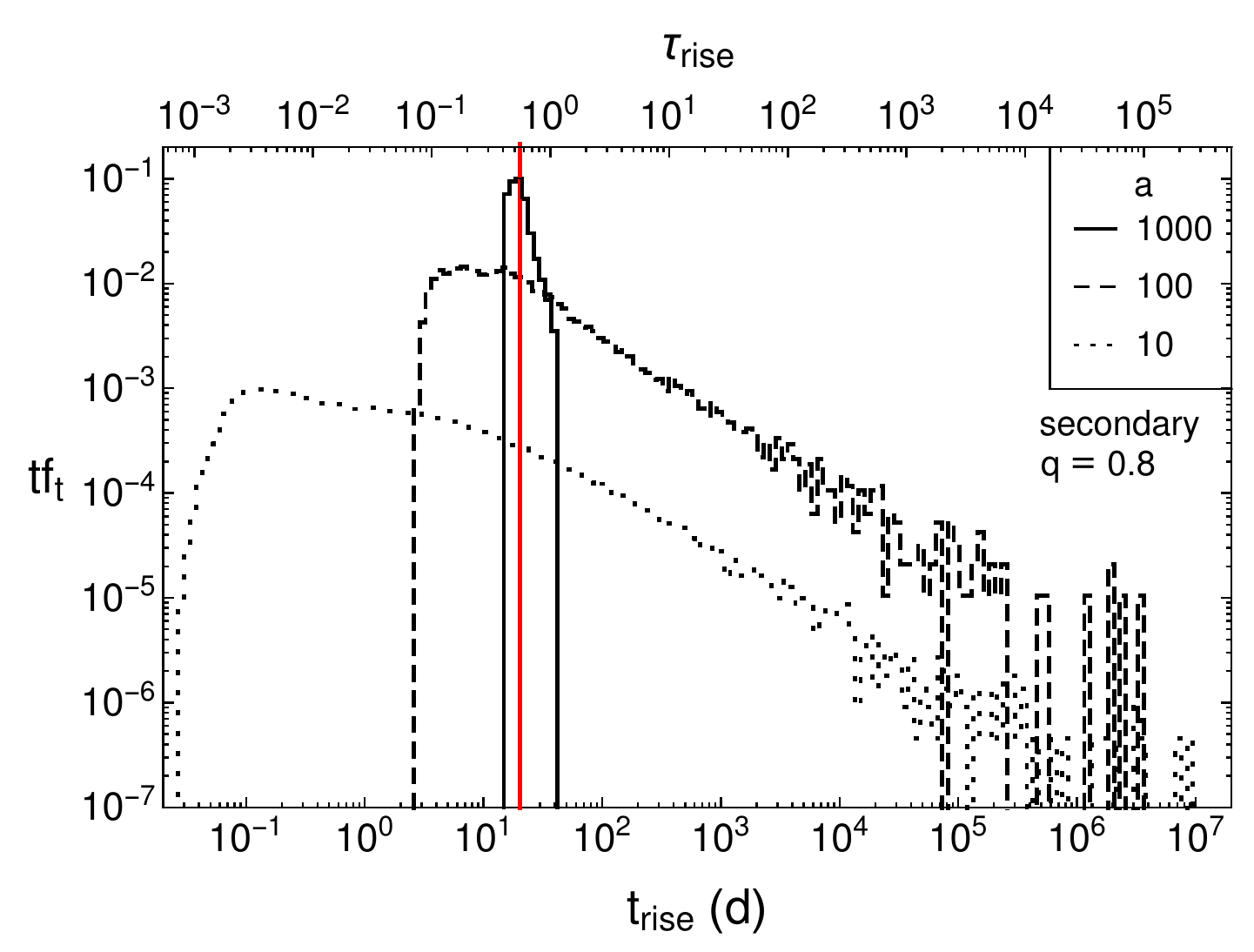}}
\caption{The probability $t f_t$ that a disruption by the binary will produce debris with a rise time $t_\textrm{rise} = \tau_0 \tau_\textrm{rise}$ for a primary mass $M_1 = 10^6 M_\odot$ and $a=1000$ (solid), $100$ (dashed), and $10$ (dotted). The probabilities for each histogram sum to $(\lambda_{ti} / \lambda_t) f_b$, where $i = 1 \ (2)$ refers to the primary (secondary), and these quantities depend on $q$ and $a$ (Figures \ref{fig:lambdat12b}, \ref{fig:fbound}). The top panels show the results for $q = 0.05$, the center ones for $q = 0.2$, and the bottom ones for $q = 0.8$. The left panels show disruptions by the primary and the right ones show those by the secondary. The logarithmic bin widths are $\Delta_t = 0.05$ and the heights are the probabilities in each bin. The red line marks the rise time $t_\textrm{rise} = \tau_0 (3\sqrt{3} - 5^{3/4})/5^{3/4}$ for a TDE from a star with $\varepsilon_c = 0$ (parabolic orbit, half of the debris is bound) disrupted by the appropriate black hole.}
\label{fig:histtrisefullprobappendix}
\end{figure*}

\begin{figure*}
\centering
\subfloat{\includegraphics[width=0.49\textwidth]{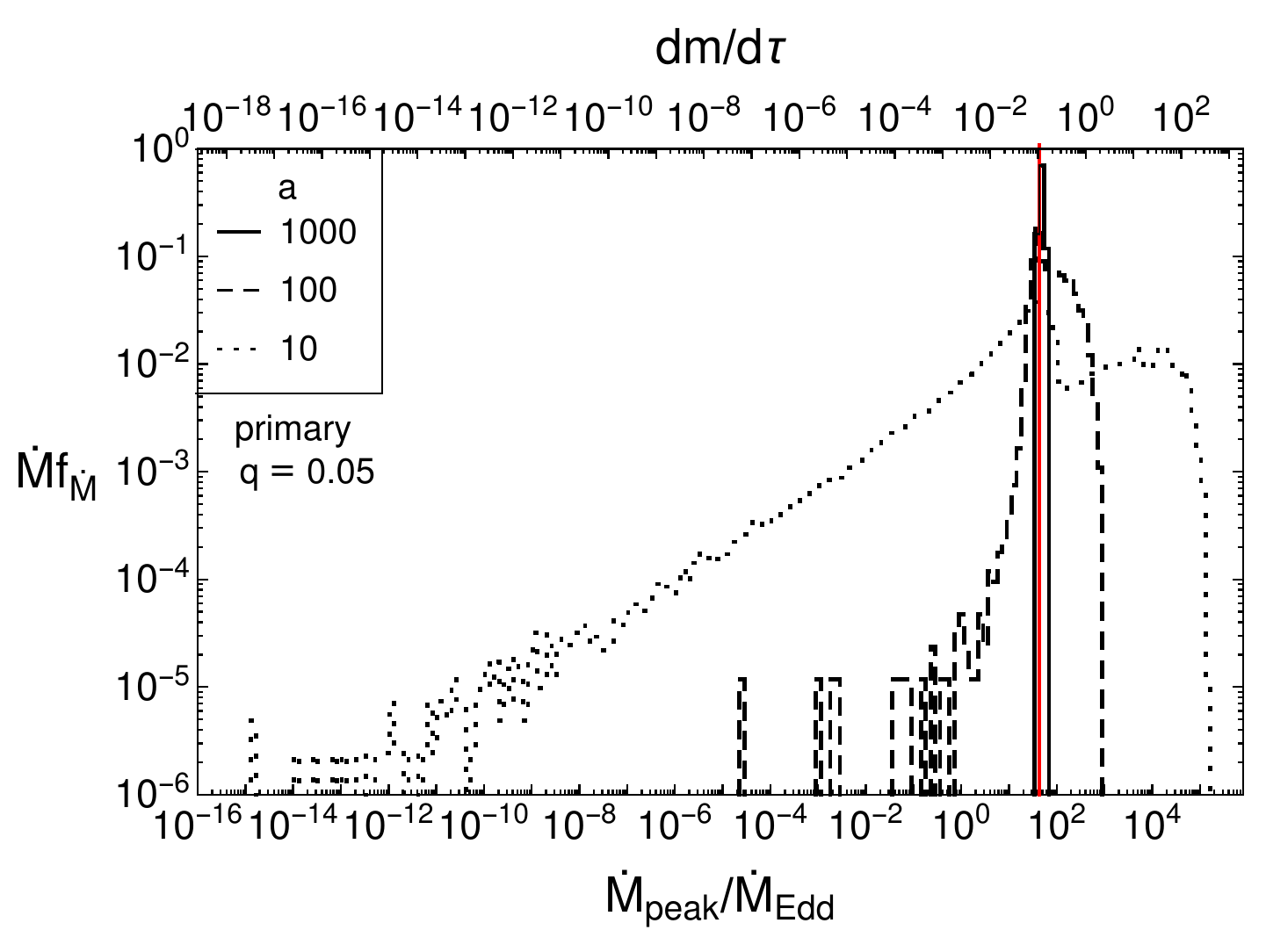}}\hfill
\subfloat{\includegraphics[width=0.49\textwidth]{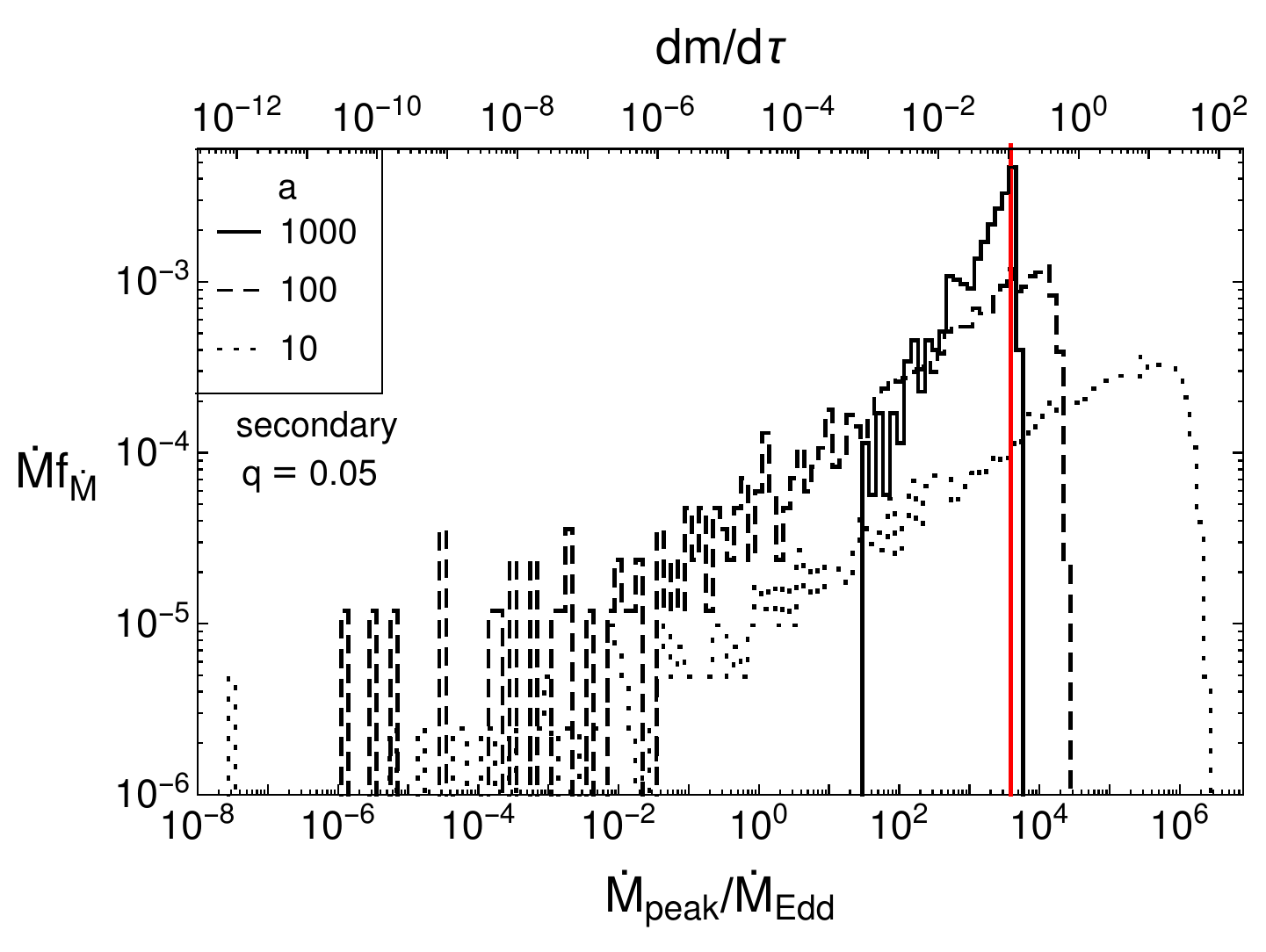}}\\
\subfloat{\includegraphics[width=0.49\textwidth]{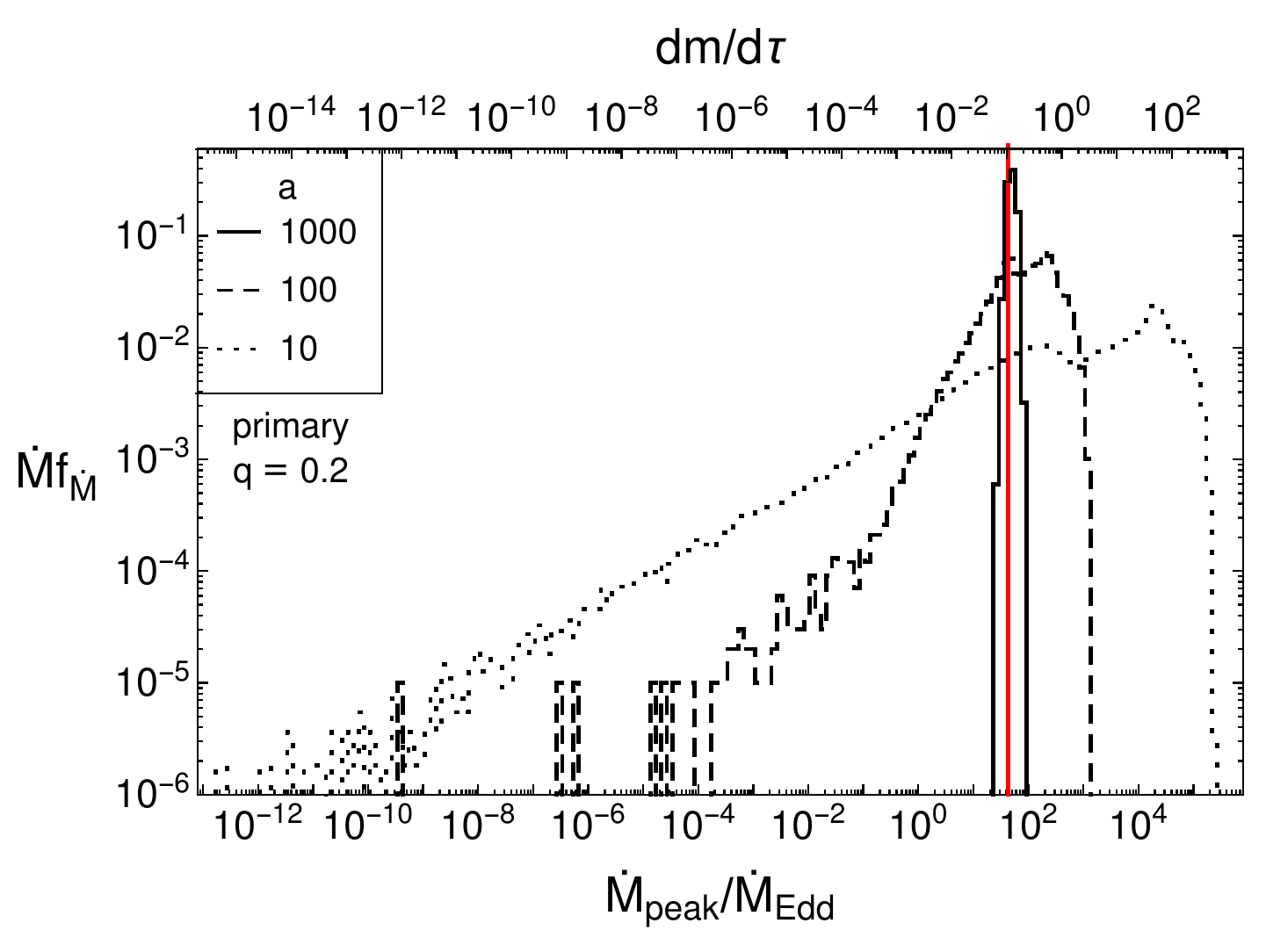}}\hfill
\subfloat{\includegraphics[width=0.49\textwidth]{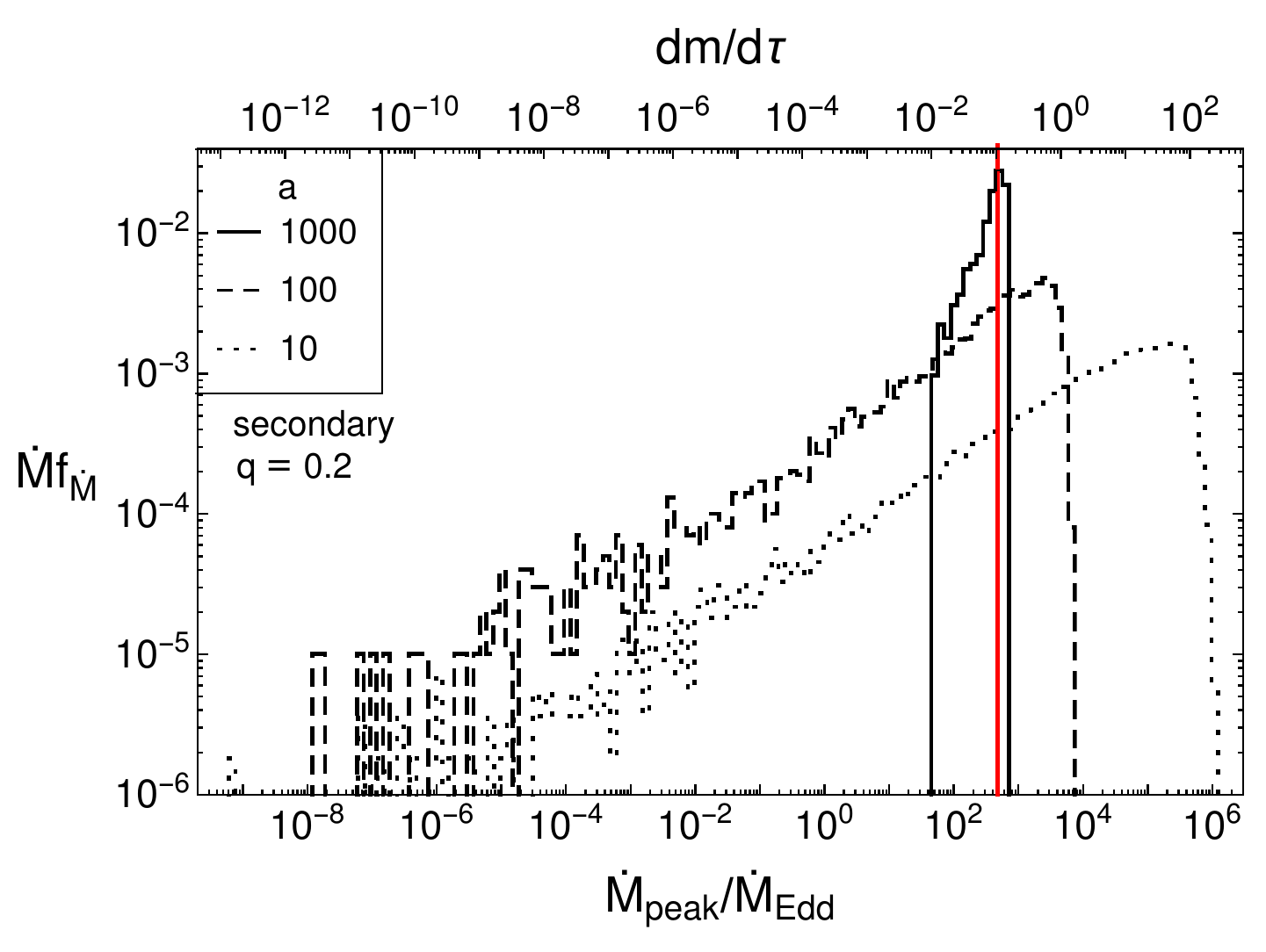}}\\
\subfloat{\includegraphics[width=0.49\textwidth]{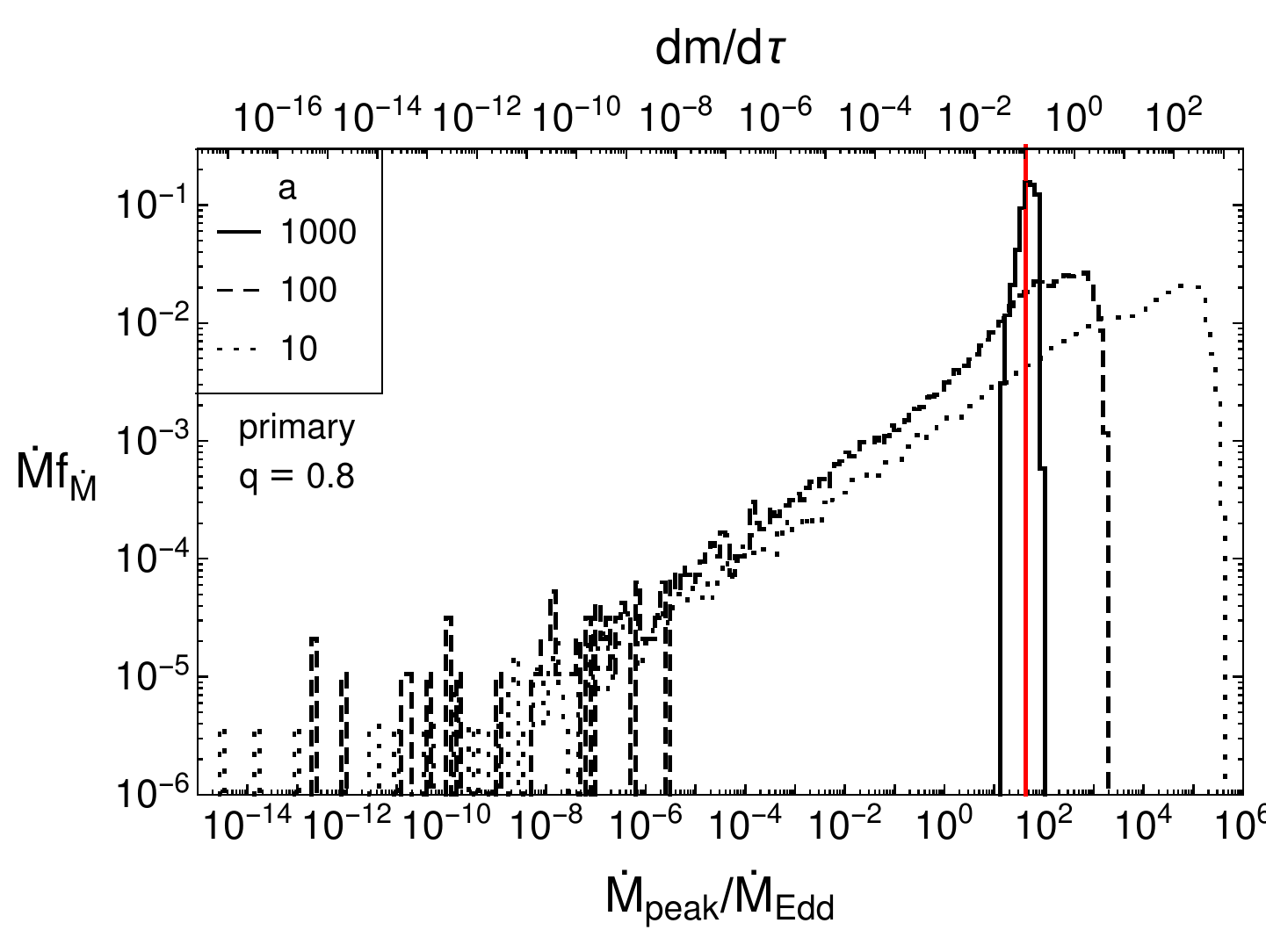}}\hfill
\subfloat{\includegraphics[width=0.49\textwidth]{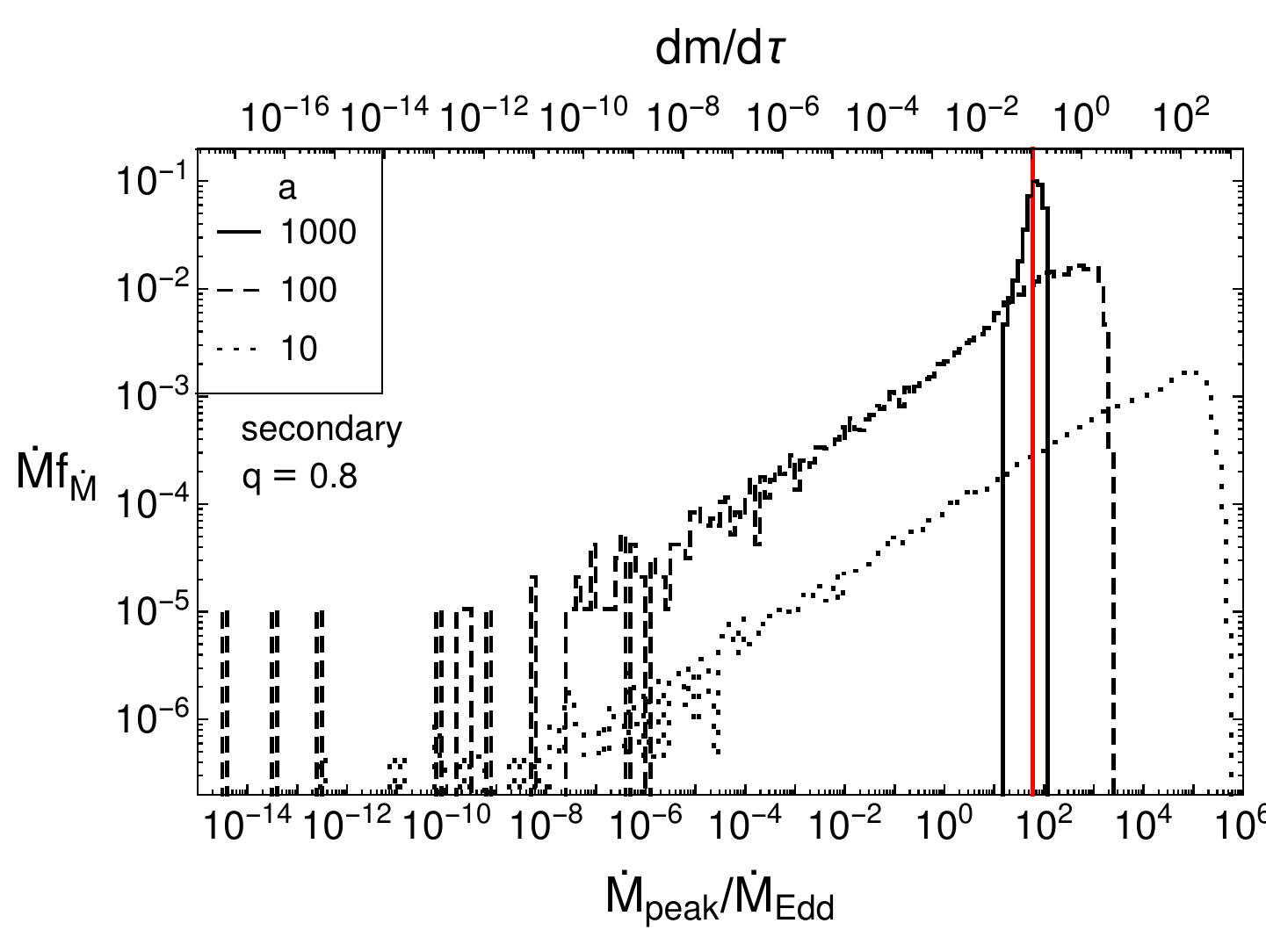}}
\caption{The probability $\dot{M} f_{\dot{M}}$ that a disruption by the binary will produce debris with a peak fallback rate $\dot{M}_\textrm{peak}/\dot{M}_\textrm{Edd}$ (Eq. \ref{eq:mdotpeakovermedd}; see surrounding text for the parameters used) for a primary mass $M_1 = 10^6 M_\odot$ and $a=1000$ (solid), $100$ (dashed), and $10$ (dotted). The probabilities for each histogram sum to $(\lambda_{ti} / \lambda_t) f_b$, where $i = 1 \ (2)$ refers to the primary (secondary), and these quantities depend on $q$ and $a$ (Figures \ref{fig:lambdat12b}, \ref{fig:fbound}). The top panels show the results for $q = 0.05$, the center ones for $q = 0.2$, and the lower ones for $q = 0.8$. The left panels show disruptions by the primary and the right ones show those by the secondary. The logarithmic bin widths are $\Delta_{\dot{M}} = 0.1$ and the heights are the probabilities in each bin. The red line shows the peak return rate for a TDE from a star with $\varepsilon_c = 0$ (parabolic orbit, half of the debris is bound) disrupted by the appropriate black hole.}
\label{fig:histmdotpeakfullprobappendix}
\end{figure*}

\begin{figure*}
\centering
\subfloat{\includegraphics[width=0.49\textwidth]{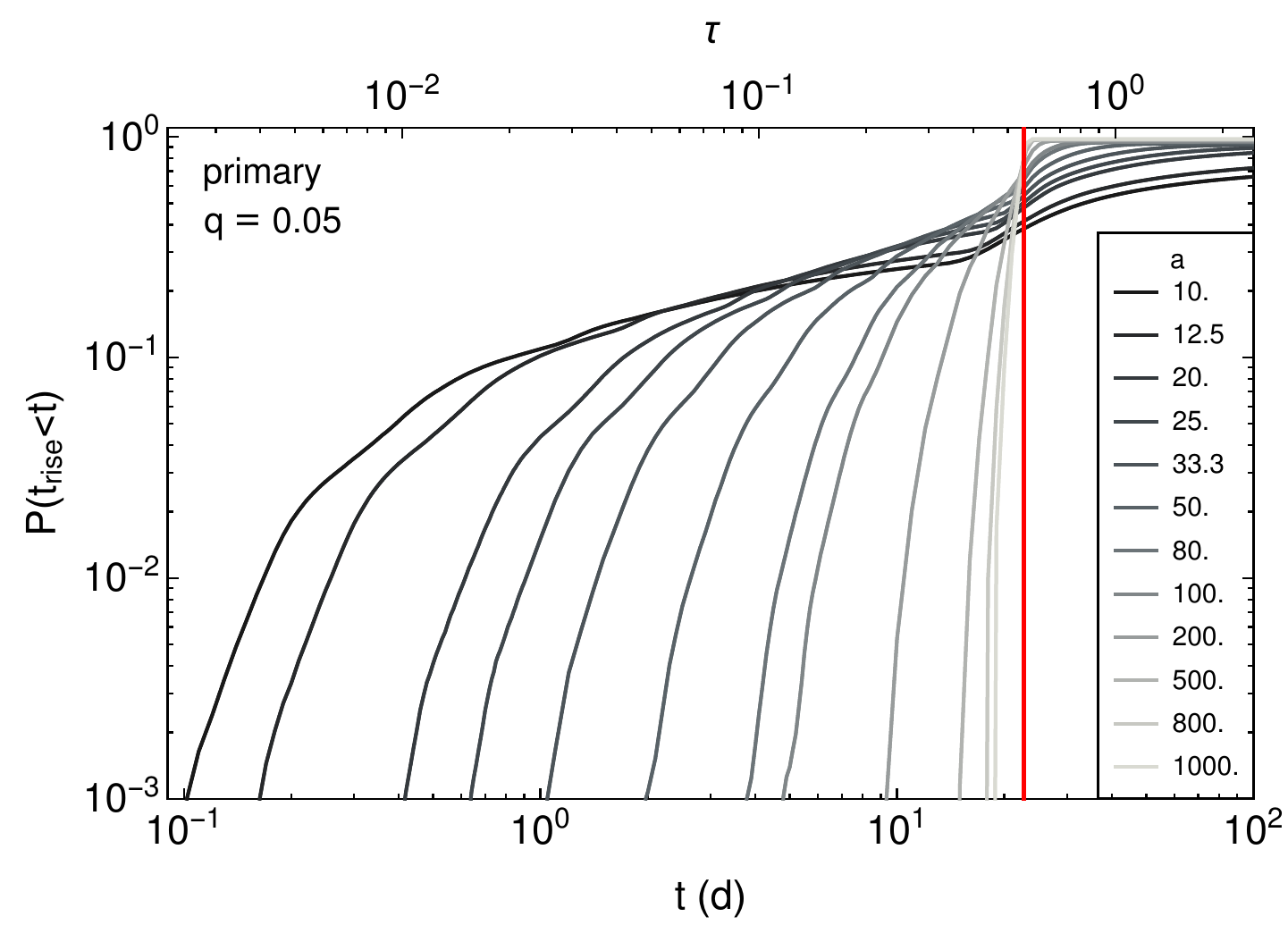}}\hfill
\subfloat{\includegraphics[width=0.49\textwidth]{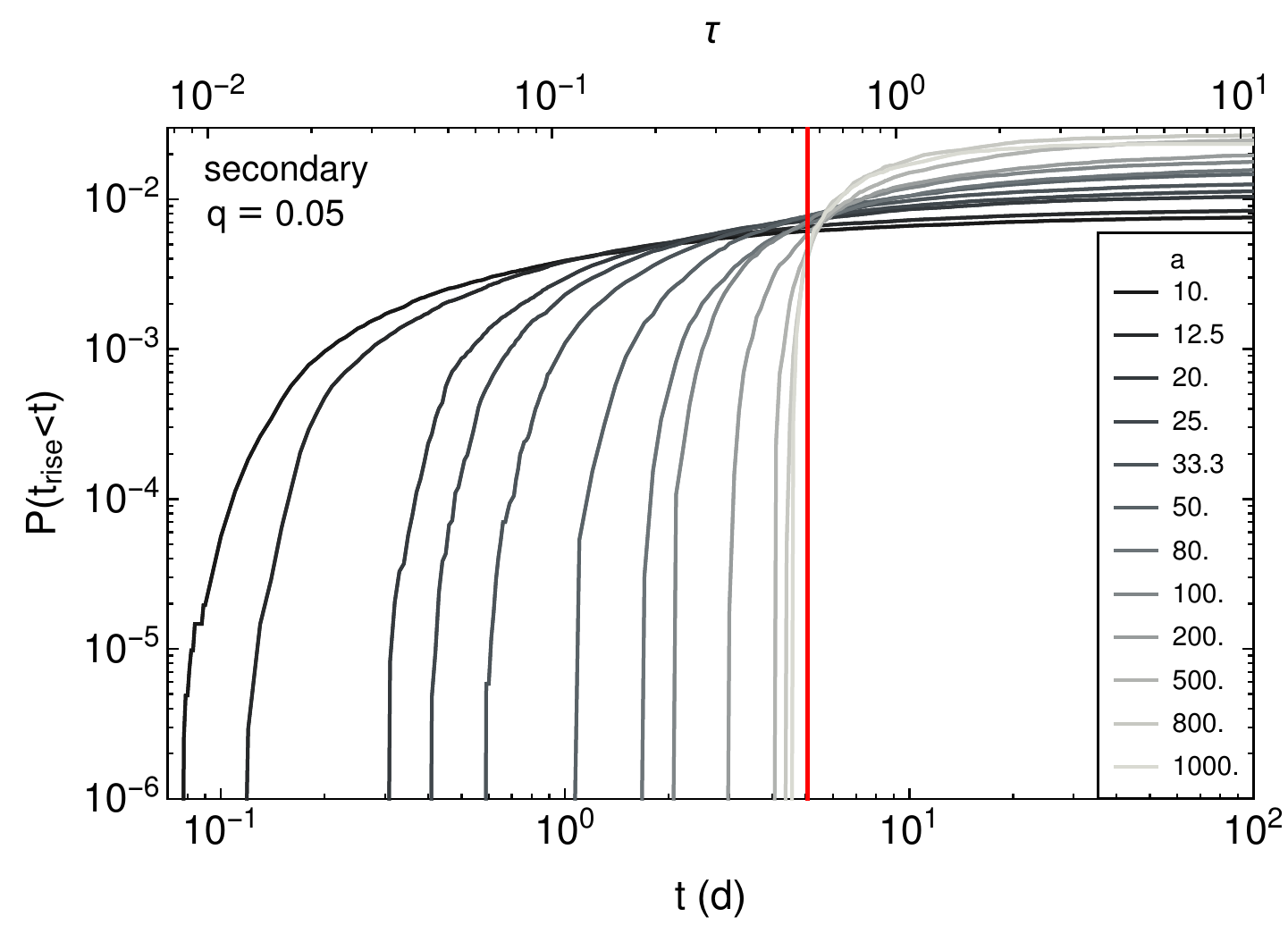}}\\
\subfloat{\includegraphics[width=0.49\textwidth]{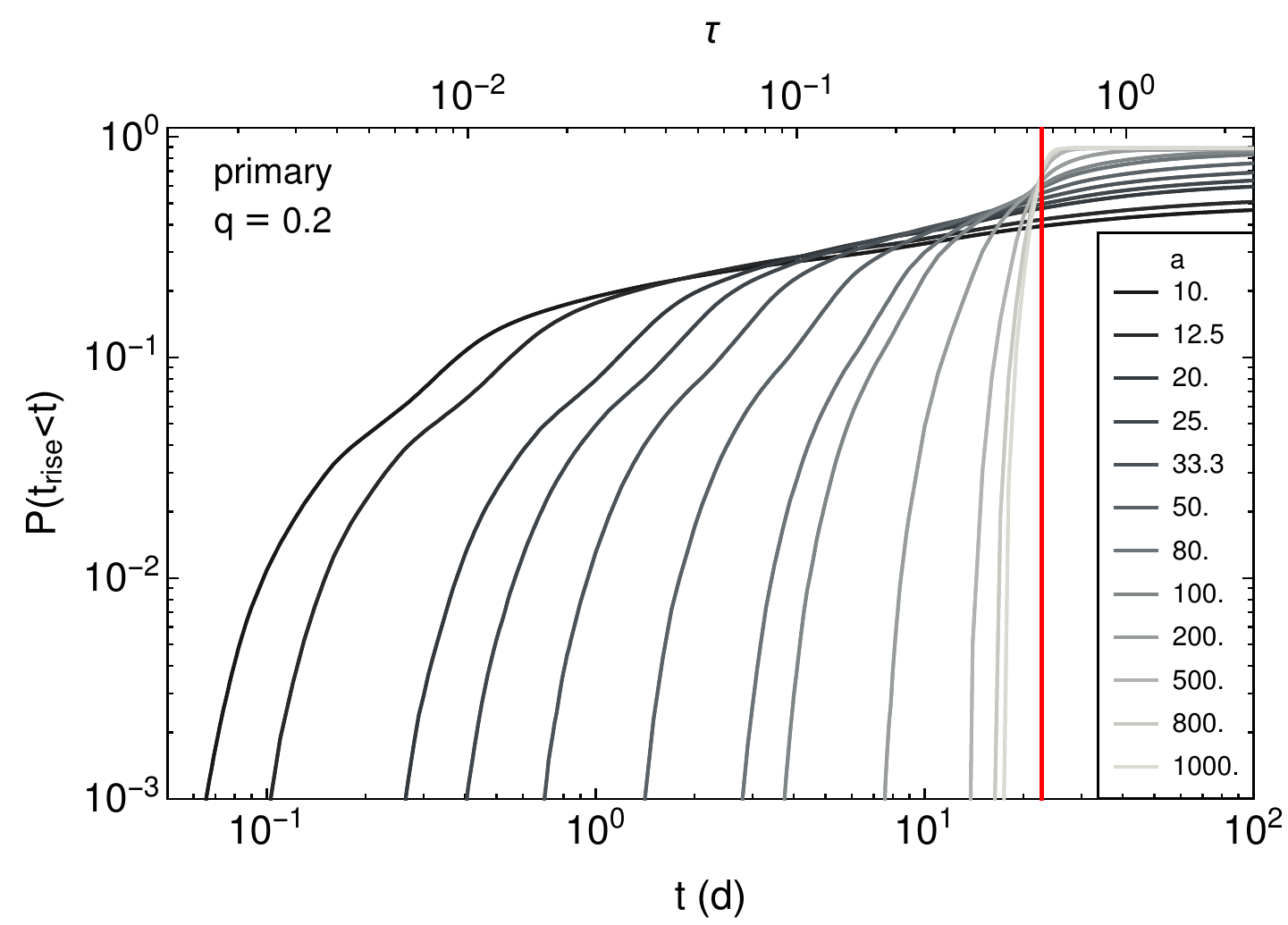}}\hfill
\subfloat{\includegraphics[width=0.49\textwidth]{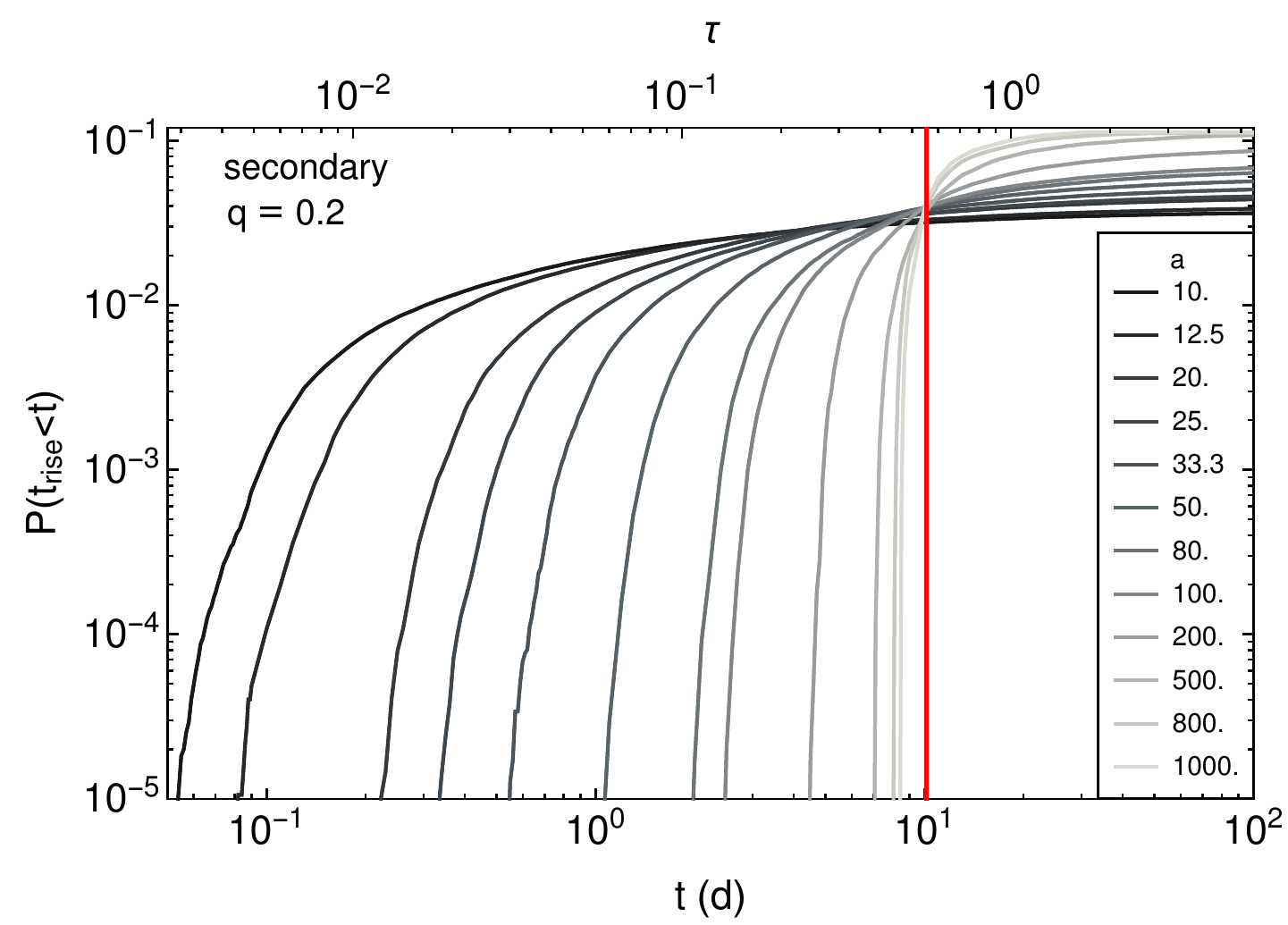}}\\
\subfloat{\includegraphics[width=0.49\textwidth]{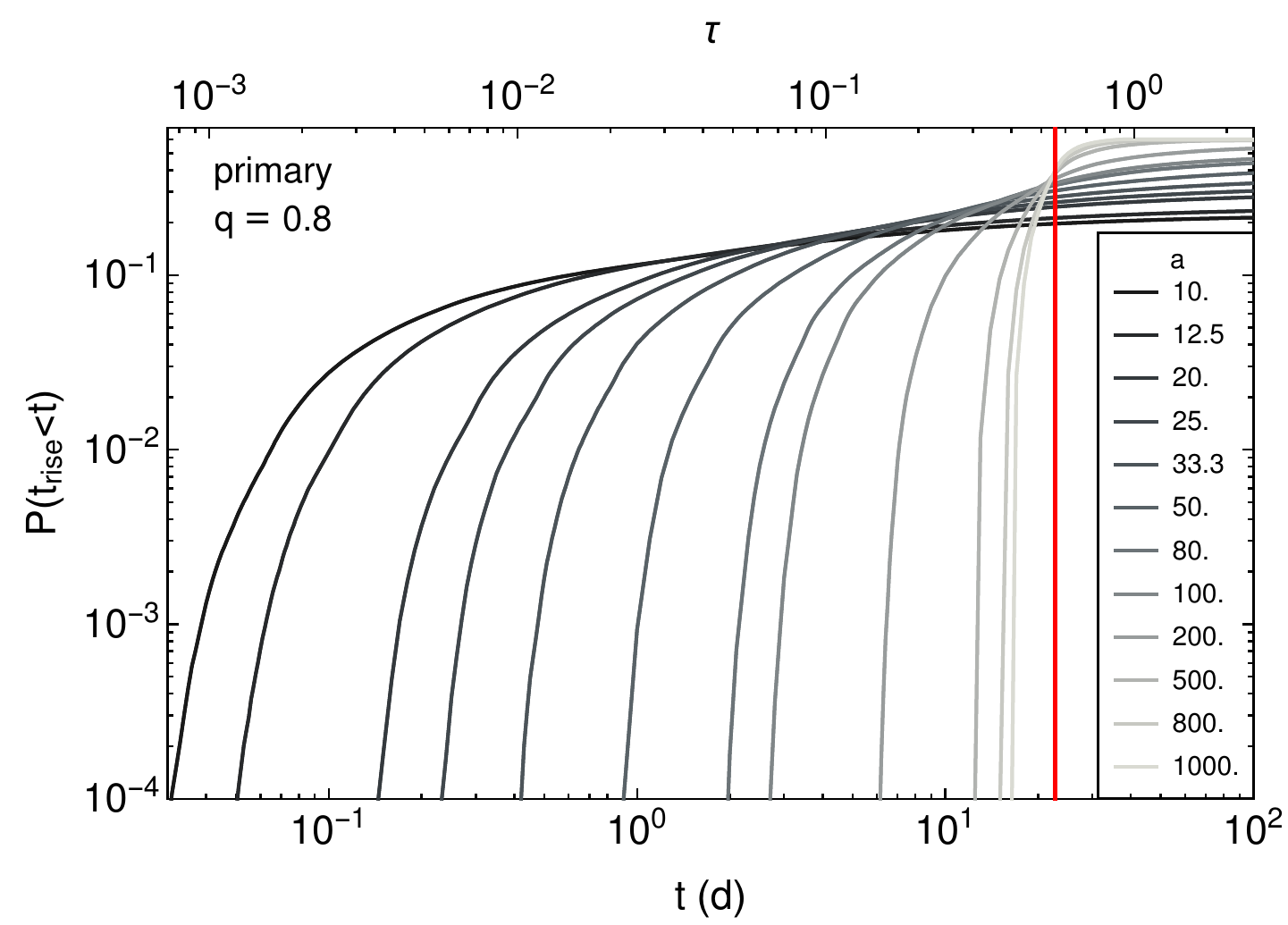}}\hfill
\subfloat{\includegraphics[width=0.49\textwidth]{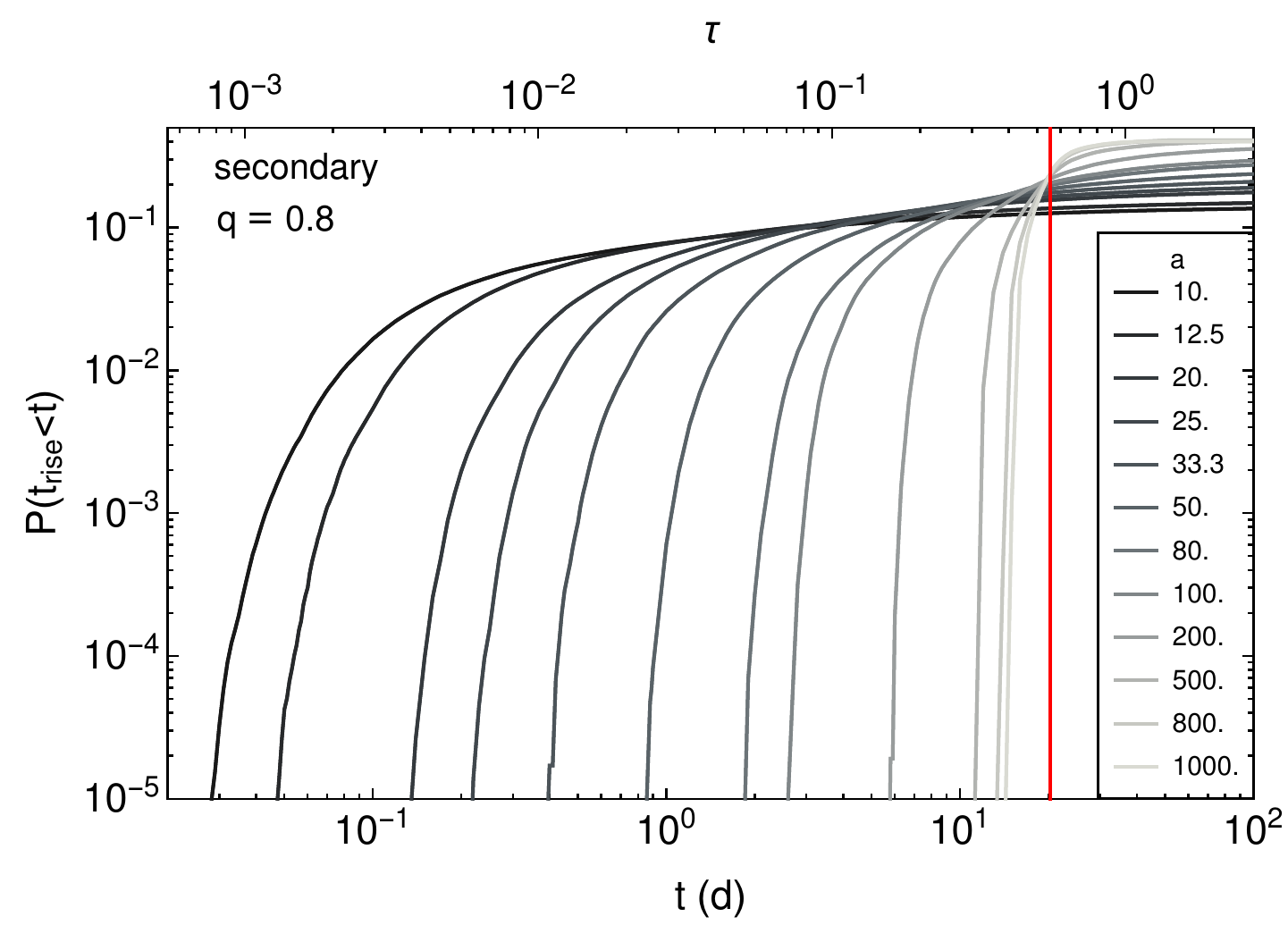}}
\caption{The probability that a disruption by the binary will produce debris with a rise time $t_\textrm{rise} < t$ (d) for a primary mass $M_1 = 10^6 M_\odot$ and different $a$. The maximum probabilities are $(\lambda_{ti} / \lambda_t) f_b$, where $i = 1 \ (2)$ refers to the primary (secondary), and these quantities depend on $q$ and $a$ (Figures \ref{fig:lambdat12b}, \ref{fig:fbound}). The top panels show the results for $q = 0.05$, the center ones for $q = 0.2$, and the bottom ones for $q = 0.8$. The left panels show disruptions by the primary and the right ones show those by the secondary. The black line marks the rise time $t_\textrm{rise} = \tau_0 (3\sqrt{3} - 5^{3/4})/5^{3/4}$ for a TDE from a star with $\varepsilon_c = 0$ (parabolic orbit, half of the debris is bound) disrupted by the appropriate black hole.}
\label{fig:ptrise1fullprobappendix}
\end{figure*}

\begin{figure*}
\centering
\subfloat{\includegraphics[width=0.49\textwidth]{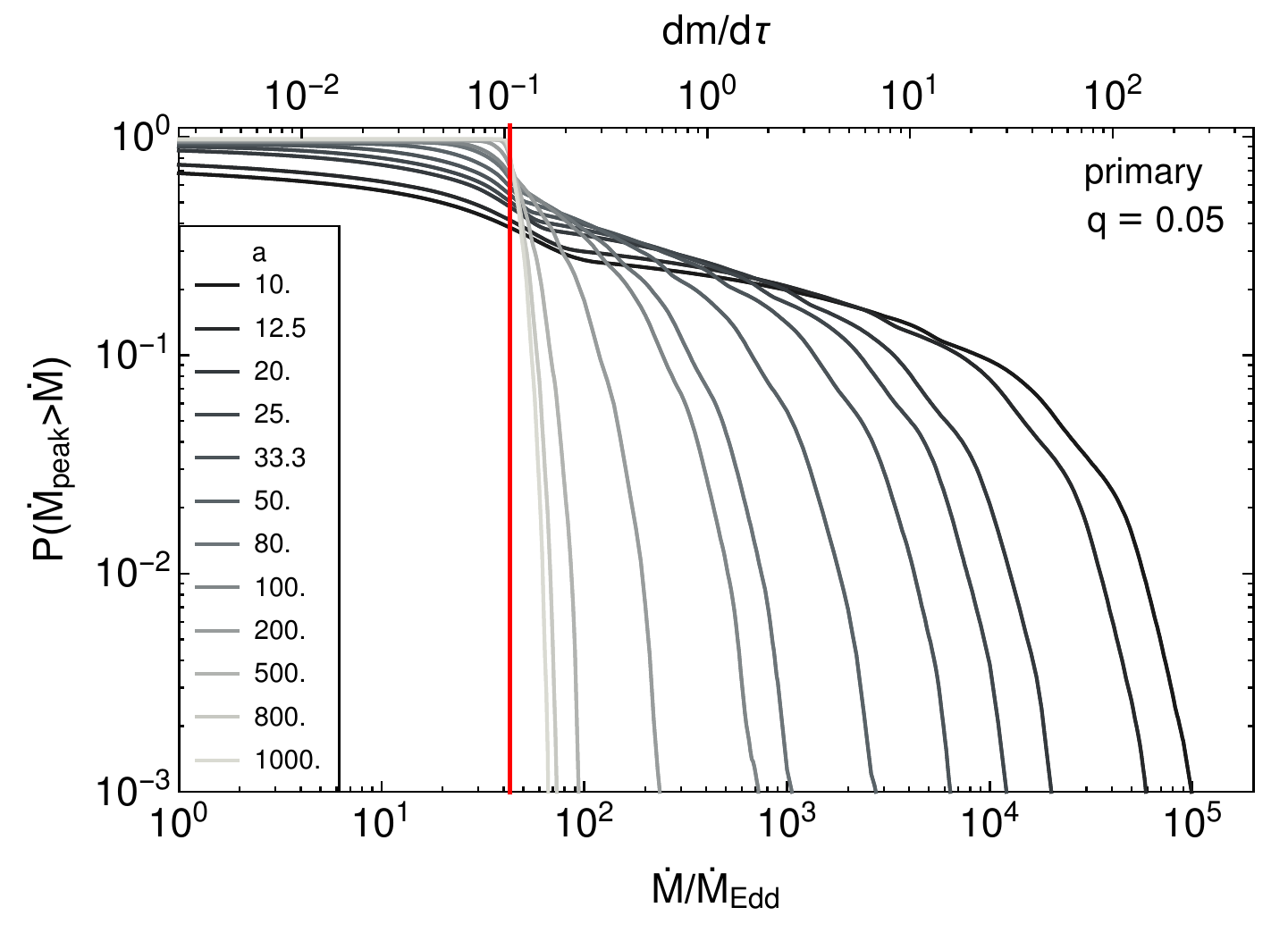}}\hfill
\subfloat{\includegraphics[width=0.49\textwidth]{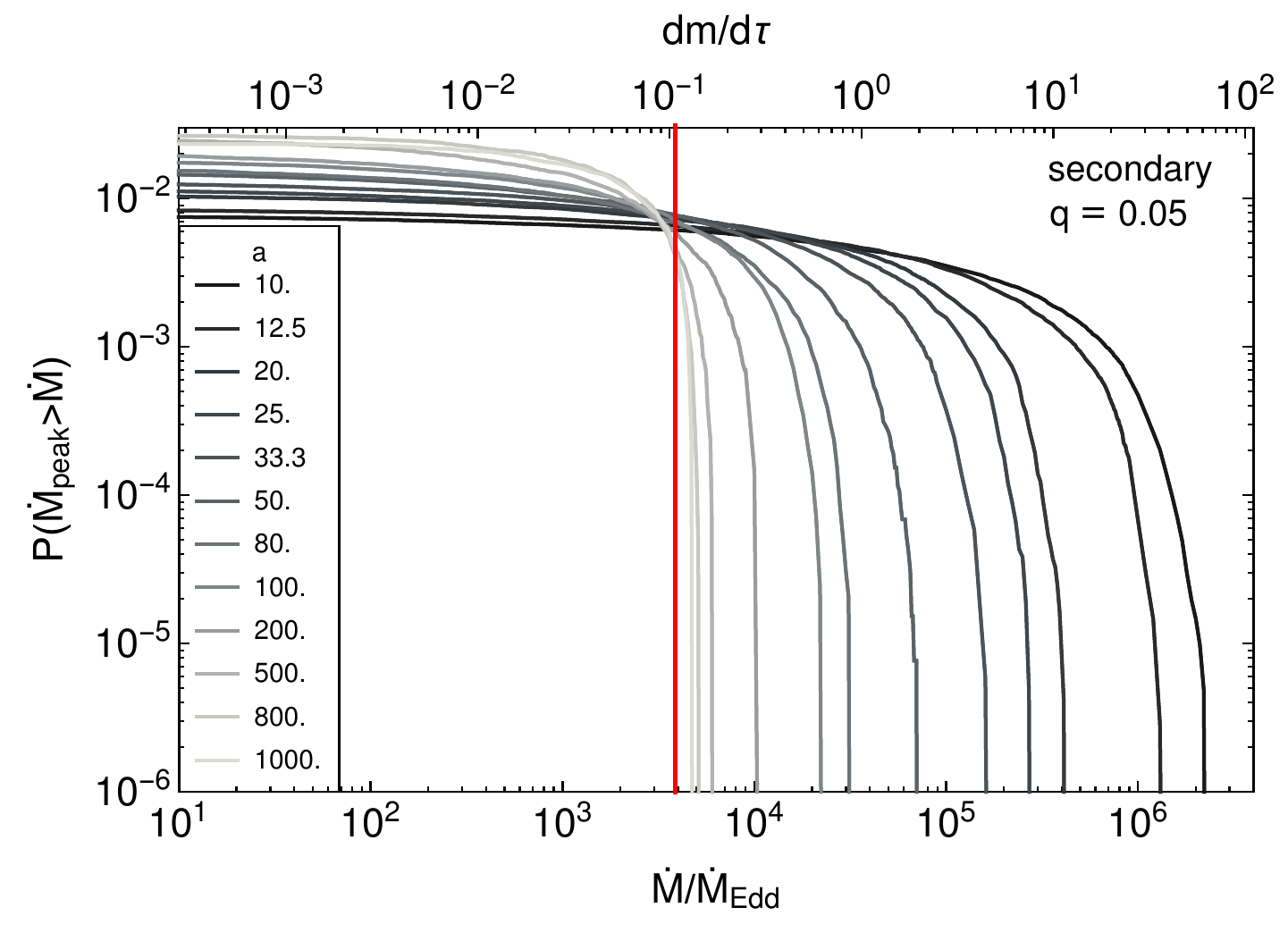}}\\
\subfloat{\includegraphics[width=0.49\textwidth]{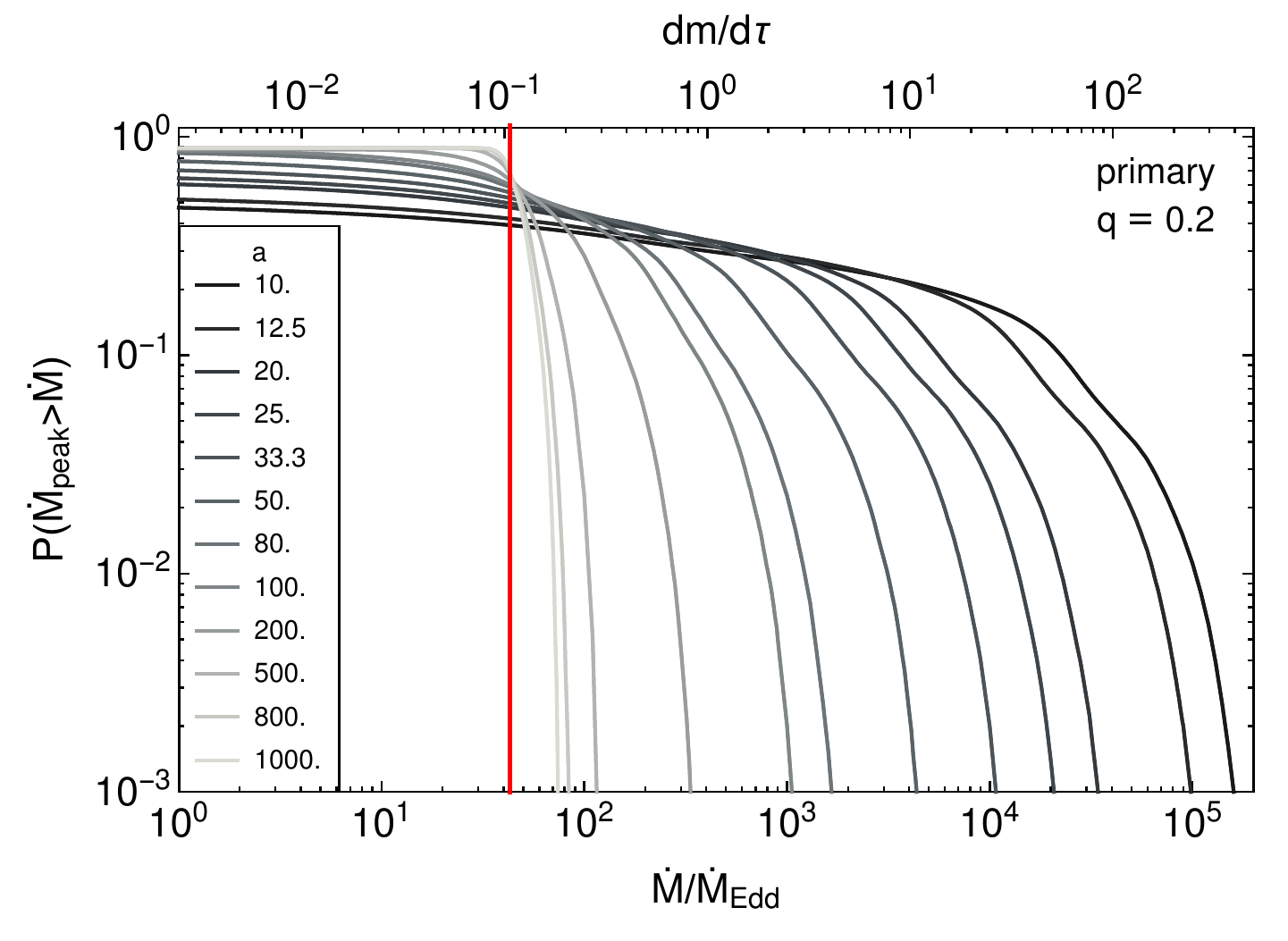}}\hfill
\subfloat{\includegraphics[width=0.49\textwidth]{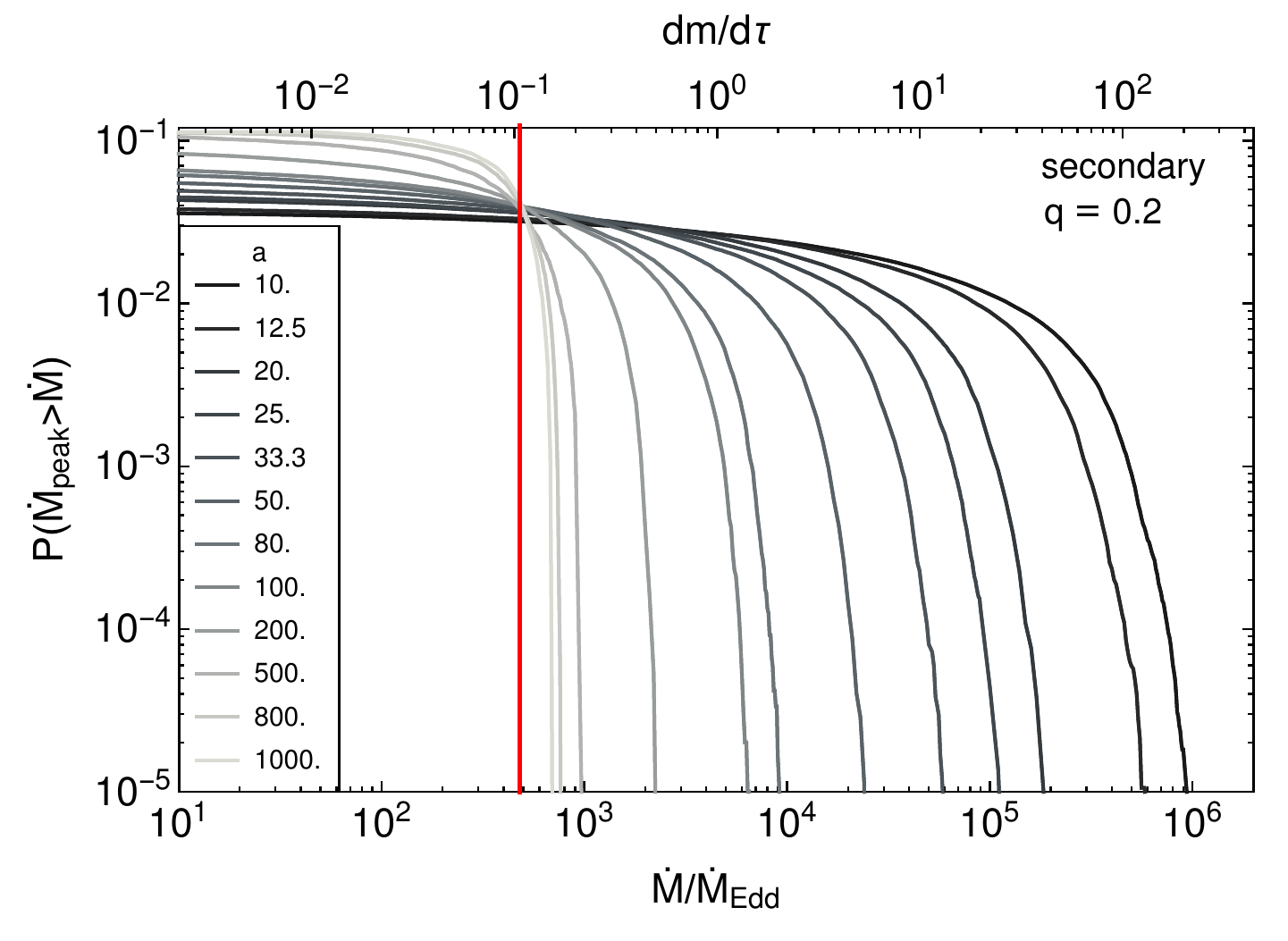}}\\
\subfloat{\includegraphics[width=0.49\textwidth]{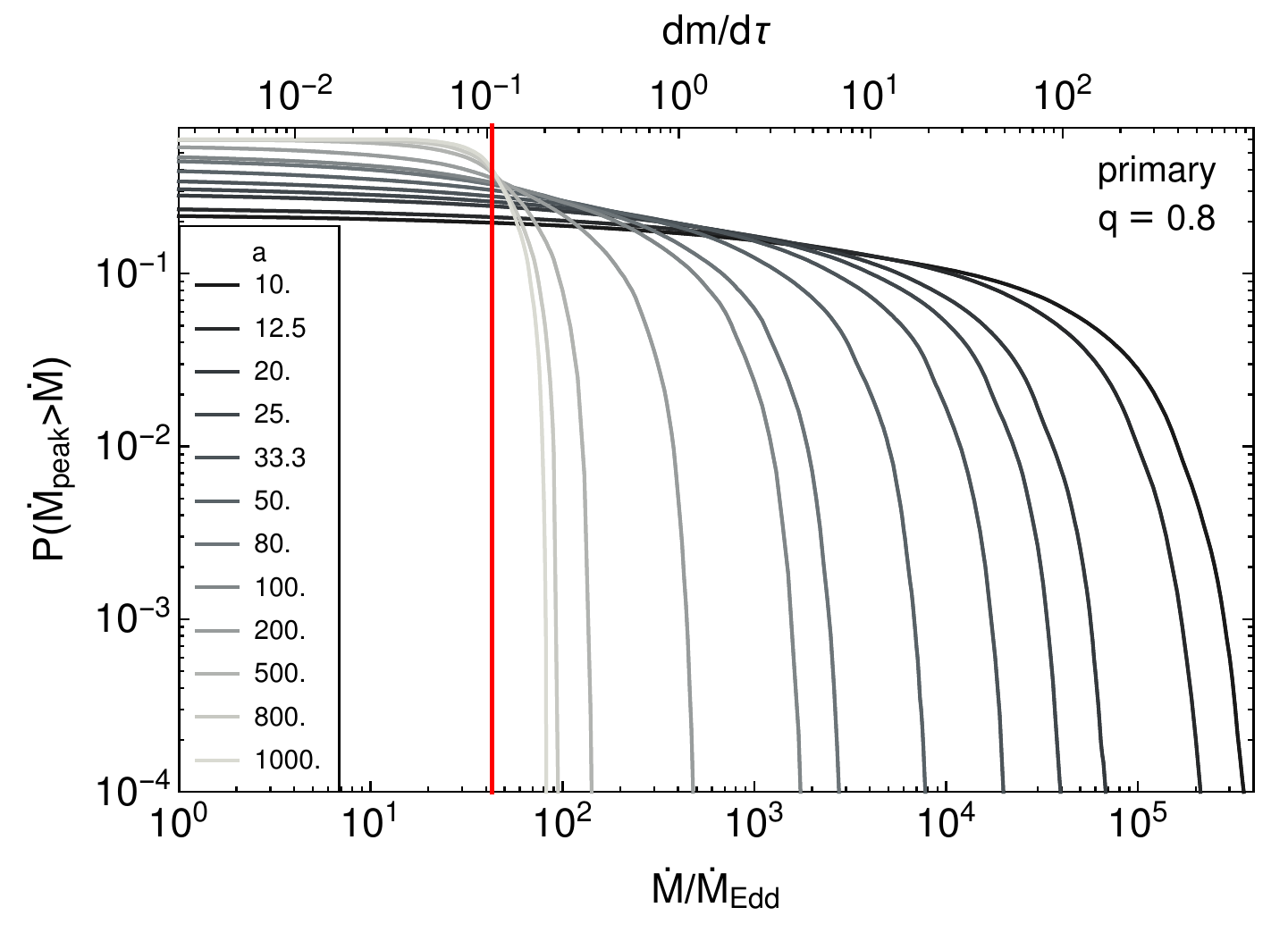}}\hfill
\subfloat{\includegraphics[width=0.49\textwidth]{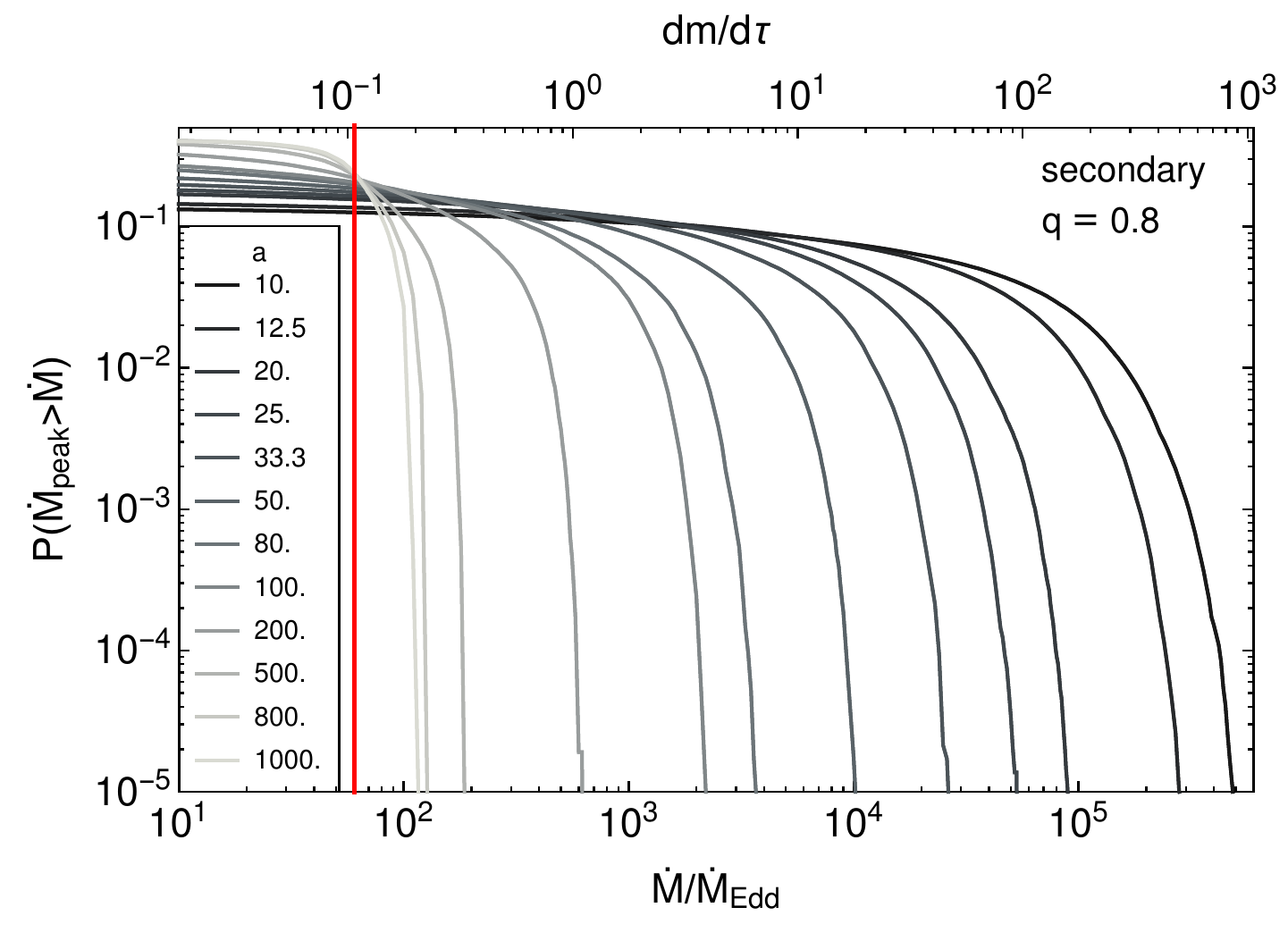}}
\caption{The probability that a disruption by the binary will produce debris with a peak fallback rate $\dot{M}_\textrm{peak} / \dot{M}_\textrm{Edd} > \dot{M} / \dot{M}_\textrm{Edd}$ for a primary mass $M_1 = 10^6 M_\odot$ and different $a$. The maximum probabilities are $(\lambda_{ti} / \lambda_t) f_b$, where $i = 1 \ (2)$ refers to the primary (secondary), and these quantities depend on $q$ and $a$ (Figures \ref{fig:lambdat12b}, \ref{fig:fbound}). The top panels show the results for $q = 0.05$, the center ones for $q = 0.2$, and the bottom ones for $q = 0.8$. The left panels show disruptions by the primary and the right ones show those by the secondary. The black line shows the peak return rate for a TDE from a star with $\varepsilon_c = 0$ (parabolic orbit, half of the debris is bound) disrupted by the appropriate black hole.}
\label{fig:pmdotpeak1fullprobappendix}
\end{figure*}

In this appendix, we expand the figures presented in the paper to compare the behavior of disruptions over the range of binary mass ratios $q$ and separations $a$ that we explored. In addition, we present the full probabilities of observing particular types of TDEs given that a disruption has occurred (Figures \ref{fig:histtrisefullprobappendix} -- \ref{fig:histmdotpeakfullprobappendix} and \ref{fig:ptrise1fullprobappendix} -- \ref{fig:pmdotpeak1fullprobappendix}).

\begin{figure*}
\centering
\subfloat{\includegraphics[width=0.49\textwidth]{figures/{ecct1-q_0_05-q_0_2-q_0_8-rt1_0_002}.pdf}}\hfill
\subfloat{\includegraphics[width=0.49\textwidth]{figures/{ecct2-q_0_05-q_0_2-q_0_8-rt1_0_002}.pdf}}\\
\subfloat{\includegraphics[width=0.49\textwidth]{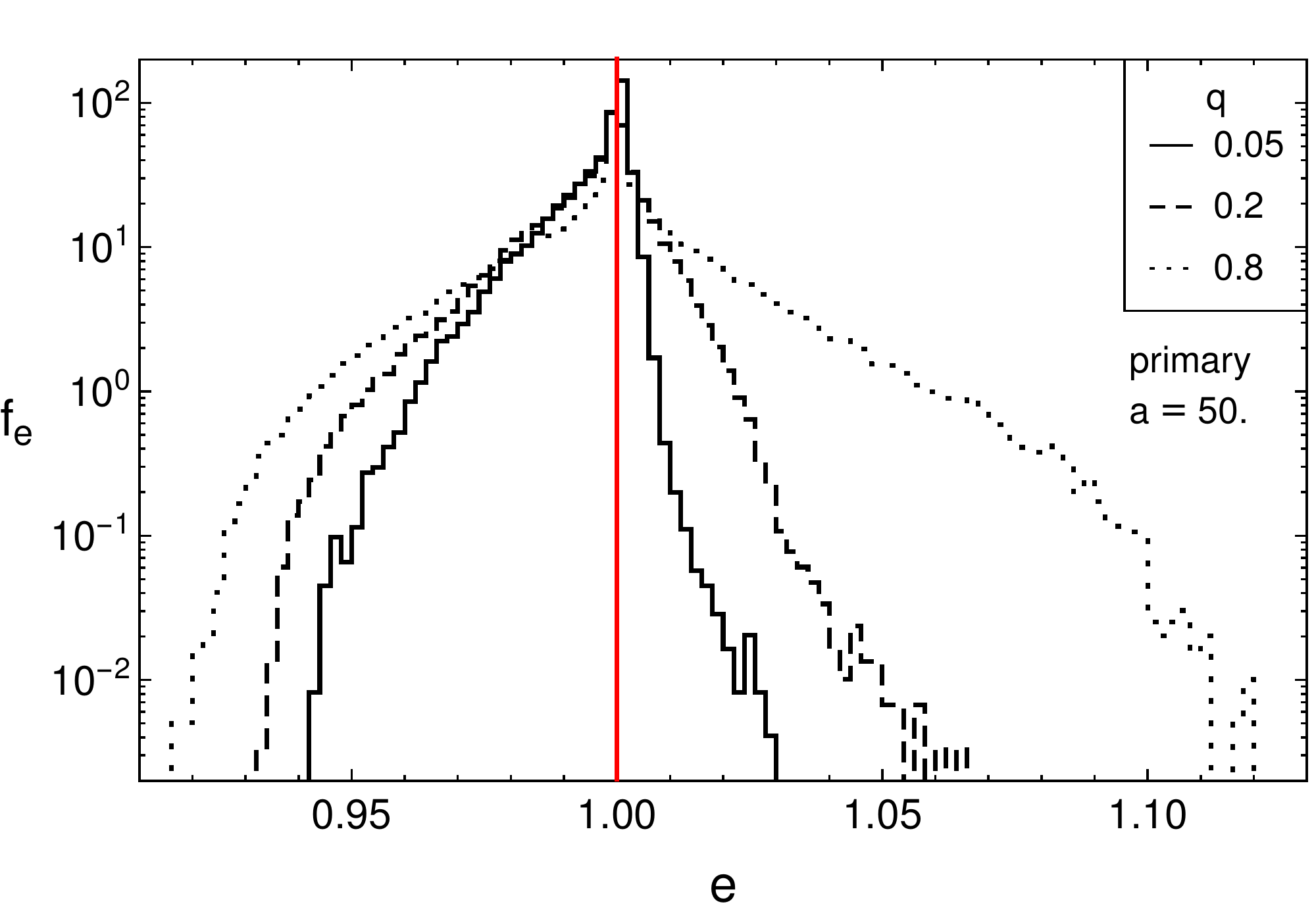}}\hfill
\subfloat{\includegraphics[width=0.49\textwidth]{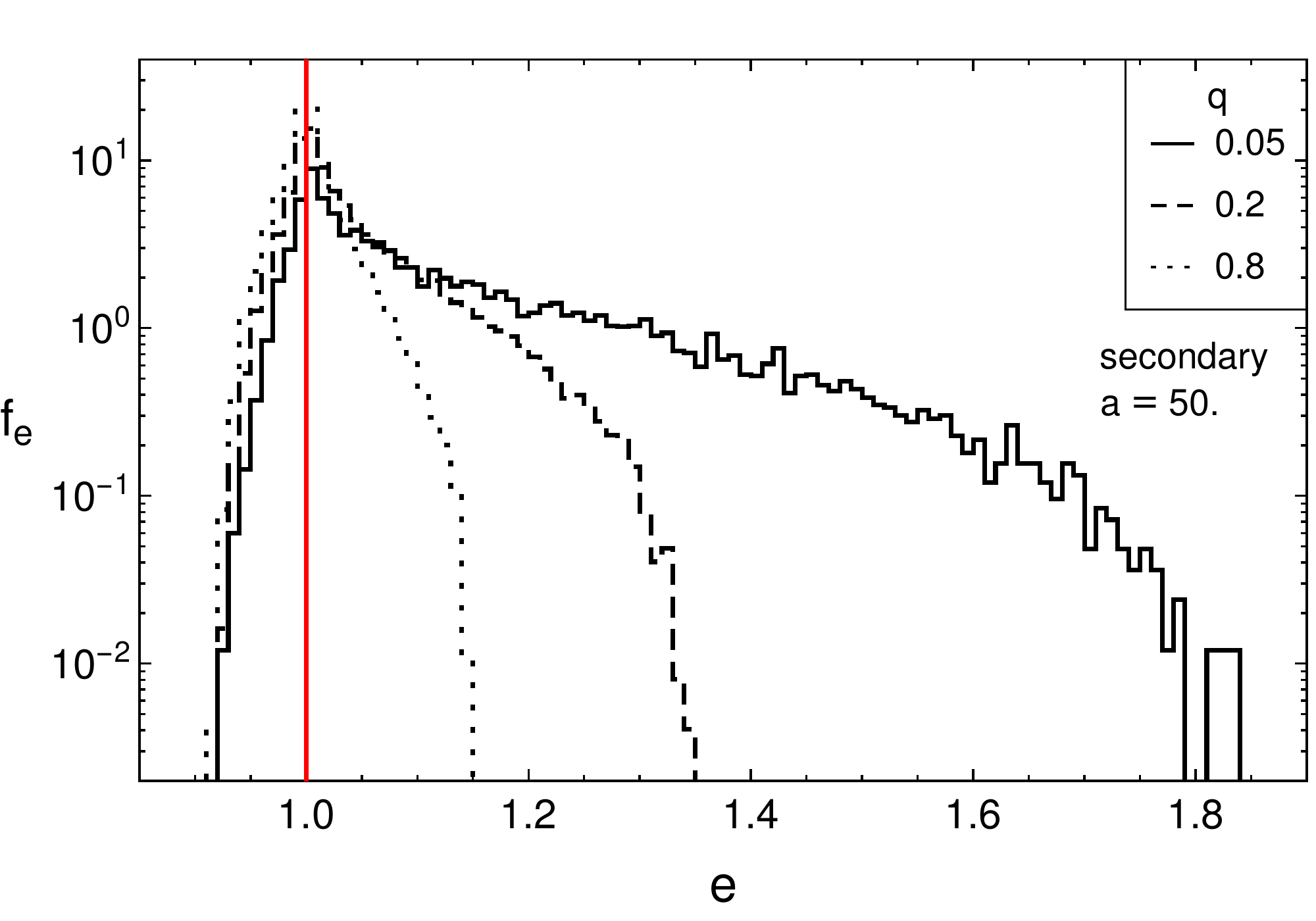}}
\caption{The PDFs for the eccentricity of disrupted stars relative to the disrupting BH calculated at the tidal radius, for $a = 500$ and $q=0.05$ (solid), $0.2$ (dashed), and $0.8$ (dotted). The upper panels show the results for $a = 500$ and the lower ones for $a = 50$. The left panel shows disruptions by the primary and the right one shows those by the secondary. For $a=500$, the bin width is $2 \times 10^{-4}$ for the primary and $10^{-3}$ for the secondary, and for $a=50$ they are 10 times larger. The relative eccentricity is given by $e = \sqrt{2 \ell^2_\textrm{cm} \epsilon_\textrm{cm} / (GM_\bullet)^2 + 1}$, where $\epsilon_\textrm{cm}$ and $\ell_\textrm{cm} = \lvert \vec{r}_\textrm{cm} \times \vec{v} \rvert$ are the specific energy and angular momentum of the center-of-mass of the star relative to the BH at its tidal radius, and $M_\bullet$ is the mass of the BH. A star on a parabolic orbit disrupted by an isolated BH has an eccentricity $e=1$, as marked by the red line.
}
\label{fig:histeccappendix}
\end{figure*}

\begin{figure*}
\centering
\subfloat{\includegraphics[width=0.49\textwidth]{figures/{epsct1-q_0_05-q_0_2-q_0_8-rt1_0_002}.pdf}}\hfill
\subfloat{\includegraphics[width=0.49\textwidth]{figures/{epsct2-q_0_05-q_0_2-q_0_8-rt1_0_002}.pdf}}\\
\subfloat{\includegraphics[width=0.49\textwidth]{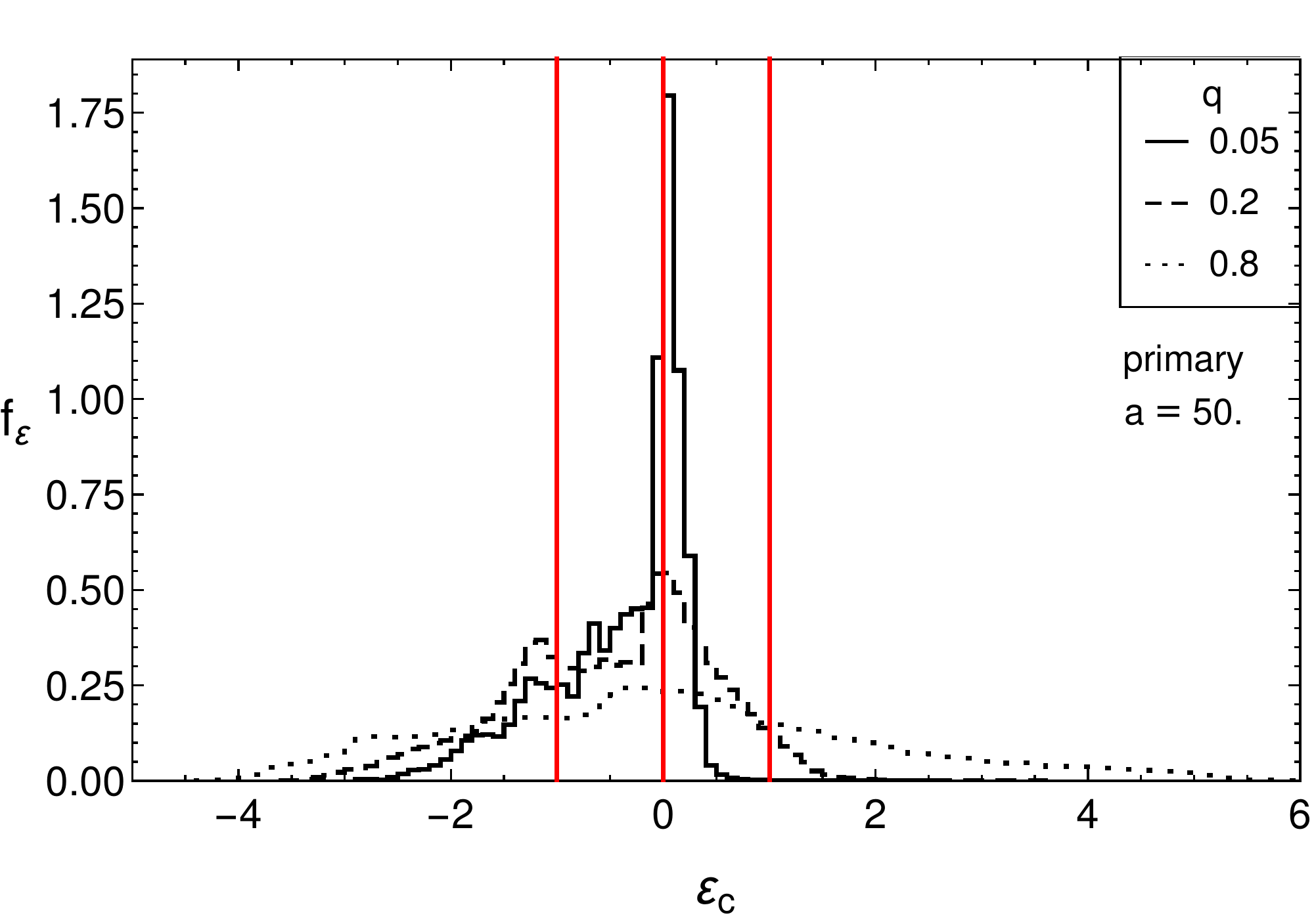}}\hfill
\subfloat{\includegraphics[width=0.49\textwidth]{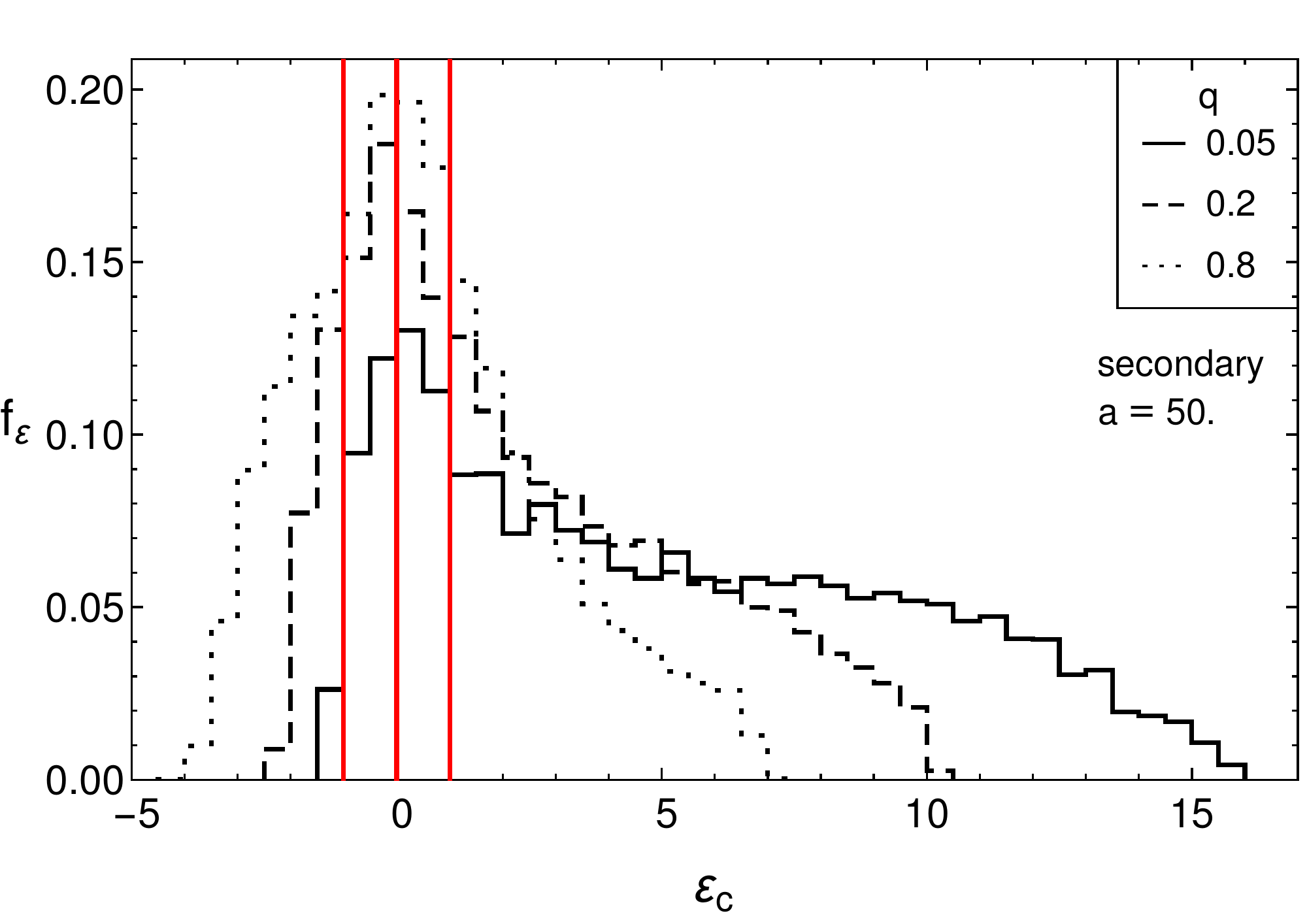}}
\caption{The PDFs for the critical energy $\varepsilon_c$ of disrupted stars relative to the disrupting BH calculated at the tidal radius, for $a=500$ and $q=0.05$ (solid), $0.2$ (dashed), and $0.8$ (dotted). The left panels show disruptions by the primary and the right ones show those by the secondary. The upper panels show the results for $a = 500$ and the lower ones for $a = 50$. The upper panels have bin widths $\Delta_\varepsilon = 0.01$ (primary) and $0.05$ (secondary), and the bottom ones have bins a factor of 10 larger. The red lines mark $\varepsilon_c = -1, 0, +1$, if they are in a histogram's domain. The stellar debris is bound to a different extent in the regions the lines delimit: if $\varepsilon_c < -1$, then all of the debris is bound to the BH; if $-1 \leq \varepsilon_c < 1$, then part of it is bound; and if $\varepsilon_c \geq 1$, then all of it is unbound. If $\varepsilon_c=0$, half of the debris remains bound and half escapes, which occurs when a star on a parabolic orbit is disrupted by an isolated BH.
}
\label{fig:histepscappendix}
\end{figure*}

\begin{figure*}
\centering
\subfloat{\includegraphics[width=0.49\textwidth]{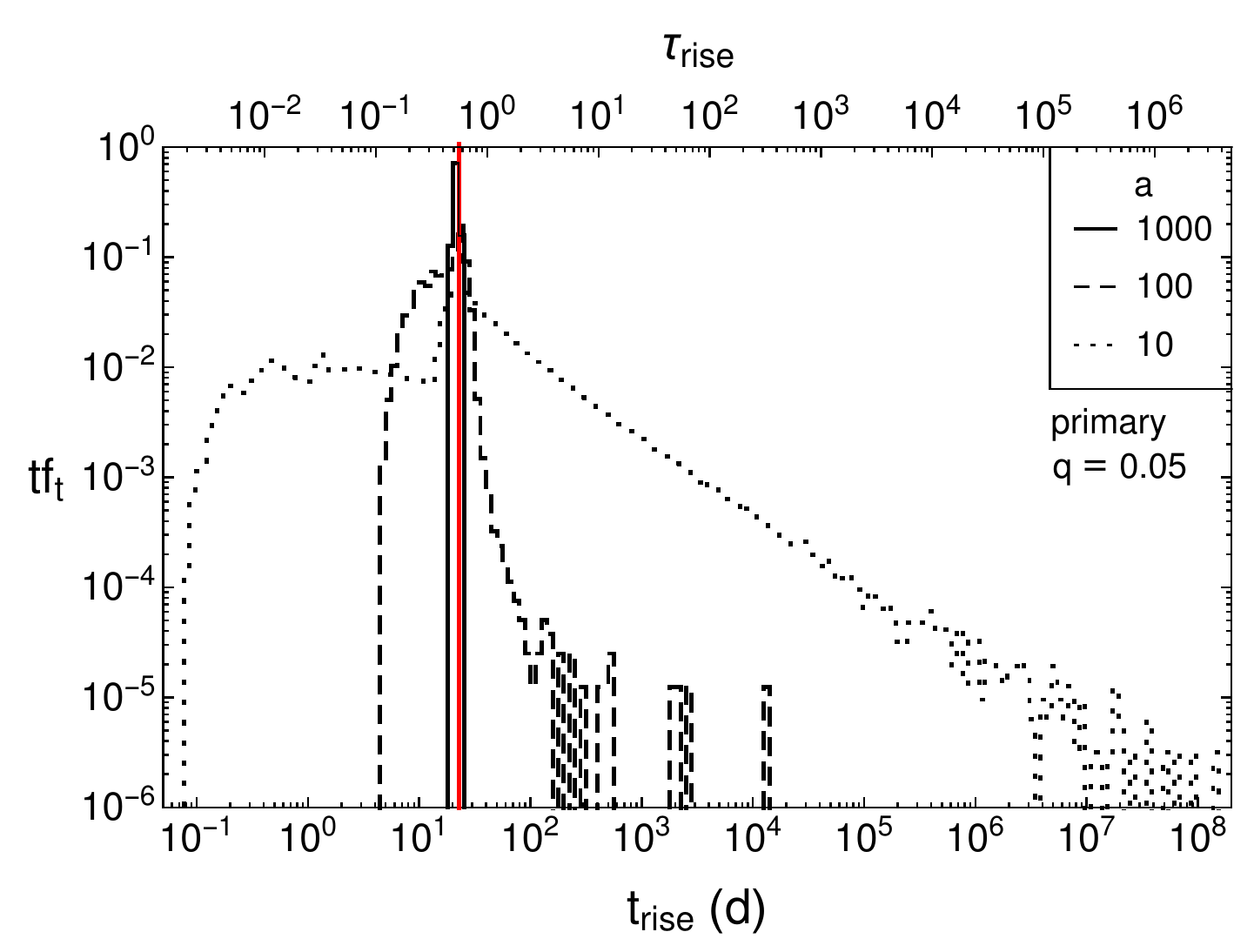}}\hfill
\subfloat{\includegraphics[width=0.49\textwidth]{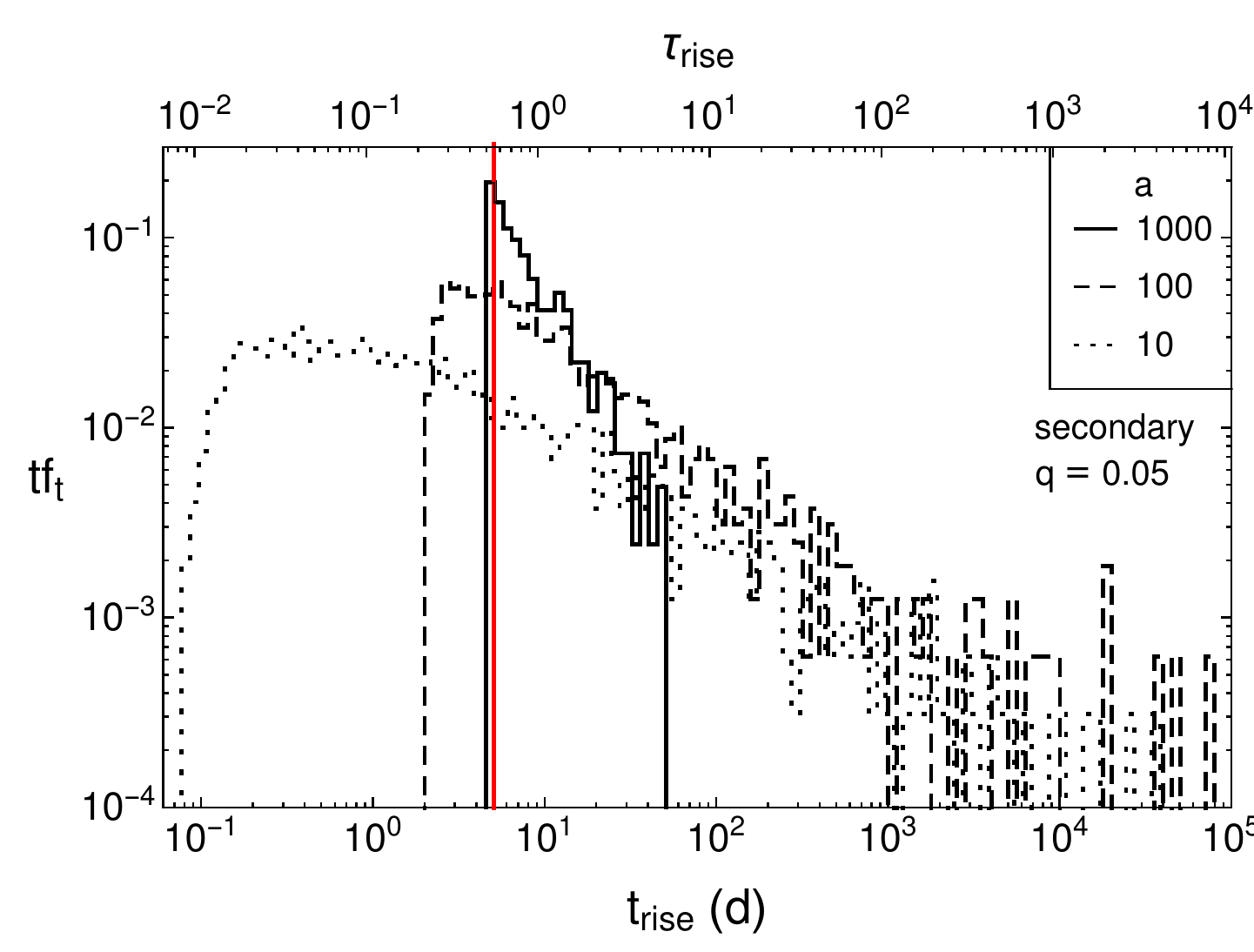}}\\
\subfloat{\includegraphics[width=0.49\textwidth]{figures/{timeriset1-q_0_2-rt1_0_001-rt1_0_01-rt1_0_1}.pdf}}\hfill
\subfloat{\includegraphics[width=0.49\textwidth]{figures/{timeriset2-q_0_2-rt1_0_001-rt1_0_01-rt1_0_1}.pdf}}\\
\subfloat{\includegraphics[width=0.49\textwidth]{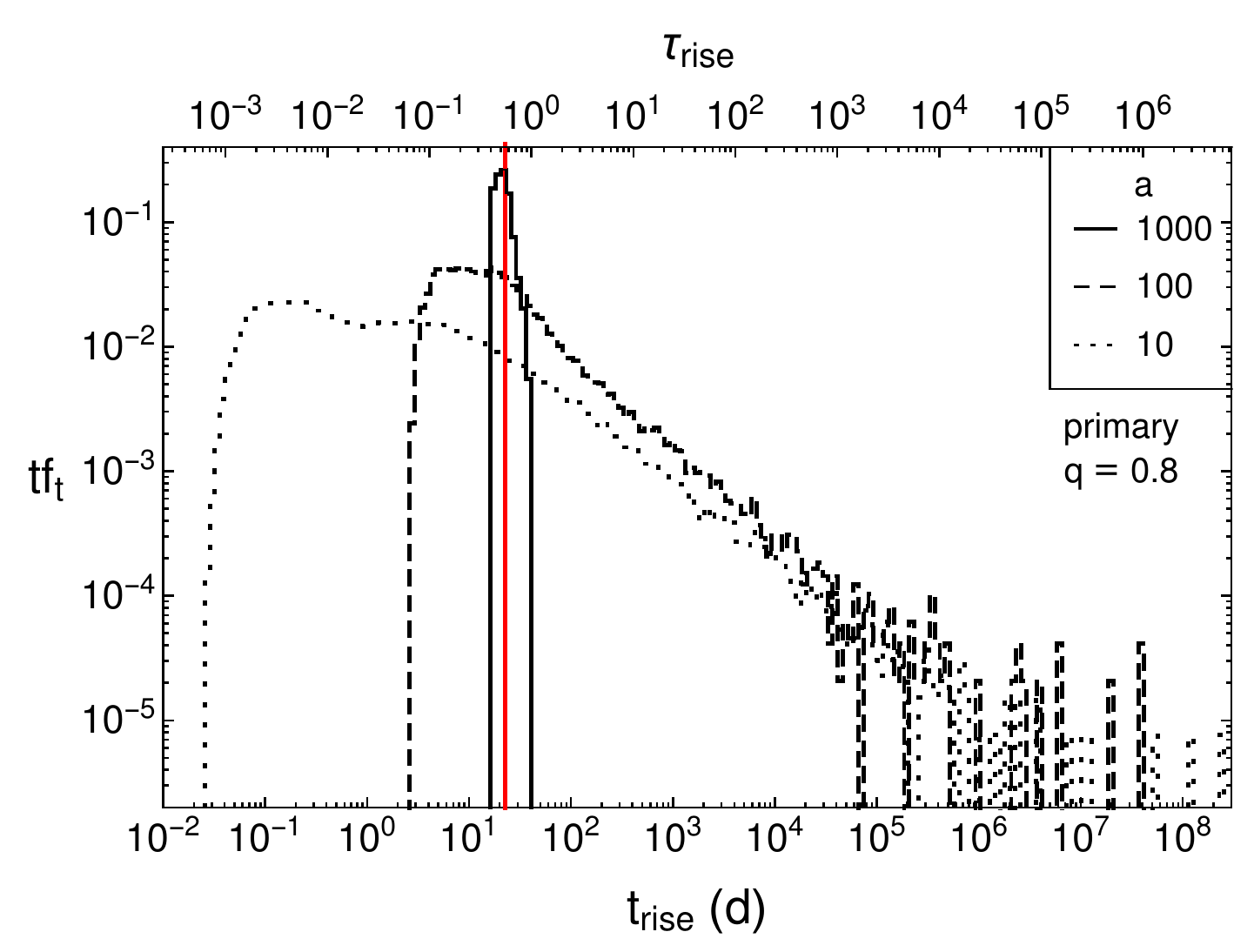}}\hfill
\subfloat{\includegraphics[width=0.49\textwidth]{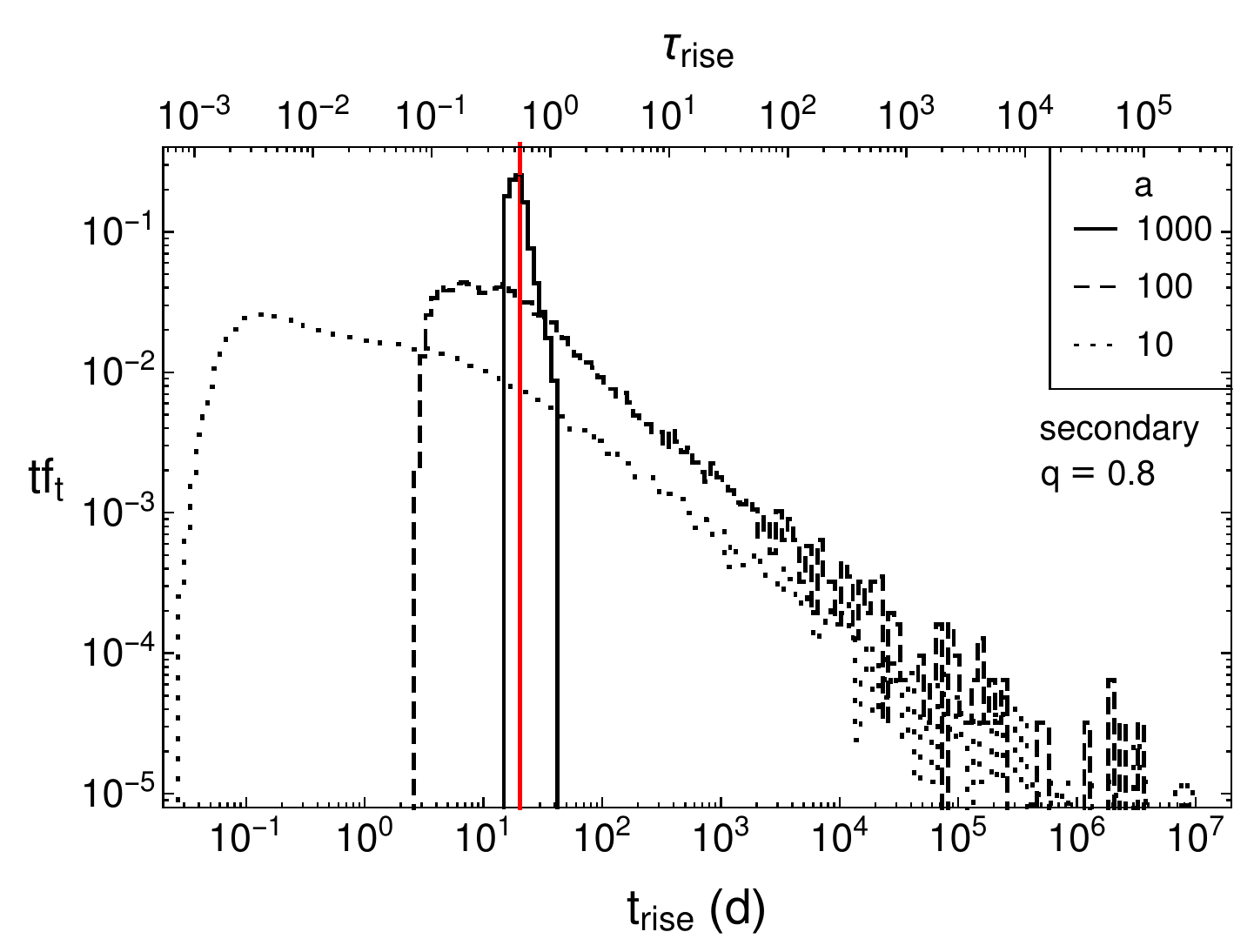}}
\caption{The probability $t f_t$ that the bound debris of the disrupting black hole will produce a rise time $t_\textrm{rise} = \tau_0 \tau_\textrm{rise}$ for a primary mass $M_1 = 10^6 M_\odot$ and $a=1000$ (solid), $100$ (dashed), and $10$ (dotted). The probabilities for each histogram sum to $1$. The top panels show the results for $q = 0.05$, the center ones for $q = 0.2$, and the bottom ones for $q = 0.8$. The left panels show disruptions by the primary and the right ones show those by the secondary. The logarithmic bin widths are $\Delta_t = 0.05$ and the heights are the probabilities in each bin. The red line marks the rise time $t_\textrm{rise} = \tau_0 (3\sqrt{3} - 5^{3/4})/5^{3/4}$ for a TDE from a star with $\varepsilon_c = 0$ (parabolic orbit, half of the debris is bound) disrupted by the appropriate black hole.}
\label{fig:histtriseappendix}
\end{figure*}

\begin{figure*}
\centering
\subfloat{\includegraphics[width=0.49\textwidth]{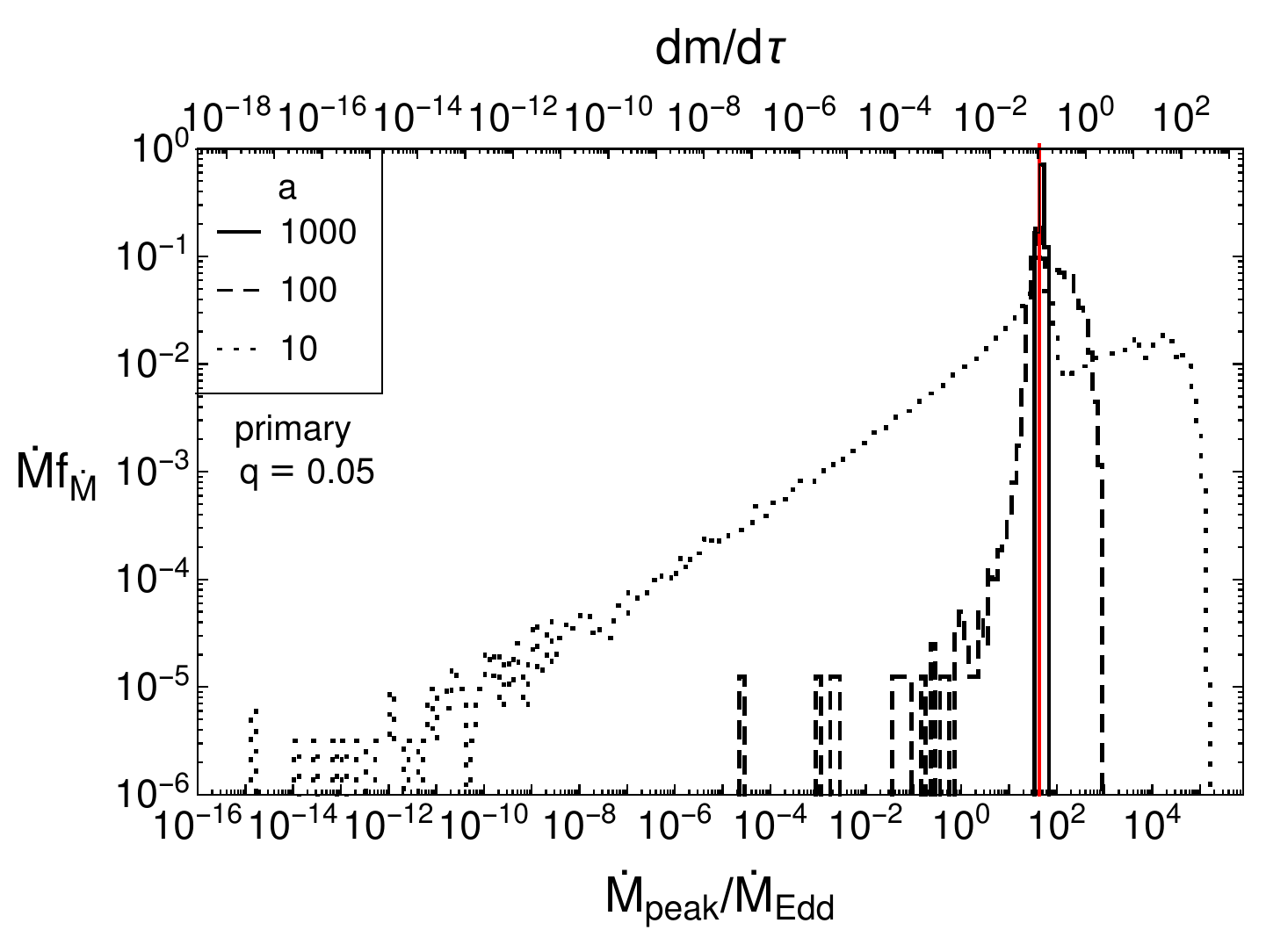}}\hfill
\subfloat{\includegraphics[width=0.49\textwidth]{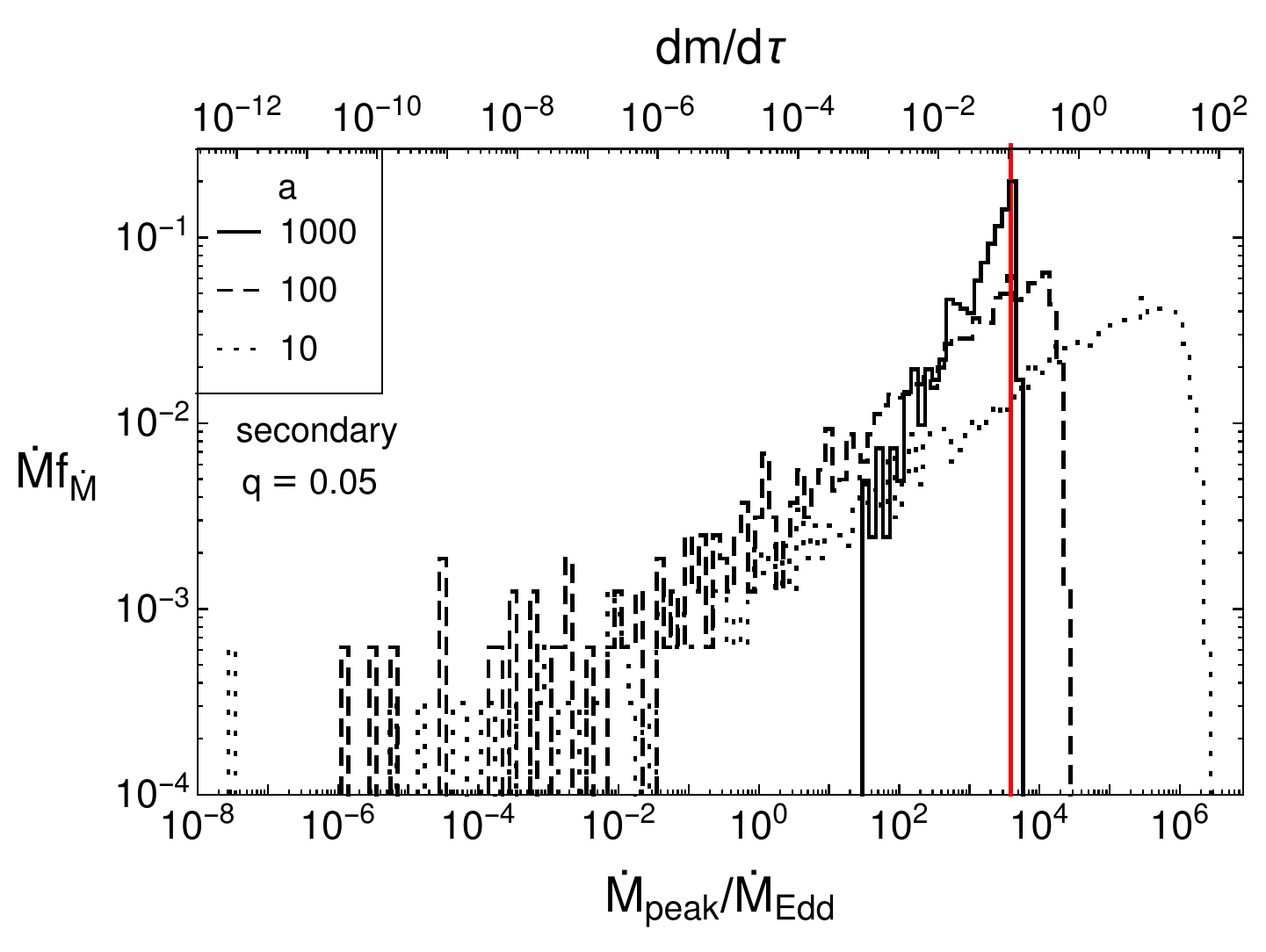}}\\
\subfloat{\includegraphics[width=0.49\textwidth]{figures/{mdotpeakt1-q_0_2-rt1_0_001-rt1_0_01-rt1_0_1}.pdf}}\hfill
\subfloat{\includegraphics[width=0.49\textwidth]{figures/{mdotpeakt2-q_0_2-rt1_0_001-rt1_0_01-rt1_0_1}.pdf}}\\
\subfloat{\includegraphics[width=0.49\textwidth]{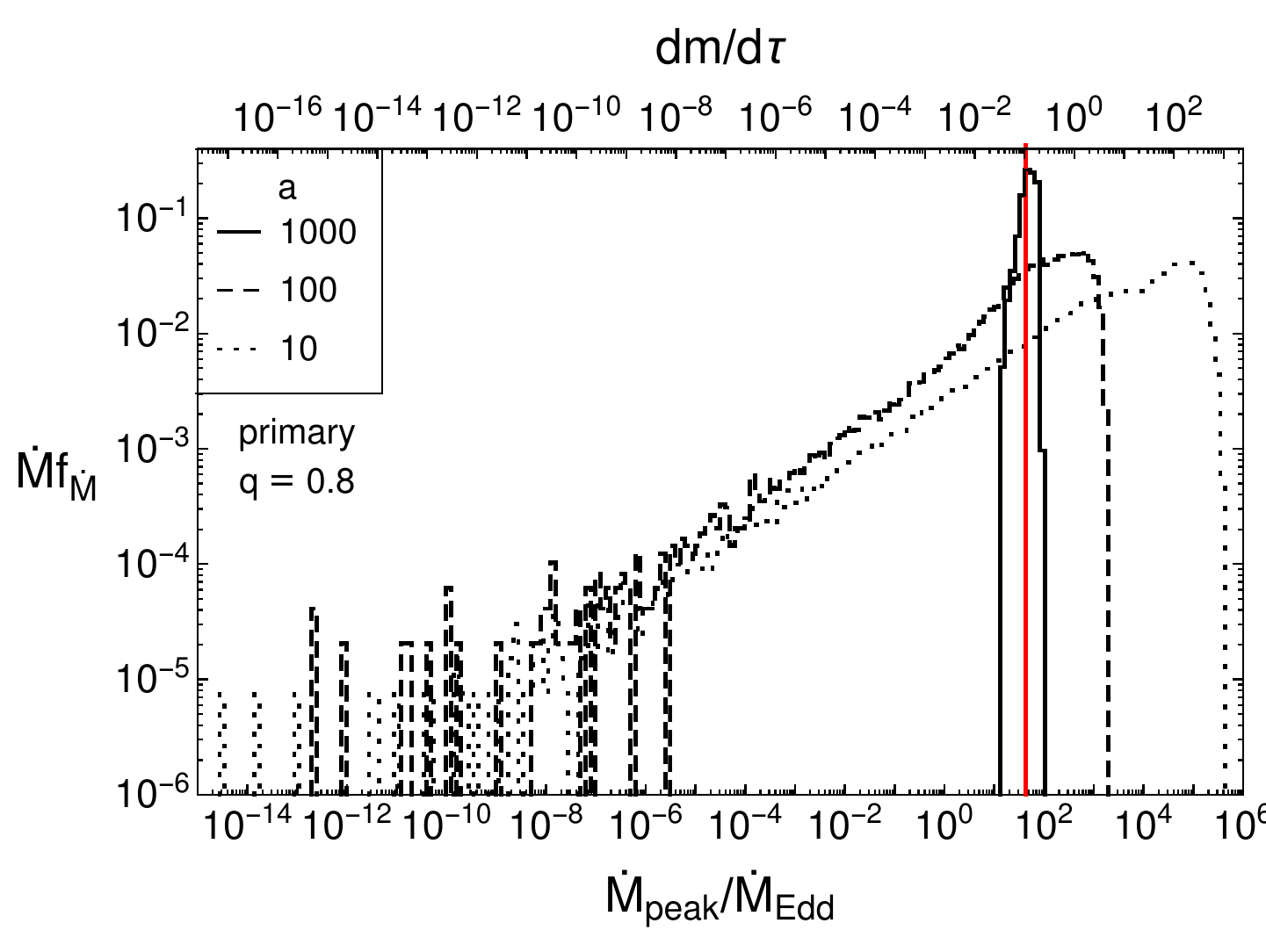}}\hfill
\subfloat{\includegraphics[width=0.49\textwidth]{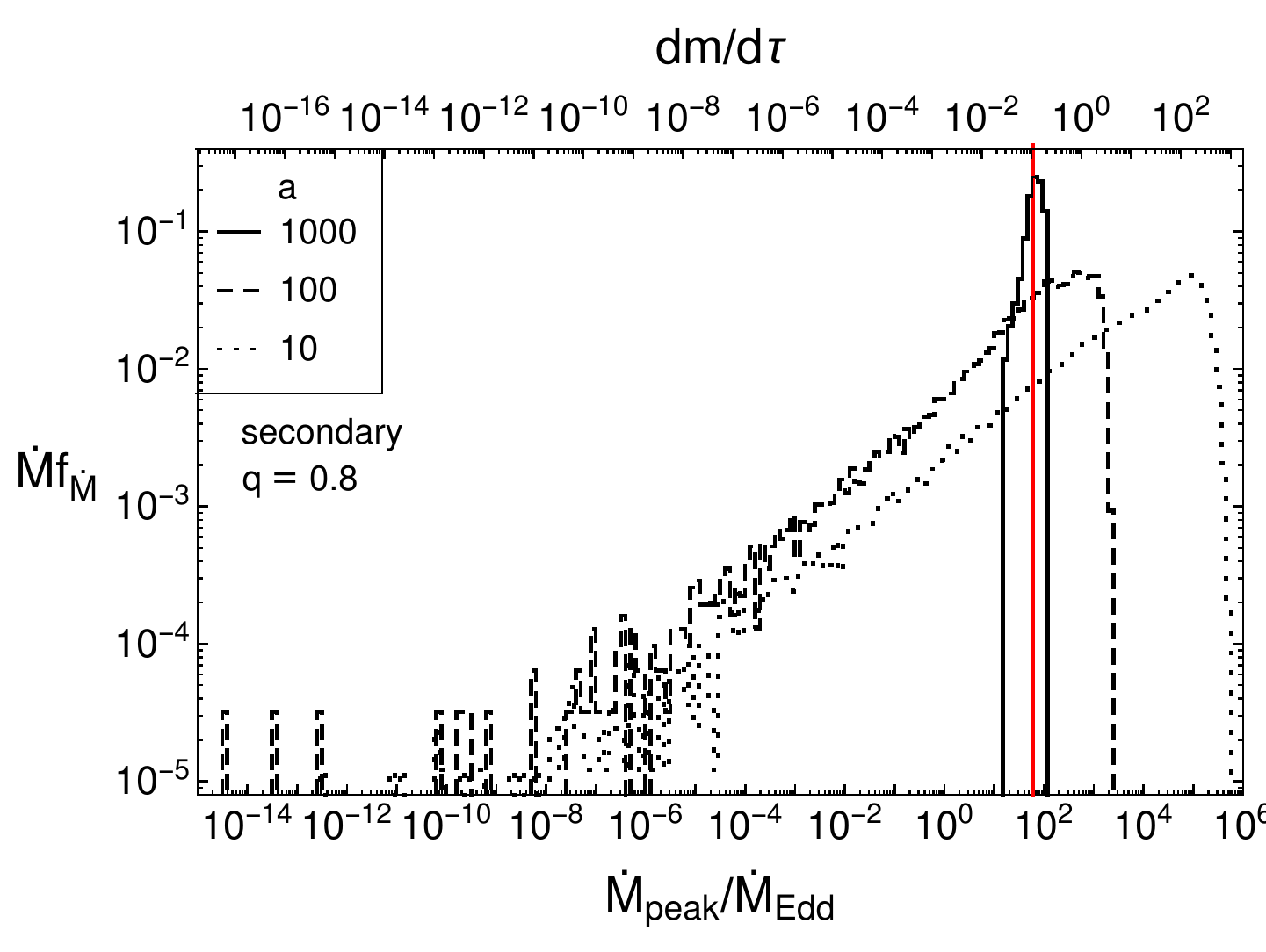}}
\caption{The probability $\dot{M} f_{\dot{M}}$ that the bound debris of the disrupting black hole will produce a peak fallback rate $\dot{M}_\textrm{peak}/\dot{M}_\textrm{Edd}$ (Eq. \ref{eq:mdotpeakovermedd}; see surrounding text for the parameters used) for a primary mass $M_1 = 10^6 M_\odot$ and $a=1000$ (solid), $100$ (dashed), and $10$ (dotted). The probabilities for each histogram sum to $1$. The top panels show the results for $q = 0.05$, the center ones for $q = 0.2$, and the lower ones for $q = 0.8$. The left panels show disruptions by the primary and the right ones show those by the secondary. The logarithmic bin widths are $\Delta_{\dot{M}} = 0.1$ and the heights are the probabilities in each bin. The red line shows the peak return rate for a TDE from a star with $\varepsilon_c = 0$ (parabolic orbit, half of the debris is bound) disrupted by the appropriate black hole.}
\label{fig:histmdotpeakappendix}
\end{figure*}

\begin{figure*}
\centering
\subfloat{\includegraphics[width=0.49\textwidth]{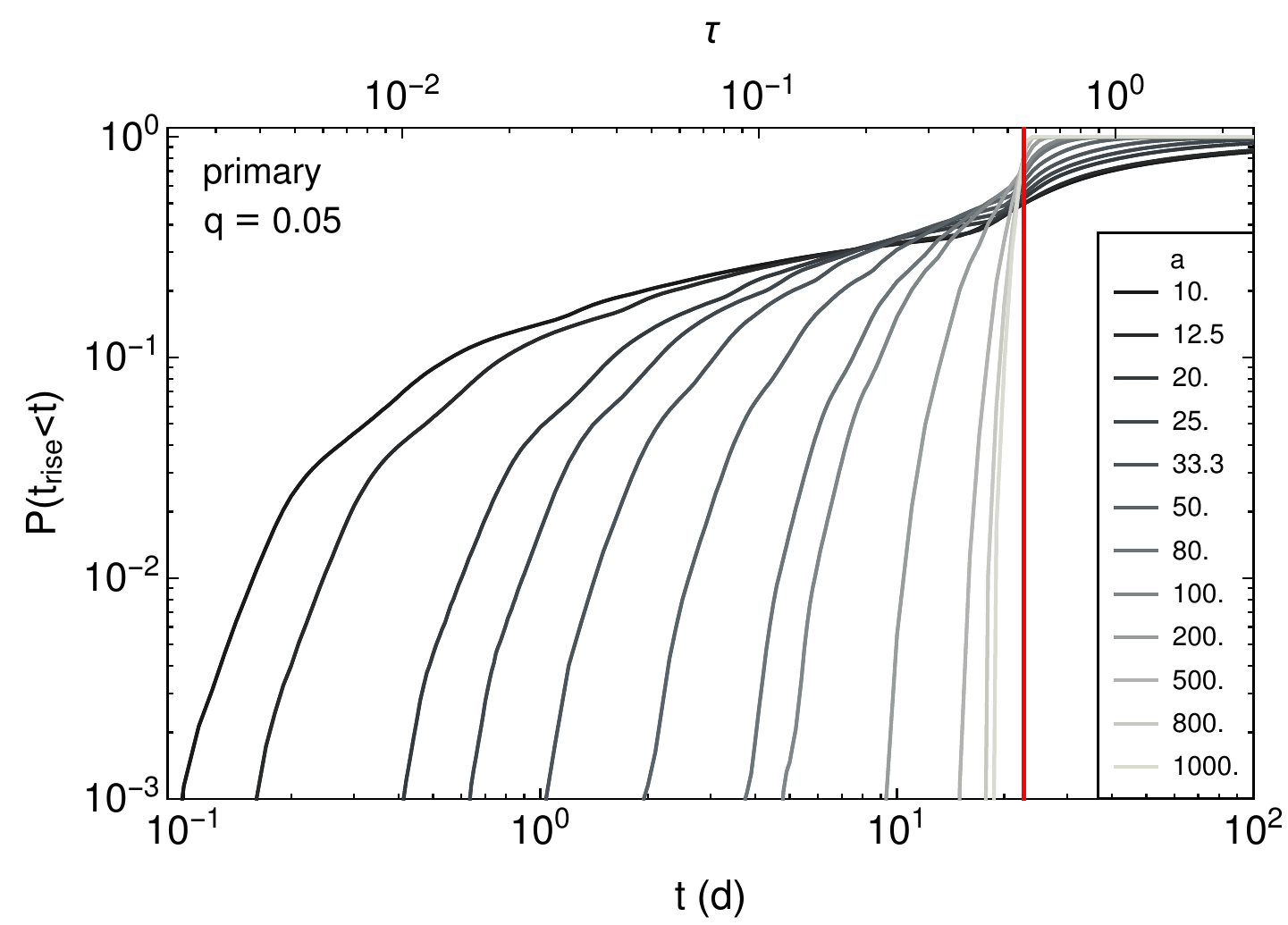}}\hfill
\subfloat{\includegraphics[width=0.49\textwidth]{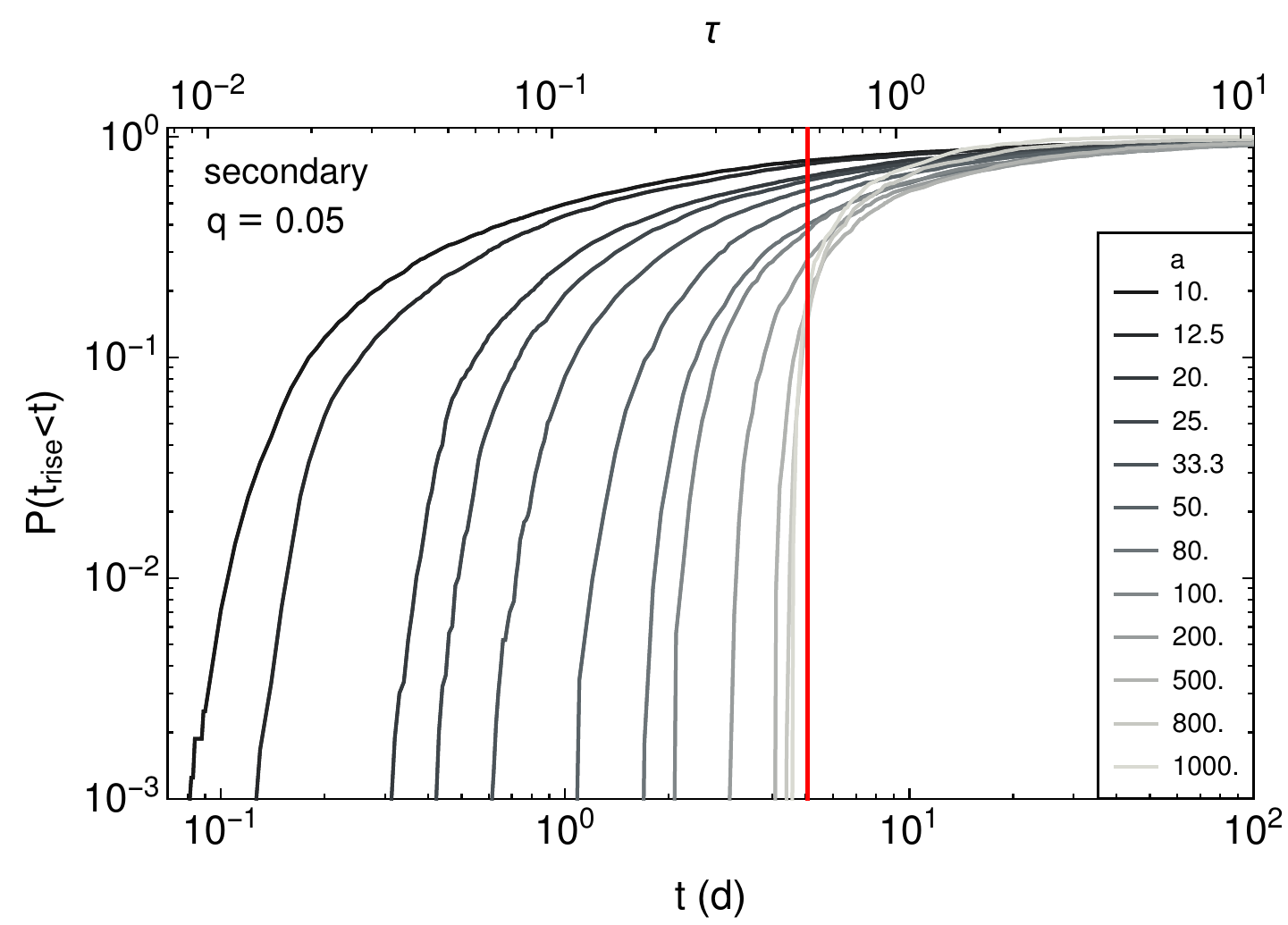}}\\
\subfloat{\includegraphics[width=0.49\textwidth]{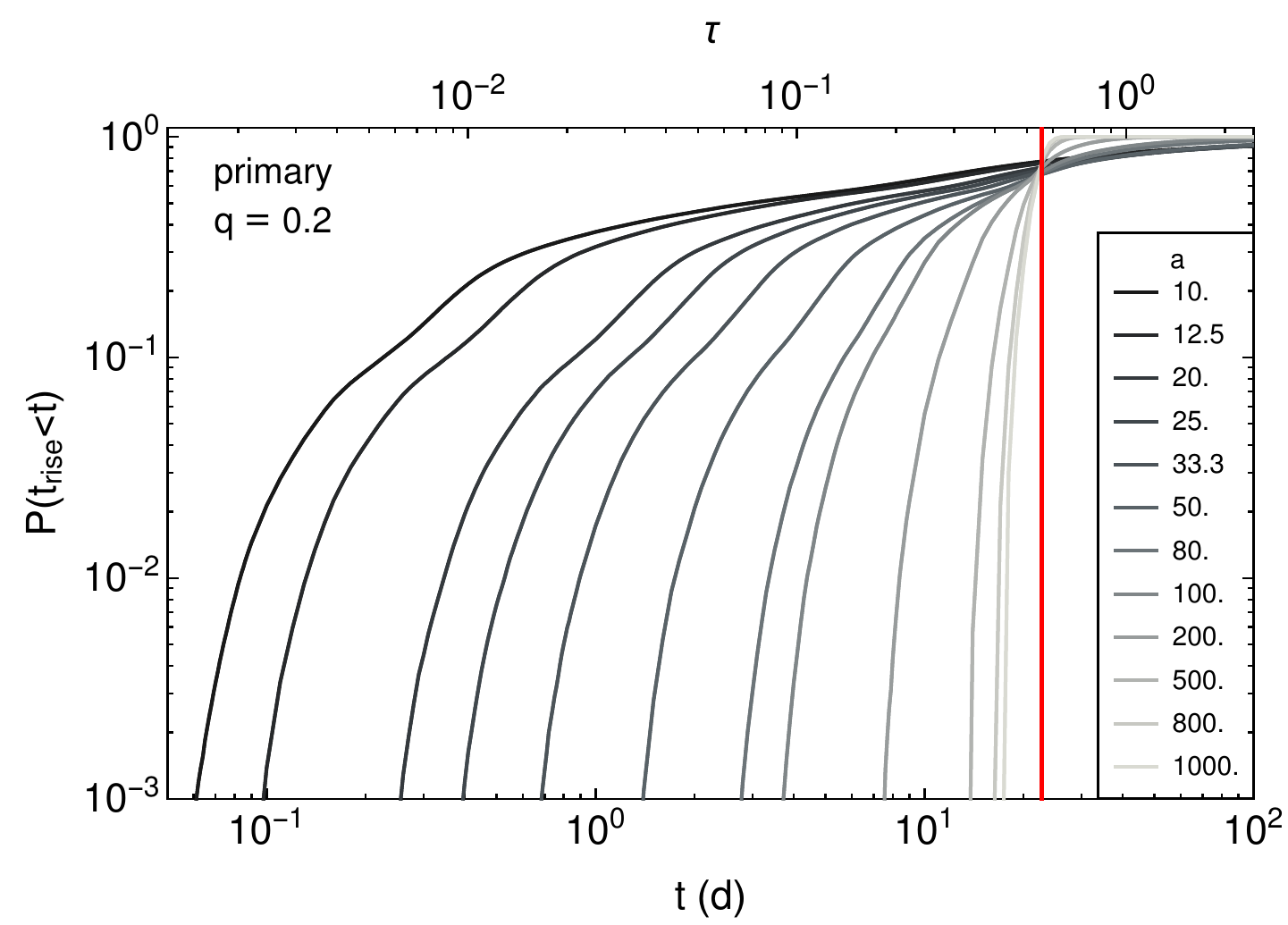}}\hfill
\subfloat{\includegraphics[width=0.49\textwidth]{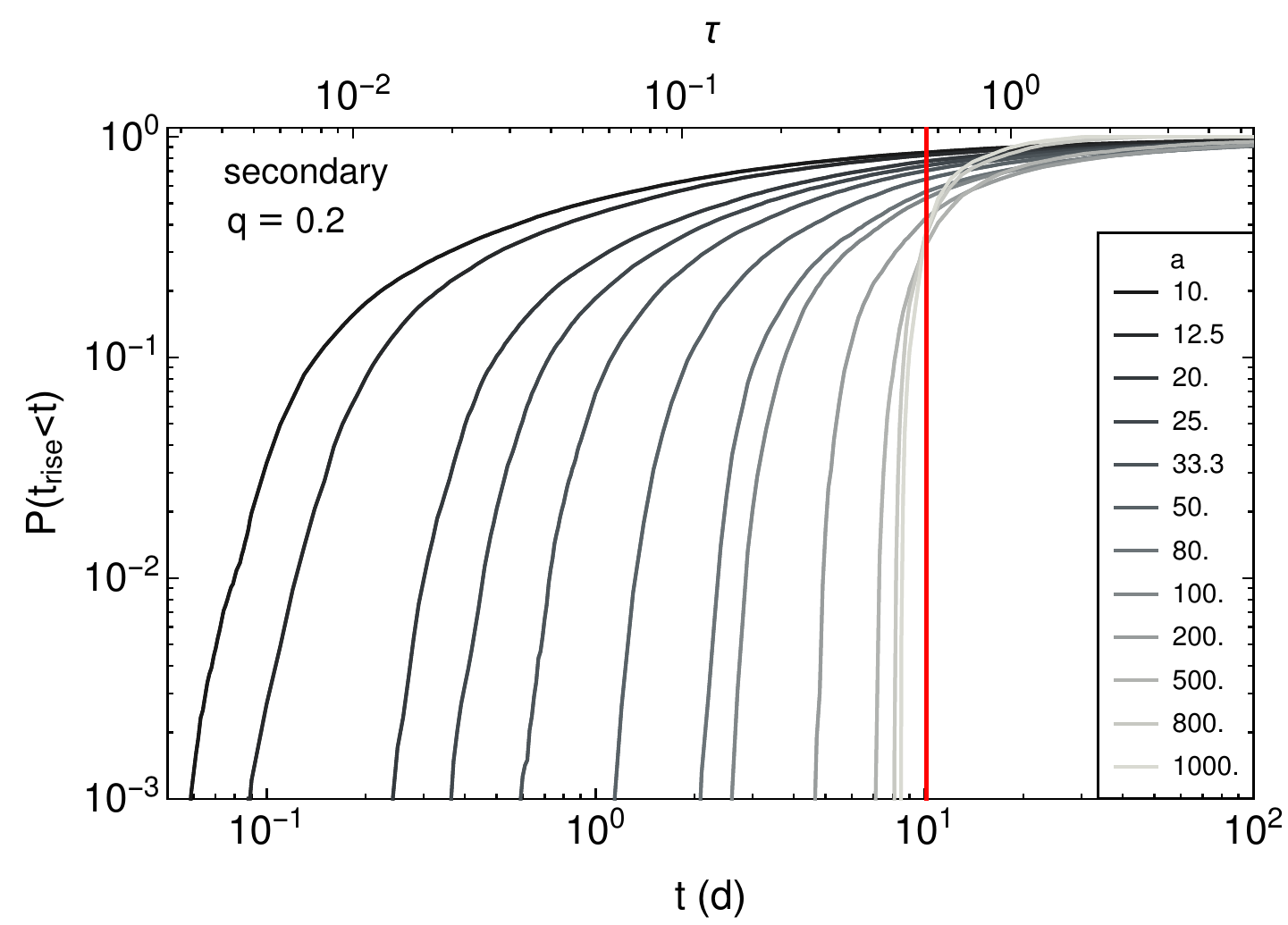}}\\
\subfloat{\includegraphics[width=0.49\textwidth]{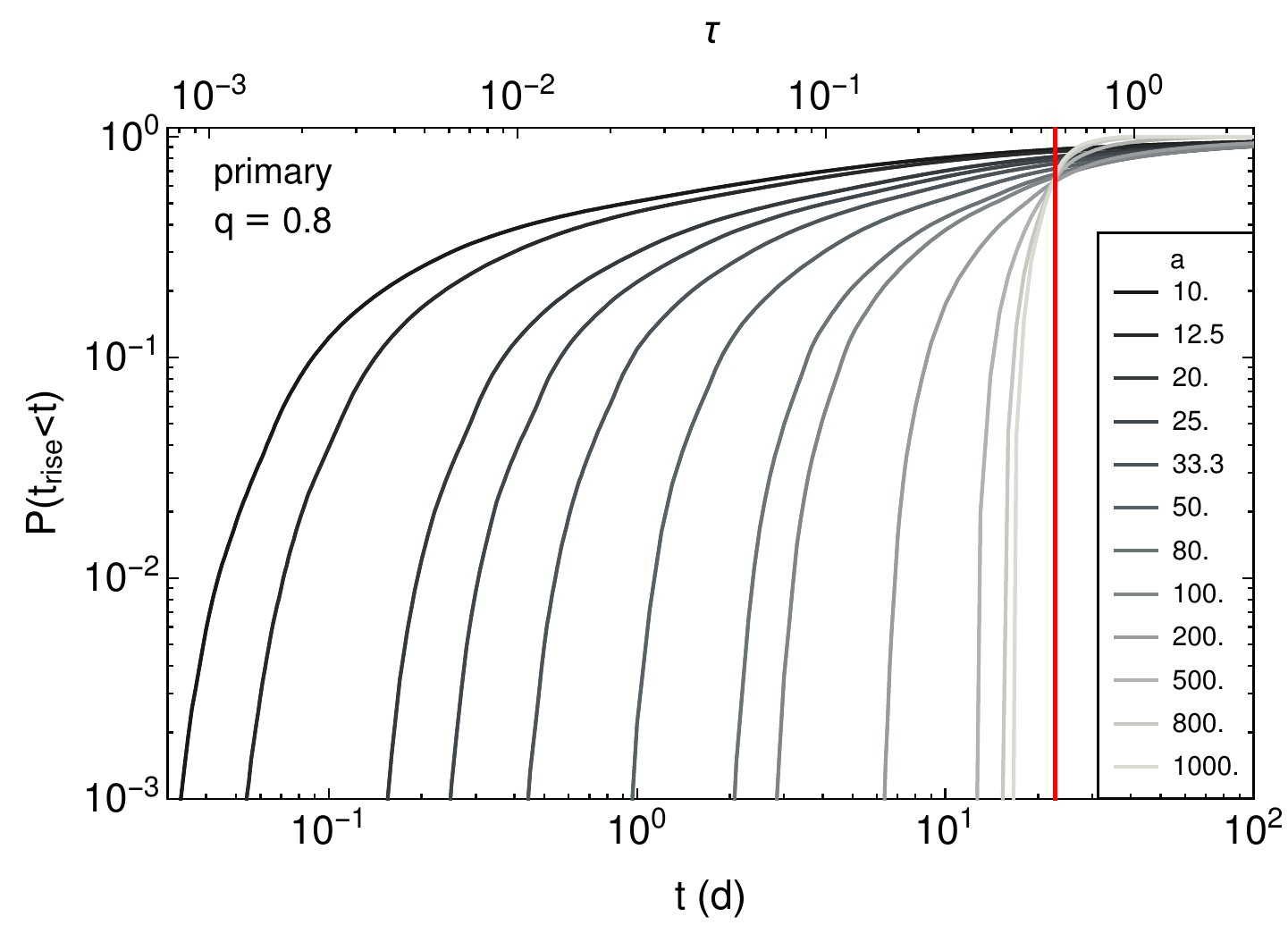}}\hfill
\subfloat{\includegraphics[width=0.49\textwidth]{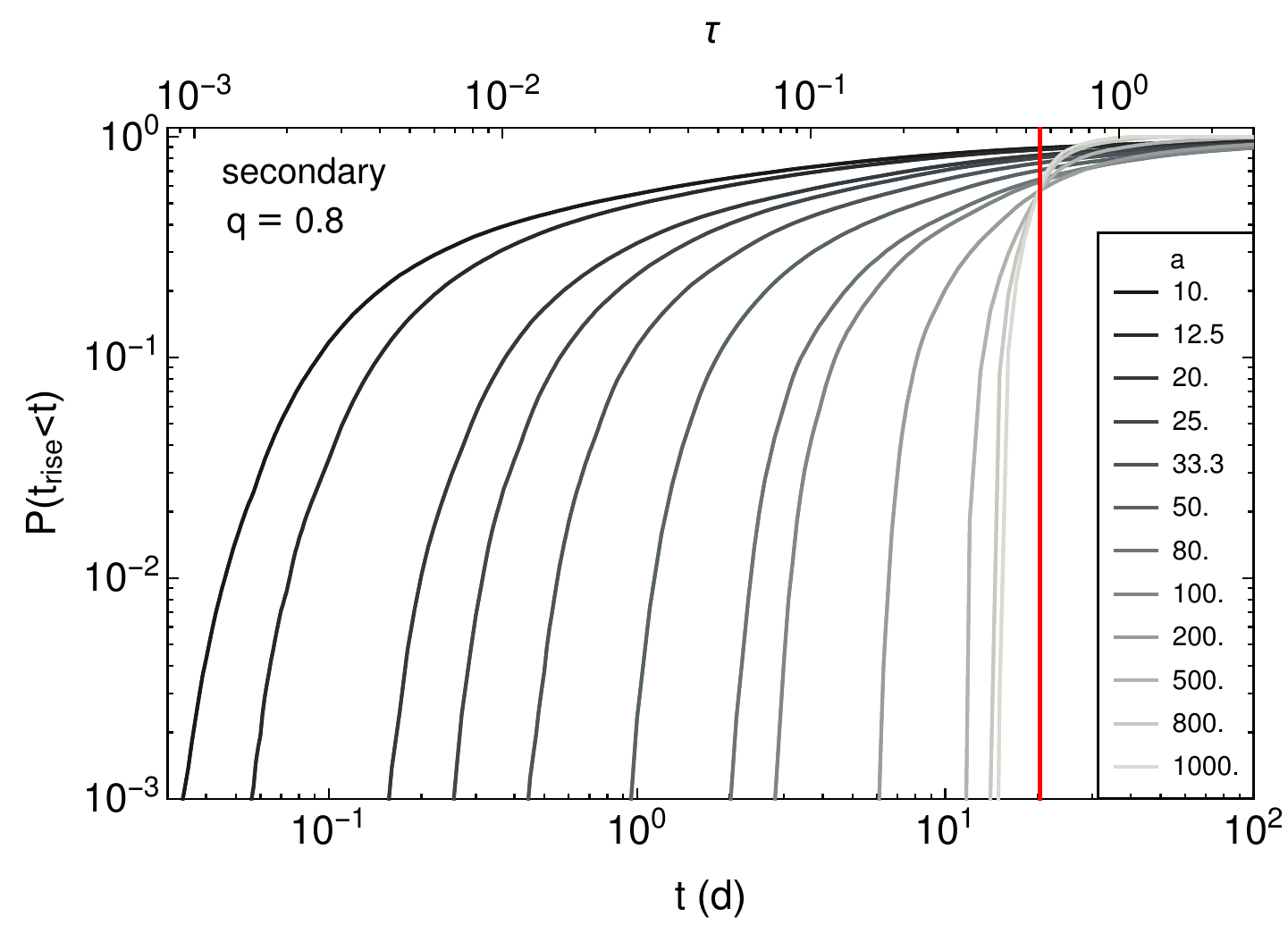}}
\caption{The probability that the bound debris of the disrupting black hole will produce a rise time $t_\textrm{rise} < t$ (d) for a primary mass $M_1 = 10^6 M_\odot$ and different $a$. The maximum probability is $1$. The top panels show the results for $q = 0.05$, the center ones for $q = 0.2$, and the bottom ones for $q = 0.8$. The left panels show disruptions by the primary and the right ones show those by the secondary. The black line marks the rise time $t_\textrm{rise} = \tau_0 (3\sqrt{3} - 5^{3/4})/5^{3/4}$ for a TDE from a star with $\varepsilon_c = 0$ (parabolic orbit, half of the debris is bound) disrupted by the appropriate black hole.}
\label{fig:ptrise1appendix}
\end{figure*}

\begin{figure*}
\centering
\subfloat{\includegraphics[width=0.49\textwidth]{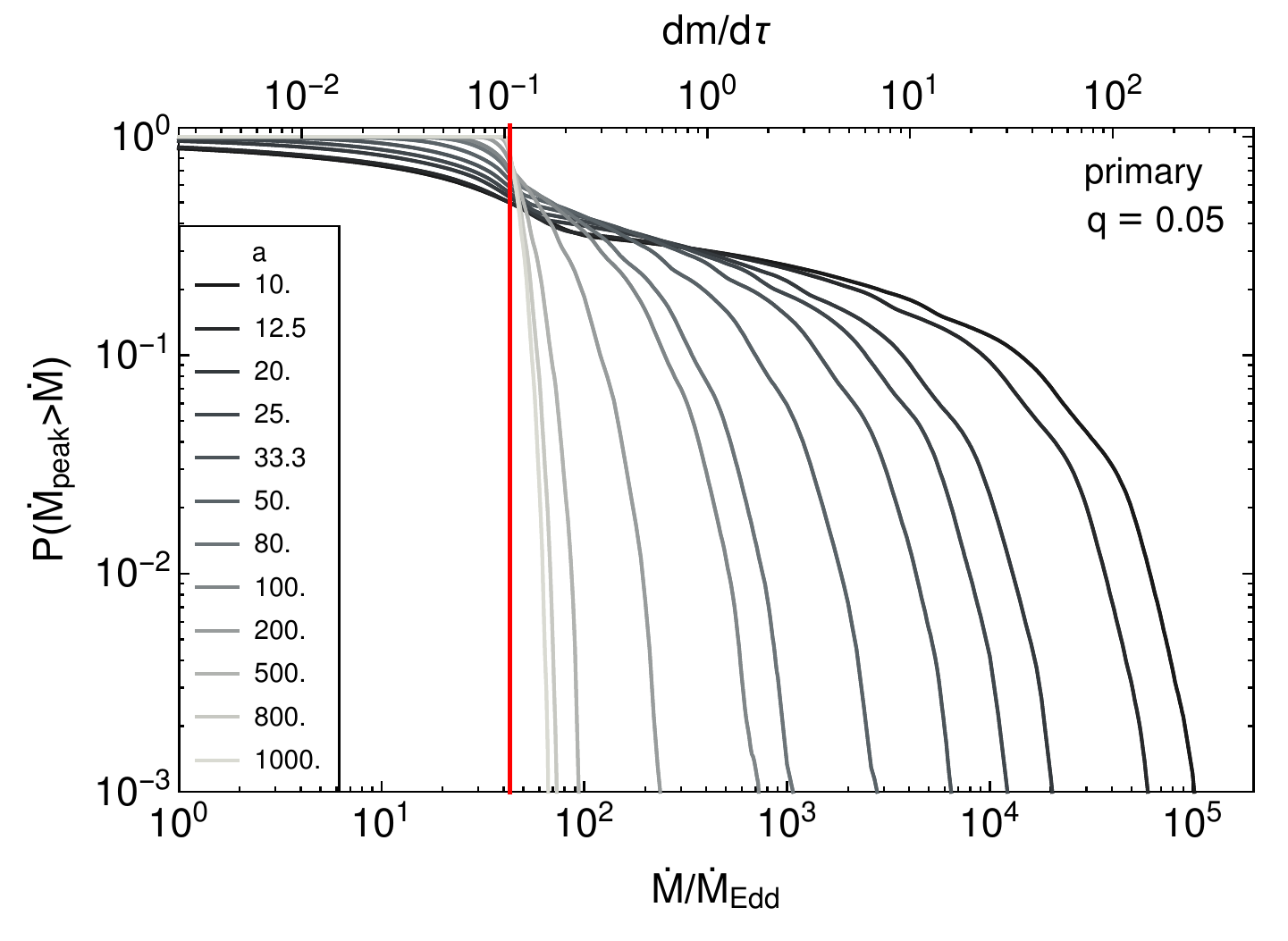}}\hfill
\subfloat{\includegraphics[width=0.49\textwidth]{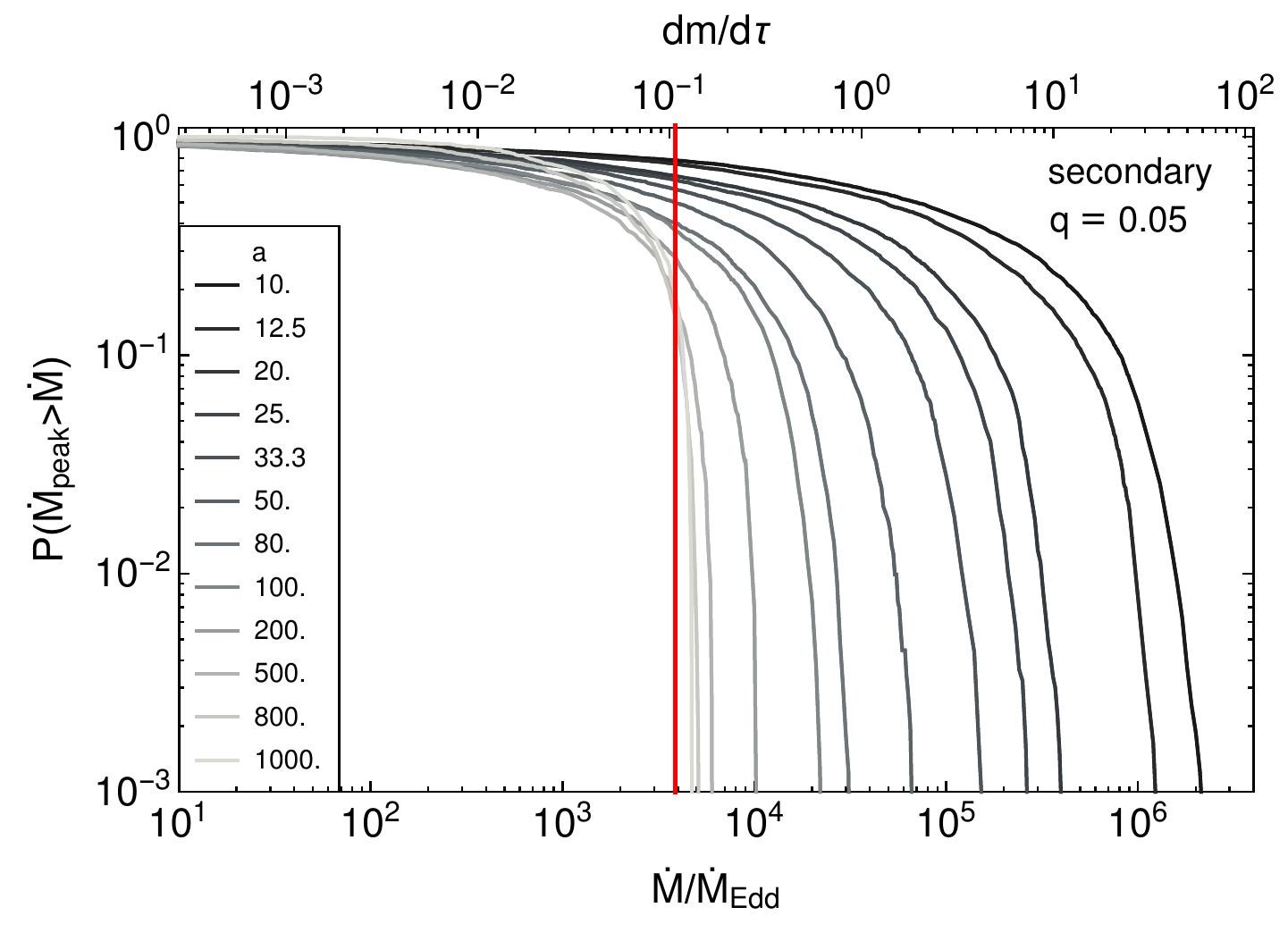}}\\
\subfloat{\includegraphics[width=0.49\textwidth]{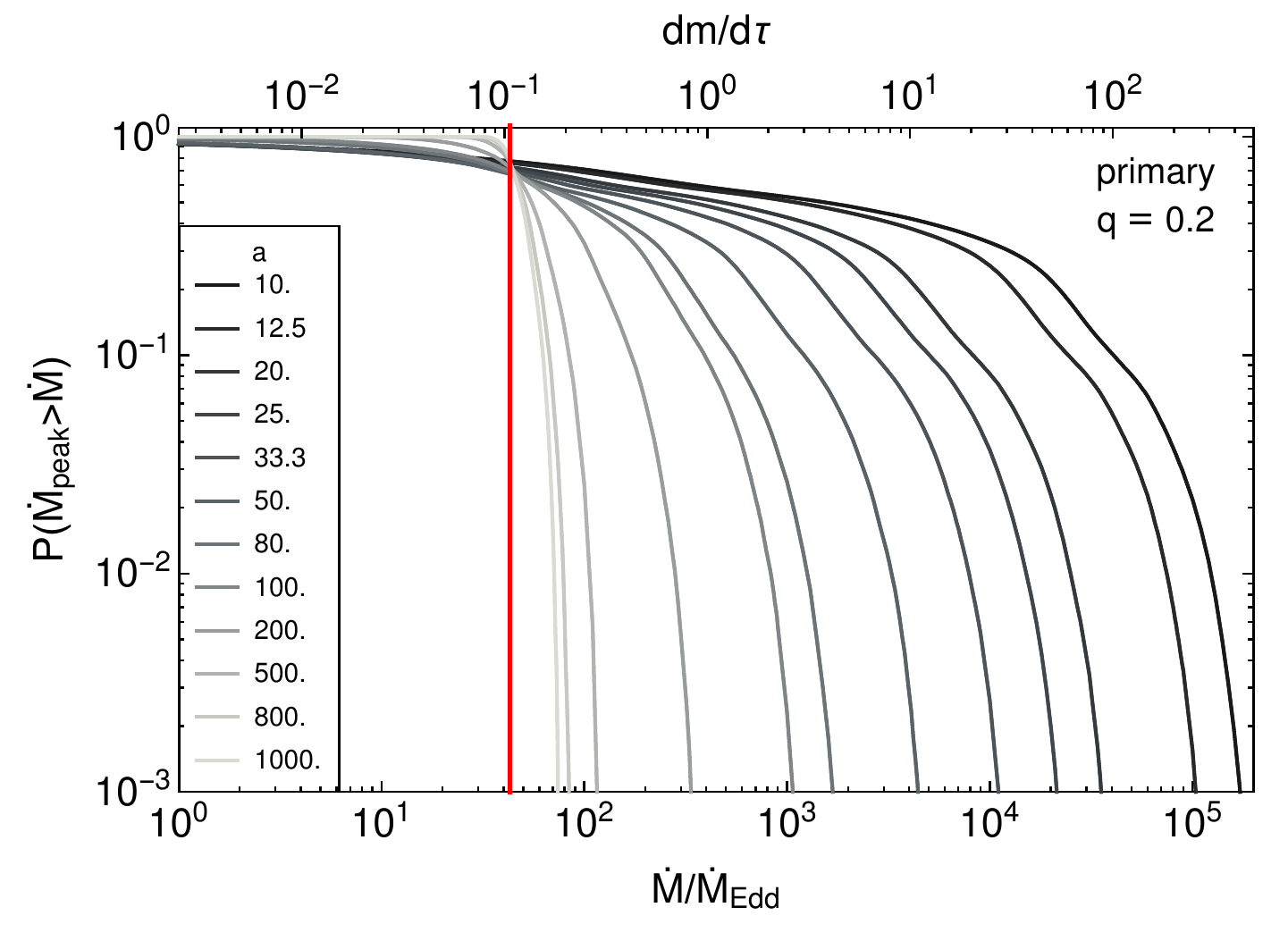}}\hfill
\subfloat{\includegraphics[width=0.49\textwidth]{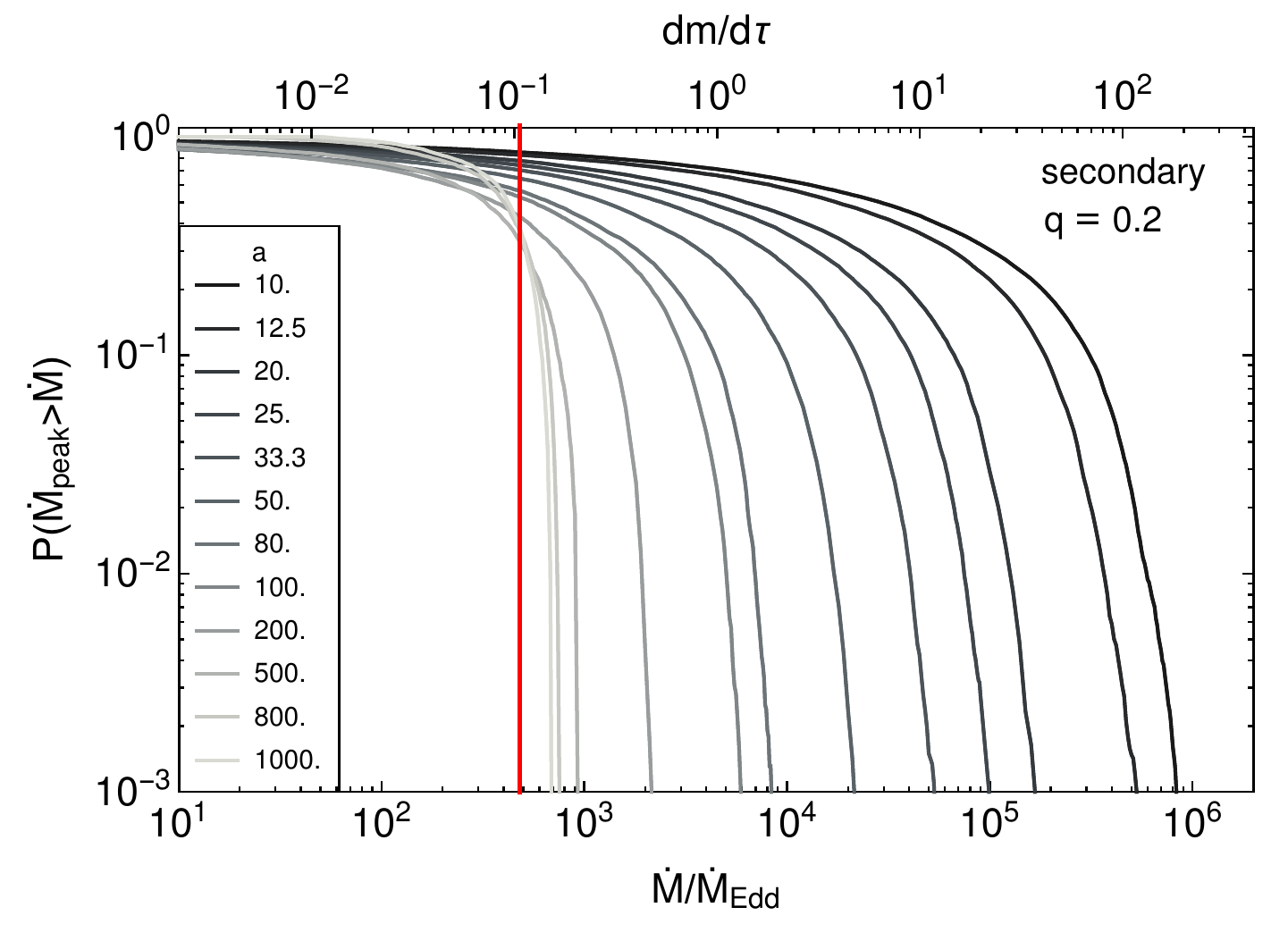}}\\
\subfloat{\includegraphics[width=0.49\textwidth]{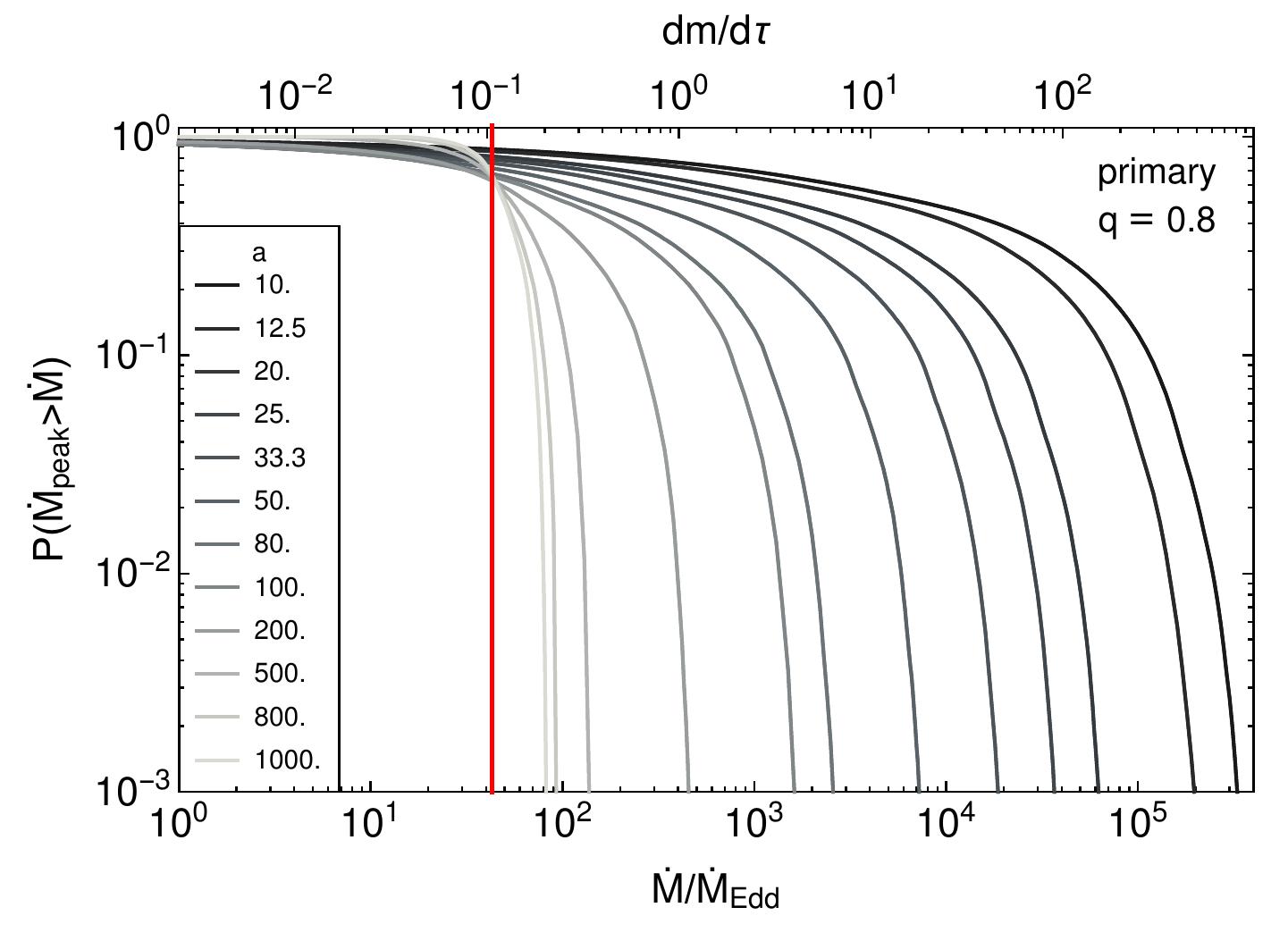}}\hfill
\subfloat{\includegraphics[width=0.49\textwidth]{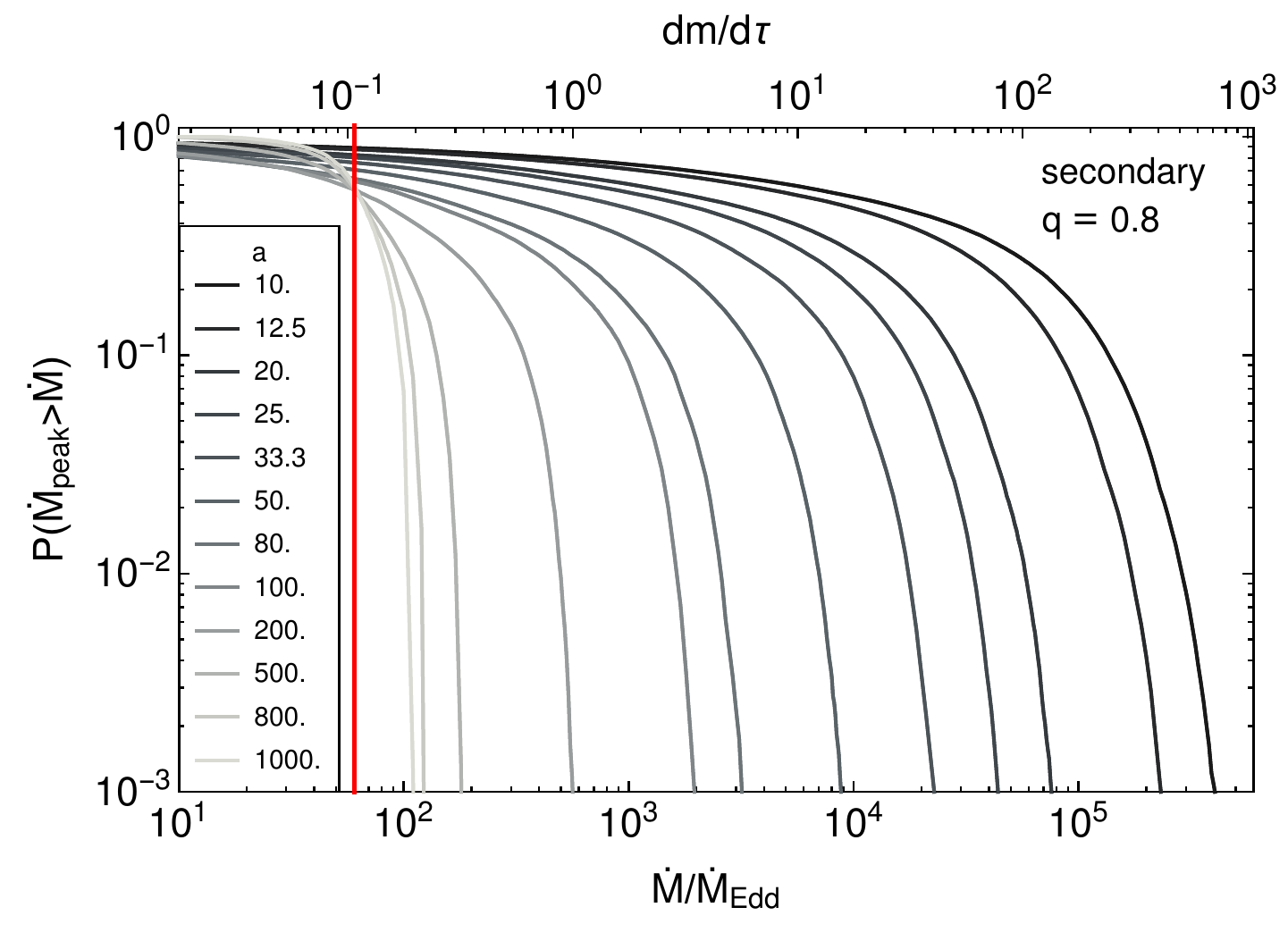}}
\caption{The probability that the bound debris of the disrupting black hole will produce a peak fallback rate $\dot{M}_\textrm{peak} / \dot{M}_\textrm{Edd} > \dot{M} / \dot{M}_\textrm{Edd}$ for a primary mass $M_1 = 10^6 M_\odot$ and different $a$. The maximum probability is $1$. The top panels show the results for $q = 0.05$, the center ones for $q = 0.2$, and the bottom ones for $q = 0.8$. The left panels show disruptions by the primary and the right ones show those by the secondary. The black line shows the peak return rate for a TDE from a star with $\varepsilon_c = 0$ (parabolic orbit, half of the debris is bound) disrupted by the appropriate black hole.}
\label{fig:pmdotpeak1appendix}
\end{figure*}

%%%%%%%%%%%%%%%%%%%%%%%%%%%%%%%%%%%%%%%%%%%%%%%%%%

% Don't change these lines
\bsp	% typesetting comment
\label{lastpage}
\end{document}